\documentclass[preprint,12pt]{elsarticle}

\usepackage{graphicx}
\usepackage{subcaption}
\usepackage{amssymb}
\usepackage{amsmath}
\usepackage{comment}
\usepackage{color}
\usepackage{soul}
\usepackage{booktabs}
\usepackage{tabularx}
\usepackage{siunitx}
\usepackage{algorithm}
\usepackage{algcompatible}
\usepackage[left=1in, right=1in]{geometry}

\newcommand{\br}{\mathbf{r}}

\begin{document}

\begin{frontmatter}



\title{A three-dimensional numerical scheme for modeling three-phase contact line pinning using Smoothed Particle Hydrodynamics}


\author[]{Subrat Kumar Nayak, Amitabh Bhattacharya, and Prapanch Nair}

\affiliation[]{organization={Department of Applied Mechanics, Indian Institute of Technology Delhi},
            addressline={Hauz Khas}, 
            city={New Delhi},
            postcode={110016}, 
            country={India}}

\begin{abstract}
In several capillary dynamics experiments, the liquid domain is confined by pinning the three-phase contact line along a sharp edge or a discontinuity on the substrate. Simulating the dynamics of pinned droplets can offer    valuable insights into capillary flow phenomena involving wetting of inhomogeneous substrates. However, Eulerian multi-phase flow solvers are usually not able to directly implement pinning of three-phase contact lines.
We present the implementation of a model for pinning the contact line of a liquid along an arbitrary pinning curve on the substrate, in an updated Lagrangian, meshless flow solver based on the smoothed particle hydrodynamics (SPH) method. We develop the pinning model for a continuum surface force scheme and assume a free surface for the liquid-gas interface. We validate the model against several capillary dynamics experiments involving pinned three phase contact lines with fixed and dynamic substrates to demonstrate the robustness and accuracy of the solver.
\end{abstract}


\end{frontmatter}

\section{Introduction}
\begin{figure}[h!]
\centering
\includegraphics[width=0.7\textwidth]{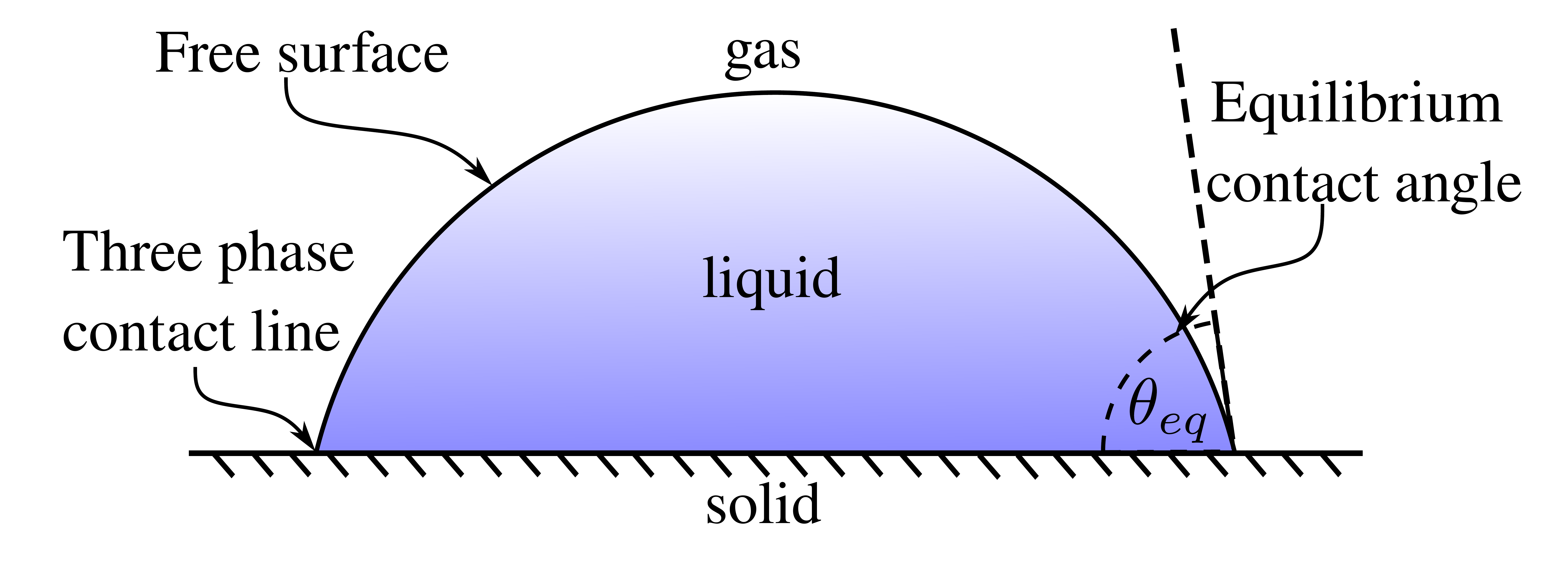}
\caption{Schematic of a liquid droplet on a substrate}
\label{fig:intro_schematic}
\end{figure}

When a liquid droplet is brought in contact with a solid substrate, it assumes a shape that balances the force of cohesion (within the liquid), adhesion (between the liquid and the solid), and external forces (e.g. gravity), resulting in a bound \emph{wetted} region on the substrate. The line enclosing the wetted region, where the substrate, the liquid and the ambient gas phases meet, is called the ``three-phase contact line" or, concisely, the contact line of the droplet. On a smooth substrate, the liquid-air interface assumes a contact angle  $\theta_{eq}$ at equilibrium (Fig. \ref{fig:intro_schematic}). Any departure of  the instantaneous droplet contact angle from  its  equilibrium value may result in the motion of the three-phase contact line. However, if the surface is heterogeneous \cite{robbins1987contact, pomeau1985contact, cox1983spreading}  in its physical or chemical properties, the motion of the contact line may cease, resulting in pinning,  as a result of local changes in surface energy. Pinned contact lines are extremely significant in several industrial processes and are extensively studied in experiments. Several engineering processes that rely on capillary flow instabilities encounter contact line pinning at substrate discontinuities. For example, the manufacturing of fibers, foams and emulsions, wicking systems, and dyeing and coating processes, encounter contact line pinning scenarios and can greatly benefit from numerical simulations. However, numerical implementation of pinning of three-phase contact lines has remained a challenge.

The physics of the wetting (and the dewetting) line has been of interest for more than half a century, thanks to major applications in many engineering processes such as spray coating techniques \cite{aziz2015spray}, oil recovery \cite{babadagli2002dynamics}, and printing techniques such as inkjet printing \cite{calvert2001inkjet}.  For gaining insight into the flow dynamics in these processes, several fundamental experimental studies were performed with a pinned contact line in order to limit the liquid's wetted area. For instance, in the study of dynamics of a sessile drop \cite{deepu2014oscillation} on an oscillating substrate the contact line of droplet  remains pinned  throughout the oscillation period for glycerol, while it partially de-pins for a less viscous liquid like water. Similarly, stretching of a liquid bridge held between parting substrates were studied by \citet{zhang1996nonlinear} where the pinning of the liquid bridge was engineered using a surface discontinuity along the desired line. Pinning is also observed in the studies of evaporation of liquid droplets held on substrates  with patterned trench  \cite{gleason2014microdroplet}, forced polymeric rings
 \cite{kalinin2009contact}, and chemical patterning  \cite{li2016pinning, li2021precise}. Pinning of the three phase contact line is quite significant in industrial processes such as inkjet printing, 3D printing, and microfluidic devices where  pinning of the three phase contact line enhances the stability of liquid bridges.

While numerical techniques for the simulation of capillary flows have been an active area of research due to the industrial relevance, simulating pinning along an arbitrary curve has remained a challenge, thanks to inherent difficulties in traditional macroscopic capillary models.
For modelling the surface tension force at a macroscopic level in numerical techniques (such as the finite volume method) the surface tension force is represented by a spatially varying force density that is applied only at the interface region. In this method, known as the continuum surface force (CSF) model, the magnitude of the volumetric surface tension force  \cite{brackbill1992continuum} depends on the product of the surface tension coefficient and the local mean curvature when surface tension remains constant. The volumetric force is directed in along the gradient of a color function representing the phases. Wetting phenomena is typically modelled in CSF \cite{afkhami2009mesh,tryggvason2011direct} by imposing a static contact angle at the three phase contact line and modifying the direction of the surface unit normal at the contact line region, accordingly  \cite{brackbill1992continuum, breinlinger2013surface, blank2024surface}. For capturing contact angle dynamics in Eulerian mesh based simulation techniques, contact angles are computed as a function of the contact line velocity (often using empirical relations) \cite{bogdanov2022level} and imposed at the contact line region. Pinning of the contact line, another scenario where contact angle changes dynamically, is achieved in these Eulerian mesh based techniques by exploiting sharp edges on the substrate \cite{ferdowsi2012combined,sunder2013numerical,  pathak2023three}. At the substrate edge, pinning of the contact line is achieved by employing Gibb's theory  \cite{gibbs1906scientific}. 
In some Eulerian simulations, pinning is also achieved by setting an approximate discontinuity in static contact angle across a region, such as that attempted in a 2D axisymmetric test case, using the VoF method in \cite{sakakeeny} to find the natural frequency of oscillating drops pinned on a substrate. 
Pinning is also implemented in the finite element method  using a large value for a fitting parameter \cite{dodds2012dynamics} in the linearized molecular kinetic model for wetting \cite{blake2006physics} and is employed in the simulation of non-axial stretching of liquid bridges.  In another work employing the finite element method in 2D axisymmetric configuration, \cite{dettmer2006computational} employed a pinning model based on the force balance at the pinning line. Thus, in Eulerian, mesh-based multiphase simulation methods, pinning is simulated either at cusps along the substrate or using phenomenological static contact angle functions. On the other hand, in Lagrangian meshless methods, where force balance at the contact line can be naturally implemented, a model for pinning the contact line along a given curve on an arbitrary substrate has not been developed to the best of our knowledge. 

Owing to its Lagrangian nature, dynamic variation in contact angles is naturally achieved in the Smoothed Particle Hydrodynamics method \cite{nair2018dynamic, blank2024surface}. In a recent work, wetting phenomena at the free surface of liquid was modeled using a modified CSF model in the SPH method \cite{blank2024surface}. The CSF model has been implemented by several researchers in SPH solvers in the past in two phase flow scenarios \cite{morris2000simulating, adami2012generalized}, along with models for wetting  \cite{breinlinger2013surface, blank2024surface}. To achieve a force balance at the contact line, the normal vector is modified based on the desired equilibrium contact angle, along which the capillary force is directed. While this allows for a several dynamic wetting phenomena, such as contact angle hysteresis and pinning at cusps in the substrates to be modeled accurately \cite{blank2024surface, breinlinger2013surface}, pinning the contact line along a specified curve has not been developed so far.
We propose a generalised model for pinning of a moving contact line along a given curve on a substrate.
To achieve this, the 
surface unit normal at the three-phase contact line are modified based on the distance from the pinning curve, so that the  interfacial forces at the contact line stay balanced,
making the target contact angle a function of  the position of the contact line with respect to the given pinning curve.
The paper is organized as follows. In section \ref{sec:sph} we discuss the SPH schemes being used away from the pinning curve. In section \ref{sec:contact} we describe the SPH scheme used to implement contact-line pinning at the specified pinning curve. In section \ref{sec:res}, we validate the solver against several canonical flow setups, and in section \ref{sec:concl}, we discuss the major conclusions from this work.

\section{\label{sec:sph} Smoothed particle hydrodynamics scheme}
In Smoothed Particle Hydrodynamics, a continuum conservation law is approximated at interpolation points (called particles) with associated properties such as mass, velocity, pressure and other transported quantities. The particle velocities and positions are evolved in time using an updated Lagrangian formulation.
\subsection{SPH Interpolation}
The value of a scalar field $A$ at a position $\mathbf{r}$ can be written as the convolution of $A$ with the Dirac $\delta$ function as:
\begin{align}
A(\mathbf{r}) &= \int_\Omega A(\mathbf{r}')\delta(|\mathbf{r-r}'|)\,d\mathbf{r}' .
\label{convolution-product}
\end{align}
Here, $\delta(|\mathbf{r} -\mathbf{r}'|)$ is the Dirac delta function ($\int_\Omega \delta(|\mathbf{r} -\mathbf{r}'|)d\mathbf{r}'=1$). In SPH, $\delta(|\mathbf{r} -\mathbf{r}'|)$ is discretized as $W(|\br-\br'|,h)$, where $h$ is a discretization parameter called the smoothing length, which defines a compact support for $W$. Then, Eq. (\ref{convolution-product}) can be approximated as:
\begin{eqnarray}
A^I(\mathbf{r}) &=& \int_{\Omega} A(\mathbf{r}')W(|\mathbf{r-r}'|,h)\,d\mathbf{r}'. \label{convolution-product-kernel} 
\end{eqnarray}
On the domain discretized by particles (interpolation points), the above integral can be approximated as a quadrature:
\begin{align}
A^s(\mathbf{r}_a) & = \sum_{b\in \mathcal{N}_a} \frac{m_b}{\rho_b} A(\mathbf{r}_b) W(|\mathbf{r}_a-\mathbf{r}_b|,h).
\label{summation} 
\end{align}
Here $m_b$ and $\rho_b$ are the mass and density  of a particle at position $\br_b$, and $\mathbf{r}_a$ is the position of the particle under consideration. The kernel influences a spherical neighborhood $\mathcal{N}_a$ with radius $2h$ around particle $a$. The normalized Wendland $C^2$ kernel function is used to represent $W$, owing to its superior spectral properties \cite{dehnen2012improving}:
\begin{align}
\displaystyle
W(|\mathbf{r}_a - \mathbf{r}_b|,h) &=
\left\{ 
  \begin{array}{ l l }
    \frac{21}{16 \pi} \frac{1}{h^3} \left( 1- \frac{{r}_{ab}}{2h} \right)^4 \left (\frac{2{r}_{ab}}{h} +1 \right) & \textrm{if } |{r}_{ab}| \leq 2h \\
    0                 &  \textrm{otherwise}.
  \end{array}
\right.
\label{wendland_kernel}
\end{align}
Here ${r}_{ab} =|\mathbf{r}_{ab} |=|\mathbf{r}_{a} -\mathbf{r}_{b} |$ is the distance between particles $a$ and $b$. 
Several approximations exist in literature \cite{violeau2012fluid} for representation of vector differential operators. The approximations used in this work are presented in the subsequent section.

\subsection{SPH model for free surface flow}
Assuming a dynamically inert gas phase, the gas-liquid flow is modelled as a single liquid phase with free surface, governed by the incompressible Navier-Stokes equations:
\begin{align}
\nabla \cdot \mathbf{u} &= 0, \label{eq:continuity}\\
\frac{d \mathbf{u}}{d t} &=\frac{1}{\rho}\left[{-{\nabla p} + {\nabla \cdot (2 \mu \mathbf{D})}}\right] + \mathbf{f}^b + \mathbf{f}^s, \label{eq:momentum} 
\end{align}
where $\mathbf{u}$ is the velocity vector field, $p$ is the pressure field and  $\rho$ is the density field, $\mu$ is the dynamic viscosity, $\mathbf{D}$ is the deviatoric part of the deformation rate tensor (i.e. $\mathbf{D}=(\nabla\mathbf{u}+\nabla\mathbf{u}^T)/2$), ${\mathbf{f}^b}$ is the externally applied body force per unit mass, ${\mathbf{f}^s}$ is the volumetric equivalent of the surface tension force per unit mass acting at the free surface of the fluid, and $t$ is the time.
A discretized version of Eq. (\ref{eq:momentum}) can be applied to an SPH particle $a$ as a linear momentum balance equation, in which the total particle acceleration is equated to the sum of accelerations due to the pressure gradient ($\mathbf{f}^p_a$), viscosity ($\mathbf{f}^\mu_a$), surface tension ($\mathbf{f}^s_a$) and the body force density ($\mathbf{f}^b_a$), and can be written as
\begin{equation}
\frac{d \mathbf{u}_a}{d t} = \mathbf{f}^p_a + \mathbf{f}^\mu_a + \mathbf{f}^b_a + \mathbf{f}^s_a. \label{eq:total_acceleration} 
\end{equation}
This equation is solved using the Incompressible Smoothed Particle Hydrodynamics (ISPH) method \cite{nair2014improved,nair2018dynamic,nair2015volume,aly2025heat} in which a divergence-free velocity field is imposed to obtain the pressure \cite{cummins1999sph} field and time integration is performed using a predictor-corrector scheme \cite{cummins1999sph}. 
At given a time step $n$, an intermediate position ($\mathbf{r}_a^*$) is predicted using the velocity $\mathbf{u}_a^n$.
\begin{equation}
    \mathbf{r}_a^* = \mathbf{r}_a^n + \mathbf{u}_a^n \Delta t.
    \label{eq:predictor_pos}
\end{equation}
At the predicted position, an intermediate velocity field $\mathbf{u}_a^*$ is computed using all the acceleration components except the pressure gradient $\mathbf{f}^p_a$, as follows:
\begin{equation}
    \mathbf{u}_a^* = \mathbf{u}_a^n + \left(\mathbf{f}^\mu_a(\mathbf{r}_a^*) + \mathbf{f}^b_a(\mathbf{r}_a^*) + \mathbf{f}^s_a(\mathbf{r}_a^*)\right)\Delta t,
    \label{eq:predictor_vel}
\end{equation}
in which $\mathbf{f}^\mu_a(\mathbf{r}_a^*)$ represents the acceleration due to viscous force computed for particle $a$ at position $\mathbf{r}_a^*$. 
Imposing the divergence-free velocity condition for the next time step ($\nabla \cdot \mathbf{u}^{n+1}|_a=0$), leads to the following pressure Poisson equation (PPE) for $p_a^{n+1}$:
\begin{equation}
    \left[\nabla\cdot\left(\frac{\nabla p^{n+1}}{\rho}\right) \right]_a= \left[\frac{\nabla \cdot \mathbf{u}^*}{\Delta t}\right]_a.
    \label{eq:pressure_poisson_eq}
\end{equation}
The discretized pressure gradient is used to project the predicted velocity field onto a divergence-free space at the time step ($n+1$), as follows:
\begin{equation}
    \mathbf{u}_a^{n+1}=\mathbf{u}_a^{*}-\left[\frac{1}{\rho_a}\nabla p^{n+1} \right]_a \Delta t.
    \label{eq:corrector_vel}
\end{equation}
The position of the corresponding particle is updated taking the average particle velocities at time step $n$ and $n+1$ \cite{cummins1999sph}, so that:
\begin{equation}
    \mathbf{r}_a^{n+1} = \mathbf{r}_a^{n} + \left(\frac{\mathbf{u}_a^{n+1}+\mathbf{u}_a^{n}}{2}\right)\Delta t.
    \label{eq:corrector_pos}
\end{equation}
Numerical stability is imposed by setting the time step as follows \cite{tartakovsky2016pairwise}:
\begin{equation}
\Delta t = 0.125 \times \text{min}\left(\frac{\Delta_x}{\mid\mathbf{u}_a\mid},\frac{\Delta_x^3 \rho}{2\pi\sigma}, \frac{\rho \Delta_x^2}{\mu} \right).
\label{eq:time_step_cfl}
\end{equation}
where $\Delta_x$ is the initial spacing between the particles.
The following SPH approximations are chosen for the spatial derivatives in Eq. \ref{eq:momentum}.
For the viscous term (assuming constant viscosity), traditional Laplacian operators \cite{morris1997modeling} introduce spurious numerical errors due to the deficiency in the kernel at the free surface \cite{schwaiger2008implicit}. We instead use the Laplacian operator introduced by \citet{schwaiger2008implicit}, which is consistent with free-surface flows, and can be stated as follows: 
\begin{equation}
\begin{split}
(\nabla \cdot 2\mu D)_a = &\frac{\text{tr}(\Gamma_{ab}^{-1})}{n}  \left[ \sum_b \frac{m_b}{\rho_b} \left(\mu_a + \mu_b\right)\left(\mathbf{u}_b - \mathbf{u}_a\right)\frac{\mathbf{r}_{ab}\cdot \nabla W_{ab}}{\mathbf{r}_{ab}\cdot \mathbf{r}_{ab}} \right. \\ 
 &\left. -\left(2\mu_a \nabla \mathbf{u}_a\right)\cdot \left(\sum_b \frac{m_b}{\rho_b} \nabla W_{ab}\right) \right],
\label{schwaiger_diffusion}
\end{split}
\end{equation}
where $\text{tr}(\Gamma_{ab}^{-1})$ is the trace of the inverse of the tensor $\Gamma_{ab}$, given as: 
\begin{eqnarray}
    \Gamma_{ab} = \sum_b \frac{m_b}{\rho_b}\left(\frac{\mathbf{r}_{ab}\cdot \nabla W_{ab}}{\mathbf{r}_{ab}\cdot\mathbf{r}_{ab}+\epsilon}\right)\left(\mathbf{r}_{ab}\otimes\mathbf{r}_{ab}\right).
    \label{gamma_ab}
\end{eqnarray}
 In our simulations, only gravity is imposed as an external body force density, so that ${\mathbf{f}^b_a}$ is a constant over space and time.
For particles with full kernel support, discretization of the left hand side \cite{cummins1999sph} and right hand side \cite{monaghan2005smoothed} of the PPE (Eq. \ref{eq:pressure_poisson_eq}) is carried out as follows:
\begin{equation}
    \sum_b \frac{m_b}{\rho_b} \frac{4}{\rho_a+\rho_b}\left(p_a-p_b\right)F_{ab} = -\sum_b\frac{m_b}{\rho_b}\frac{\mathbf{u}_{ab}^*\cdot\nabla W_{ab}}{\Delta t}.
    \label{eq:PPE}
\end{equation}
For the particles near free surface ($a\in \Omega_{fs}$), for which full kernel support is not present, the ambient pressure experienced at the free surface is $p_a^o$. For such particles, the following modified version of Eq. \ref{eq:PPE} is used to calculate $p_a^{n+1}$:
\begin{equation}
 \begin{split}
    \left(p_a-p_a^o\right)\underbrace{\sum_b \frac{m_b}{\rho_b}\frac{4}{\rho_a+\rho_b}F_{ab}}_\beta - \sum_b\frac{m_b}{\rho_b}\frac{4p_b}{\rho_a+\rho_b}F_{ab} \\= \sum_b\frac{m_b}{\rho_b}\left(-\frac{\mathbf{u}_{ab}^*\cdot\nabla W_{ab}}{\Delta t}-\frac{4p_a^o}{\rho_a+\rho_b}F_{ab}\right).
 \end{split}
    \label{eq:PPE_bc}
\end{equation}
The term $\beta$ remains constant for a given uniform discretization and replaces the leading diagonal terms in the resulting linear system.
The discretized PPE is solved using an iterative solver such as the BiCGSTAB \cite{van1992bi} or GMRES \cite{saad1986gmres}.
 At the free surface, a Dirichlet boundary condition for pressure needs to be set for solving PPE (Eq. \ref{eq:pressure_poisson_eq}). 
 The application of the Dirichlet boundary condition for pressure at the free surface for the PPE in Eq. \ref{eq:pressure_poisson_eq} is introduced in \citet{nair2014improved} and is elaborated in \citet{nair2018dynamic},\citet{blank2023cmame} and \citet{blank2024surface}. In contrast to its attribution in works such as  \citet{violeau2016smoothed} and \citet{bavsic2022}, the method does not use the Shepherd filter \cite{shepard1968two} to identify free surface particles. Instead a threshold on the value of the Shepherd filter 
 \begin{equation}
     S_a = \sum_b \frac{m_b}{\rho_b}W_{ab}
\label{shepard_filter} 
\end{equation} is used only to identify particles that are in proximity of the free surface.  Thereafter, the Dirichlet boundary condition for pressure in PPE is applied semi-analytically by modifying the diagonal terms in the resulting linear system. 


The threshold on the Shepard filter is set as:
\begin{eqnarray}
a \in
\left\{ 
  \begin{array}{ c l l }
     \Omega_{fs} & \quad \textrm{if } S_a \leq 0.95, \\
     \Omega_{b} & \quad \textrm{otherwise}.
  \end{array}
\right.
\label{shephered_threshold}
\end{eqnarray}
Here, $\Omega_{fs}$ represents the region in proximity to the free surface, and $\Omega_{b}$ represents the bulk of the liquid.
The acceleration for the liquid particles \cite{blank2024surface} due to pressure gradient is given as:
\begin{eqnarray}
\mathbf{f}_a^p &=
\left\{ 
  \begin{array}{ c l l } \displaystyle
    -\sum_b m_b \left(\frac{p_a}{\rho_a^2}+\frac{p_b}{\rho_b^2}\right)\nabla W_{ab} \textrm{; } a\in \Omega_{b}, \\ \displaystyle
    -\sum_b m_b \left(\frac{p_b-p_a^o}{\rho_b^2}\right)\nabla W_{ab} \textrm{; } a\in \Omega_{fs}.
  \end{array}
\right.
\label{eq:pressure_acceleration}
\end{eqnarray}
The surface tension force is modeled as a continuum surface force (CSF) at the free surface following \cite{blank2024surface}. In the CSF model, the surface tension force is modeled as a body force on particles in the interface region. The acceleration due to surface tension ($\mathbf{f}_a^s$) in the CSF model \cite{brackbill1992continuum} is given as:
 \begin{equation}
     \mathbf{f}_a^s = \frac{1}{\rho_a} \mathbf{F}_a^s,
     \label{eq:surface_tension_acc}
 \end{equation}
 where
 \begin{equation}
     \mathbf{F}_a^s = \sigma\kappa_a\hat{\mathbf{n}}_a\delta_a^s.
     \label{eq:surface_tension_force}
 \end{equation}
Here, $\mathbf{F}_a^s$ is the surface force per unit volume at the interface expressed as the product of surface tension coefficient ($\sigma$), mean curvature ($\kappa_a$), local unit normal ($\hat{\mathbf{n}}_a$) and the interface thickness scale ($\delta_a^s$) \cite{brackbill1992continuum}.

\begin{figure}[h!]
\begin{center}
    \begin{subfigure}[b]{0.62\textwidth}
    \includegraphics[width=\textwidth]{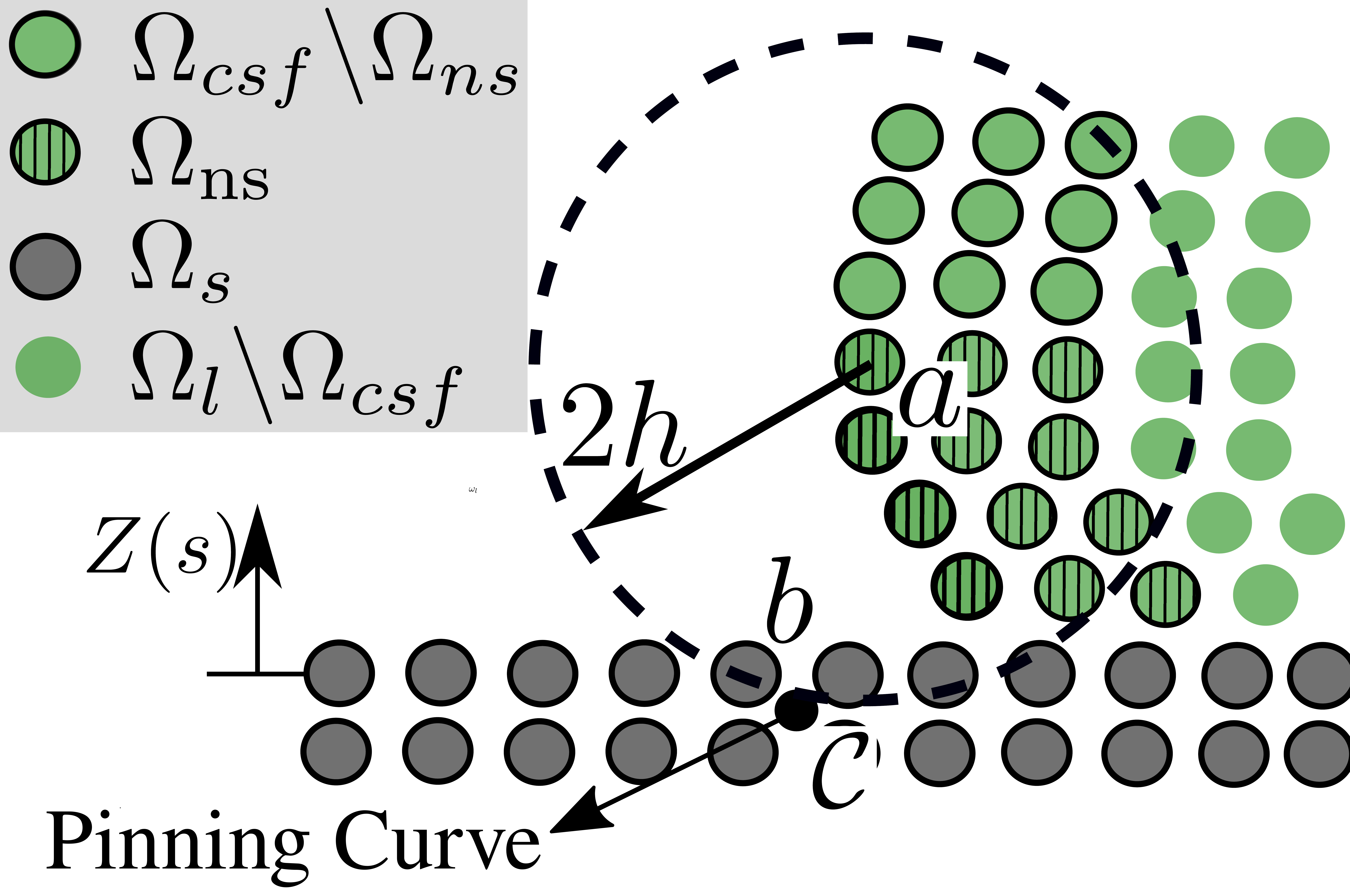}
    \caption{}
    \label{fig:define_regions}
\end{subfigure}
\begin{subfigure}[b]{0.37\textwidth}
    \includegraphics[width=\textwidth]{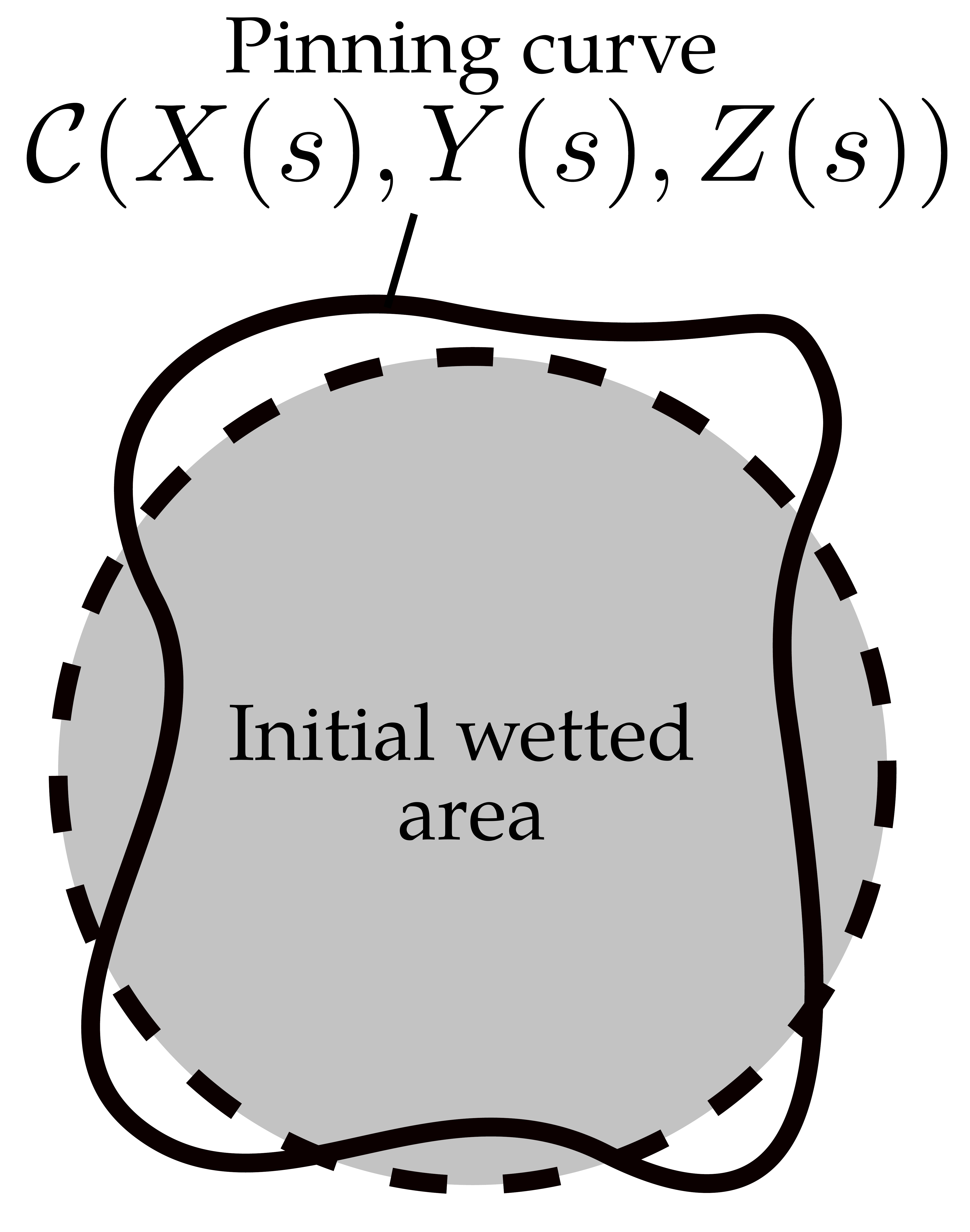}
    \caption{}
    \label{fig:pinning_curve}
\end{subfigure}
  \caption{(a) Cross sectional side view of the neighborhood of three-phase contact line, demarcating the particles into subdomains as required for the pinning model's implementation. The liquid particles ($\Omega_l$) are shaded green and the substrate particles ($\Omega_{s}$) are shaded grey. Among the liquid particles the CSF particles $\Omega_{csf}$ are marked using black outlines. Of these CSF particles ($\in \Omega_{csf}$), the ones within $2h$ distance from the substrate are the near substrate particles ($\in \Omega_{ns}$) marked with a green shade, black outline and vertical lines.   (b) Top view projection of pinning curve, defined as $\mathcal{C}(X(s),Y(s),Z(s))$ (denoted by the black solid line). The dash-lined circle encloses the area on the substrate initially wetted by the liquid. }
  \end{center}
  \label{fig:regionsAndPC}
\end{figure}

 A constant value of gas-liquid surface tension $\sigma$ is assumed throughout this paper. The preliminary unit normal vector in Eq. \ref{eq:surface_tension_force}  is first estimated as
 \begin{equation}
     \mathbf{n}_a = -\sum_b \frac{m_b}{\rho_b} \nabla W_{ab},
     \label{eq:surface_normal}
 \end{equation}
and then smoothed to reduce spatial fluctuations: 
 \begin{equation}
  \tilde{\mathbf{n}}_a= \frac{1}{S_a}\sum_b \frac{m_b}{\rho_b}\mathbf{n}_b W_{ab}.
  \label{eq:surface_normal_avg}
\end{equation}
The computation of the surface normal is restricted to the CSF particles $\Omega_{csf}$ (Fig.  \ref{fig:define_regions}), for which $|\tilde{\mathbf{n}}_a|<\epsilon$ is satisfied, where $\epsilon = \lambda/h$. A value of $\lambda=0.1$ is used to ensure sufficient resolution of the interface region by the SPH particles. 

The filtered preliminary normal $\hat{\tilde{\mathbf{n}}}_a$ is then normalized, so that
\begin{equation}
  \hat{\tilde{\mathbf{n}}}_a=\begin{cases}
    0,& \text{if } |\mathbf{n}_a| \leq \epsilon\\
    \tilde{\mathbf{n}}_a/|\tilde{\mathbf{n}}_a|, & \text{otherwise.}
  \end{cases}
  \label{eq:surface_unit_normal}
\end{equation}
The surface unit normal ($\hat{\mathbf{n}}_a$) obtained from the gradient of color function is hereon denoted as:
\begin{equation}
    \label{eq:n_csf} 
    \hat{\mathbf{n}}_a = \hat{\tilde{\mathbf{n}}}_a.
\end{equation}
The interface thickness scale $\delta_a^s$ for any liquid particle ($a$) is given \cite{morris2000simulating} as:
\begin{equation}
    \delta_a^s = |\mathbf{n}_a|.
    \label{eq:delta_function}
\end{equation}
The mean curvature of the interface is given by the divergence of the surface unit normal ($\hat{\mathbf{n}}_a$) \cite{monaghan1992smoothed, morris2000simulating}:
\begin{equation}
    \kappa_a = \sum_b \frac{m_b}{\rho_b}(\hat{\mathbf{n}}_b - \hat{\mathbf{n}}_a) \cdot \nabla W_{ab}.
    \label{eq:curvature}
\end{equation}
For accurate estimation of the surface tension force, it is necessary that the curvature is estimated using an accurate divergence operator. Therefore, a modified kernel gradient $\tilde{\nabla}W_{ab} = \mathbf{L}_a \cdot \nabla W_{ab}$ is used \cite{oger2007improved}, in which 
\begin{equation}
    \mathbf{L}_a = \left(\sum_b \frac{m_b}{\rho_b}\mathbf{r}_{ab} \otimes \nabla W_{ab}\right)^{-1}
    \label{eq:correction_matrix}
\end{equation}
is the correction matrix. The curvature estimate is re-scaled as \cite{breinlinger2013surface, blank2024surface}:
\begin{equation}
    \tilde{\kappa}_a = \frac{1}{2} \sum_b \frac{m_b}{\rho_b} \left(\hat{\mathbf{n}}_b - \hat{\mathbf{n}}_a\right) \cdot \tilde{\nabla}W_{ab},
    \label{eq:curvature_corrected}
\end{equation}
so that Eq. \ref{eq:surface_tension_acc} is modified as
\begin{equation}
    \mathbf{f}_a^s = 2\frac{\sigma_a}{\rho_a}\tilde{\kappa}_a\hat{\mathbf{n}}_a\delta_a^s.
\end{equation}
The constant $2$ is replaced \cite{blank2023surface} with a function of the Shepard filter that increases the accuracy and robustness of the CSF scheme:
\begin{equation}
    \mathbf{f}_a^s = \left( 1+\frac{1}{S_a^n} \right)\frac{\sigma_a}{\rho_a}\tilde{\kappa}_a\hat{\mathbf{n}}_a\delta_a^s.
    \label{eq:surface_tension_Blank}
\end{equation}
The Shepard filter($S_a^n$) is calculated by considering only particles belonging to $\Omega_{csf}$. 

\subsection{Wall boundary conditions}
The liquid particles very close to the substrate should satisfy the non-slip and non-penetration condition \cite{violeau2012fluid}. The solid walls are discretized as dummy particles, which have properties identical to the particles in the liquid phase. Following the approach taken by \citet{adami2012generalized}, the pressure Poisson equation is also solved for the solid wall particles in order to satisfy the zero pressure gradient in the normal direction and to ensure no penetration of the liquid particle into the solid region. In the discretization of the PPE for solid particles, only the neighbourhood liquid particles are considered, so that, for $\mathbf{r}_a\in\Omega_s$,
\begin{equation}
    \sum_{b\in \Omega_l}\frac{m_b}{\rho_b}\frac{4}{\rho_a+\rho_b}\left(p_a^{n+1}-p_b^{n+1}\right)F_{ab}=\sum_{b\in\Omega_l} -\frac{m_b}{\rho_b}\frac{\mathbf{u}_{ab}^* \cdot \nabla W_{ab}}{\Delta t}.
    \label{eq:PPE_solid}
\end{equation}

\section{\label{sec:contact} Contact line pinning scheme}
In this section, the contact line pinning along an arbitrary curve at the substrate is discussed. In the discussion below, we will use a local Cartesian coordinate system ($x,y,z$) for which $z$ is normal to the surface of the substrate and $z=Z_s$ defines the plane corresponding to the surface of the surface.
\subsection{Defining the contact line region}
A contact line region consisting of liquid particles in the proximity of both the free surface ($\Omega_{csf}$) and the substrate is first defined; thus, any CSF particle $a$ that is in the neighborhood of substrate particle $b$ will be treated as contact line particles, as shown in Fig. \ref{fig:define_regions}.
The three phase contact line region($\Omega_{ns}$) can be defined as
\begin{equation}
\Omega_{ns} = \{a:\,\textrm{ s.t. } \mathcal{N}_a\in\Omega_s \textrm{ and } a\in\Omega_{csf}\}.
\end{equation}
Here, $\mathcal{N}_a$ is the spherical neighbourhood of $a$ with radius $2h$.
A pinning curve is defined as $\mathcal{C}(X(s),X(s),Z(s))$ in the substrate, as shown in Fig. \ref{fig:pinning_curve}. The curve is positioned just below the top layer of the substrate, for example, for a horizontal substrate, $(z(s)=Z_s-\Delta_x/2)$, $z(s)$ is the vertical position of the pinning curve, $Z_s$ is the vertical coordinate of the surface of the substrate, and $\Delta_x$ is the initial particle spacing. The surface unit normals in $\Omega_{ns}$ are corrected to account for the pinning force, as discussed in the section below.

\subsection{Treatment of the pinning normal}
\begin{figure}[h!]
\centering  
  \centering
  \includegraphics[width=3in]{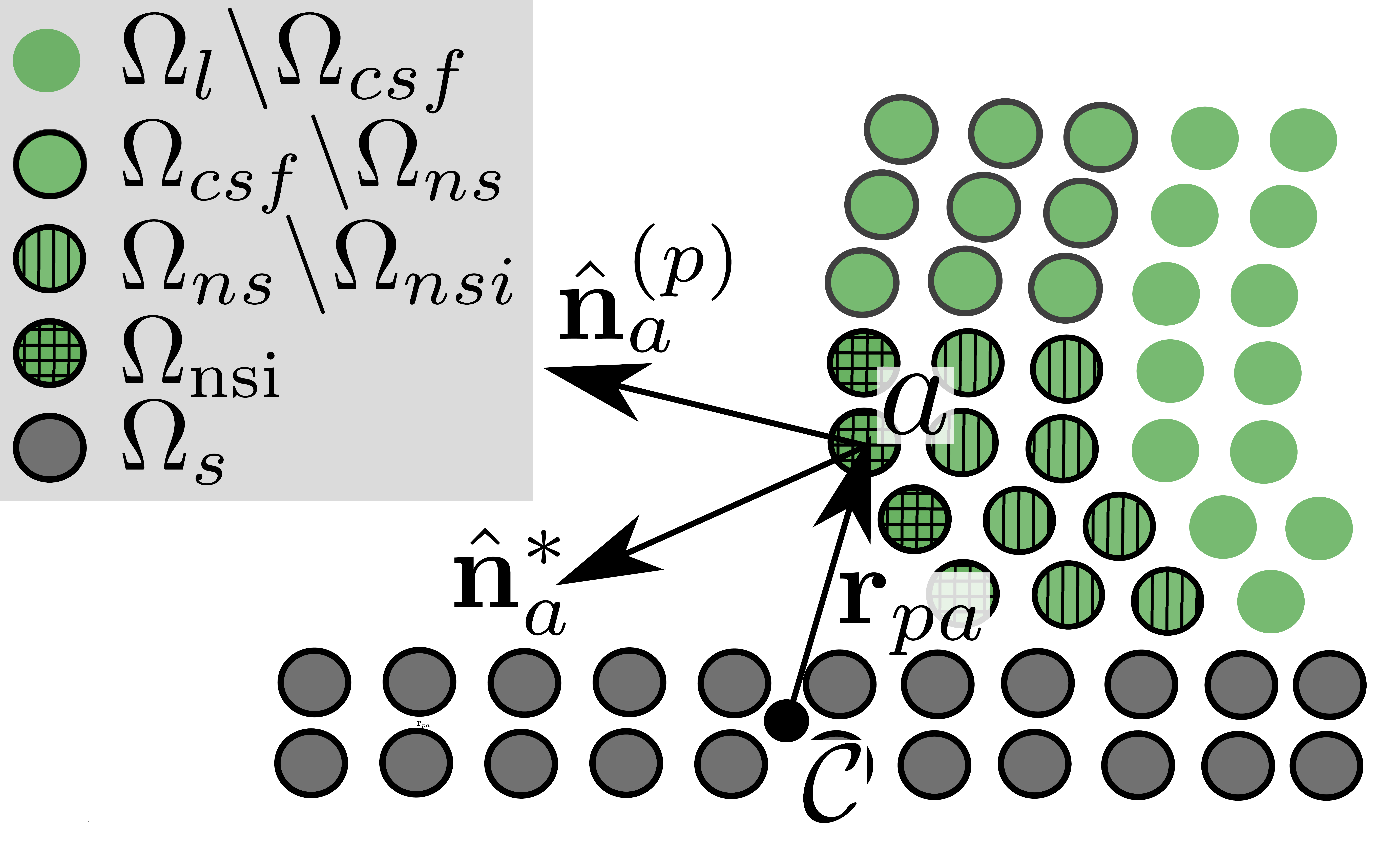}
  \caption{Treatment of the surface unit normal at the three-phase contact line region. This figure is a modified figure of Fig.\ref{fig:define_regions}, where the regions are shown inside larger particles for a clearer representation of the normal treatment. The shaded brown-colored particles are the free surface particles in the contact line region, $\mathcal{C}$ is the pinning curve, ${\mathbf{r}_{pa}}$ is the position vector of $a$ with respect to nearest point on the pinning curve $\mathcal{C}$, $\hat{\mathbf{n}}_a^*$ is the intermediate surface unit normal vector calculated from color function, and $\hat{\mathbf{n}}_p$ is the pinning unit normal obtained from treatment.} 
  \label{fig:pinning_treatment}
\end{figure}

The surface unit normals ($\{\hat{\mathbf{n}}_a:\, a\in\Omega_{ns} \}$) at the three-phase contact line are updated to account for pinning. We therefore refer to $\hat{\mathbf{n}}_a^*$ as the intermediate value of the unit normal vectors calculated using the  gradient of the color function for particles within $\Omega_{ns}$, after the application of Eqs. \ref{eq:surface_normal}, \ref{eq:surface_normal_avg}, and \ref{eq:surface_unit_normal}. A unit pinning normal $\hat{\mathbf{n}}_a^{(p)}$ for particles $a\in\Omega_{ns}$ can be defined as follows:
\begin{eqnarray}
\hat{\mathbf{n}}_a^{(p)}=\frac{\hat{\mathbf{n}}_a^* - (\hat{\mathbf{r}}_{pa} \cdot\hat{\mathbf{n}}_a^*)\hat{\mathbf{r}}_{pa}}{|\hat{\mathbf{n}}_a^* - (\hat{\mathbf{r}}_{pa}\cdot\hat{\mathbf{n}}_a^*)\hat{\mathbf{r}}_{pa}|}.
\label{eq:pin_treatment} 
 \end{eqnarray}
Here, $\hat{\mathbf{r}}_{pa}$ is the unit vector along the shortest distance $\mathbf{r}_{pa}$ from pinning curve $\mathcal{C}$ to particle $a$, shown in Fig. \ref{fig:pinning_treatment}.  The unit pinning normal $\hat{\mathbf{n}}_p$ is orthogonal to $\hat{\mathbf{r}}_{pa}$ and directs the pinning force at the contact line, so that the effective surface tension force is hydrophilic if the contact line is approaching the pinning curve (i.e. before the liquid wets the pinning curve) and hydrophobic once the contact line has crossed the pinning curve (i.e. after the liquid wets the pinning curve).

A numerical experiment we performed on a sessile drop showed unstable results for extreme contact angles ($\theta_i>155^\circ$ and $\theta_i<30^\circ$). Hence, we  limit the range of the instantaneous contact angle $\theta_i$ to these values to avoid instabilities in dynamic scenarios. Limits on the unit pinning normals ($\hat{\mathbf{n}}_p$) are set according to:
\begin{equation}
    \hat{\mathbf{n}}_a^{(p)}=\begin{cases} \displaystyle
    \hat{\mathbf{n}}_{\theta_s =30},& \text{if } \theta_i \leq 30^o, \\ \displaystyle
    \hat{\mathbf{n}}_{\theta_s =155},& \text{if } \theta_i \geq 155^o, \\ \displaystyle
    \hat{\mathbf{n}}_a^{(p)}, & \text{otherwise.}
  \end{cases}
  \label{eq:pinning_normal_limittaion}
\end{equation}
Here, $\theta_i$ is calculated as:
\begin{equation}
    \theta_i = \sin^{-1}\left({\hat{\mathbf{r}}_{pa}}\cdot \hat{k}\right),
    \label{eq:instantaneous_CA}
\end{equation}
Where, $\hat{\mathbf{n}}_{\theta_s}$ is the normal calculated based on the equilibrium contact angle $\theta_s$ as proposed by \citet{blank2024surface} as:
\begin{equation}
    \hat{\mathbf{n}}_{\theta_s} = \frac{f^s \hat{\mathbf{n}}_a^* + (1-f^s)\hat{\mathbf{n}}_a^{\infty}}{|f^s \hat{\mathbf{n}}_a^* + (1-f^s)\hat{\mathbf{n}}_a^{\infty}|},
    \label{eq:de/wetting_normal}
\end{equation}
here, $f^s$ is the linear blending function \cite{blank2024surface}, and $\hat{\mathbf{n}}_a^{\infty}$ is the normal calculated based on the equilibrium contact angle$(\theta_s)$, calculated as:
\begin{equation}
    \hat{\mathbf{n}}_a^{\infty} = \hat{\mathbf{t}}_a^{sf} \sin\theta_s - \hat{\mathbf{n}}_a^{sf}\cos\theta_s.
    \label{eq:CA_normal}
\end{equation}
Here, $\hat{\mathbf{t}}_a^{sf}$ is the smoothed normalized tangent vector between the solid and liquid, and $\hat{\mathbf{n}}_a^{sf}$ is the smoothed normalized normal vector pointing from the solid to the liquid particle.
The above scheme for $\hat{\mathbf{n}}_a^{(p)}$ leads to depinning of the three-phase contact line for $\theta_s\leq 30^\circ$ or $\theta_s\geq 155^\circ$.

\begin{figure}
    \begin{subfigure}[b]{0.495\textwidth}
        \centering
        \includegraphics[width=1.1\textwidth]{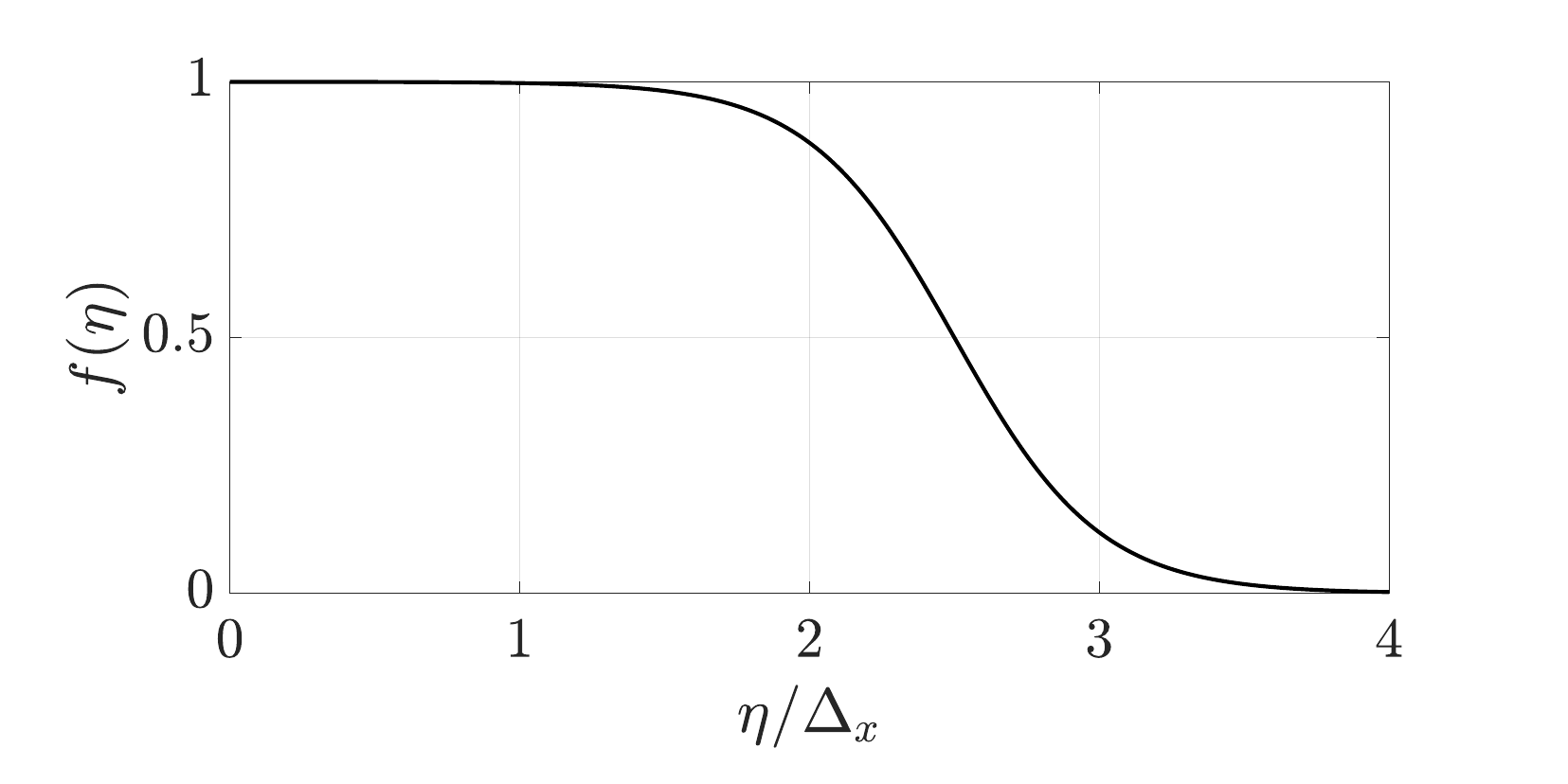}
        \caption{}
        \label{fig:blending_function}
    \end{subfigure}
    \begin{subfigure}[b]{0.495\textwidth}
        \includegraphics[width=0.957\textwidth]{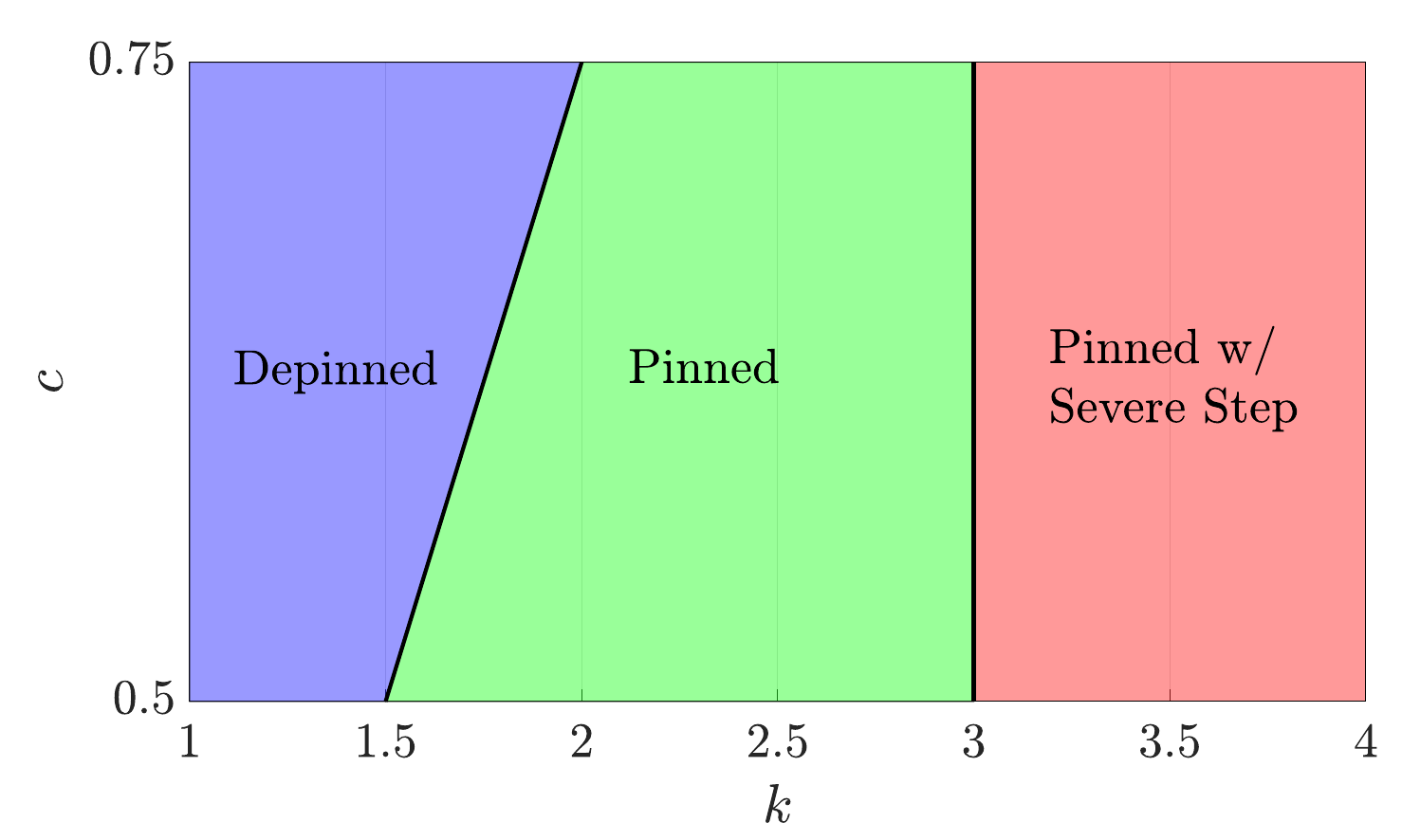}
        \caption{}
        \label{fig:k_c_sensitivity}
    \end{subfigure}
  \caption{(a) Variation of blending function($f(\eta)=(1/2)\left( 1-\tanh \left( \frac{\eta-k\Delta_x}{c\Delta_x} \right) \right)$ over $\eta\,\Delta_x \in[0,\,4]$. (b) Phase plot showing different pinning regimes: depinned (blue), pinned (green) and pinned with appearance of a severe surface step (red), for various values of $k$ and $c$.}
\end{figure}

Very close to the pinning line, where $|\mathbf{r}_{pa}|$ is small, a small change in $\mathbf{r}_{pa}$ can lead to large changes in $\hat{\mathbf{n}}_a^{(p)}$. Divergence of $\hat{\mathbf{n}}_a^{(p)}$ can therefore result in unphysically large values of curvature near the contact line. This issue is partly avoided by placing the pinning curve at a distance $\Delta_x/2$ below the surface of the substrate. 

Additionally, the pinning normal is only assigned to particles $\Omega_{nsi}$ that are present at the free surface  (Fig. \ref{fig:pinning_treatment}), which satisfy the  condition
\begin{equation}
        \Omega_{nsi} = \{a: \, a \in \Omega_{ns} \cap\Omega_{csf} \textrm{ and } S_a \leq 0.9 \}.
\end{equation}
The particles in the contact line area that don't belong to the free surface are assigned a unit normal via the following scheme:
\begin{equation}
    \hat{\mathbf{n}}_a^{(p)}  = \frac{\sum_{b \in \Omega_{nsi}} b\frac{m_b}{\rho_b}\hat{\mathbf{n}}_b^{(p)} W_{ab} }{|\sum_{b \in \Omega_{nsi}} \frac{m_b}{\rho_b}\hat{\mathbf{n}}_b^{(p)} W_{ab}|} \qquad \forall\, a \in \Omega_{ns} \backslash \Omega_{nsi}.
    \label{eqn:inside_sph_sum}
\end{equation}

The above treatment ensures that $\hat{\mathbf{n}}_a^{(p)}$ is extrapolated smoothly from particles $\Omega_{nsi}$ at the free surface to the interior particles $\Omega_{ns} \backslash \Omega_{nsi}$ near the free surface. However, using a separate treatment for unit normals in $\Omega_{ns}$ and $\Omega_{csf}\backslash\Omega_{ns}$ can lead to a discontinuity between the values of $\hat{n}_p$ and $\hat{n}_a$ along the direction normal to the surface, which can in turn lead to spurious values of curvature. To reduce the jump in curvature, the unit normal is blended over a \emph{blending region}, $z-Z_s\in[0,\,4 \Delta_x]$,
as
\begin{equation}
    \hat{\mathbf{n}}_a^{(p)} = \frac{f(z_a-Z_s) \hat{\mathbf{n}}_a^{(p)} + (1-f(z_a-Z_s)) \mathbf{\hat{n}}_a^*}{|f(z_a-Z_s) \hat{\mathbf{n}}_a^{(p)} + (1-f(z_a-Z_s)) \mathbf{\hat{n}}_a^*|},
    \label{eq:blending}
\end{equation}
where $z_a$ is the $z$ coordinate of particle $a$ and $\mathbf{\hat{n}}_p$ is the version of unit normal with pinning treatment presented in Eq. \ref{eqn:inside_sph_sum}. The blending function $f$ is defined as 
\begin{equation}
    f(\eta) = \frac{1}{2}\left[ 1-\tanh \left( \frac{\eta-k\Delta x}{c\Delta_x} \right) \right],
    \label{eq:blending_function}
\end{equation}
where $k=2.5$ and $c=0.5$ are chosen constant values. The effect of different values of $k \in [1, 4]$ and $c\in [0.5, 0.75]$, observed through numerical experimentation, is  shown in Fig. \ref{fig:k_c_sensitivity}.

The above blending function ensures a smooth transition for the unit normal $\hat{\mathbf{n}}_a^{(p)}$ within $\Omega_{ns}$ as  shown in Fig. \ref{fig:blending_function}. Finally, $\hat{\mathbf{n}}_a^{(p)}$ is assigned to unit normals of particles in the near-substrate region:
\begin{equation}
\hat{\mathbf{n}}_a=\hat{\mathbf{n}}_a^{(p)}.
\end{equation}
To avoid any discontinuity in the calculation of the mean curvature in the contact line region, the values of unit normals are extrapolated to particles in the substrate ($a \in \Omega_{s}$) as 
\begin{equation}
    \hat{\mathbf{n}}_a = \frac{\sum_{b\in\Omega_{ns}} \frac{m_b}{\rho_b}\hat{\mathbf{n}}_b W_{ab}}{|\sum_{b\in\Omega_{ns}} \frac{m_b}{\rho_b}\hat{\mathbf{n}}_b W_{ab}|} , \,\forall b\in\Omega_{ns}.
    \label{eq:solid_assign_normal}
\end{equation}
In-spite of ensuring a smooth variation of unit normal vectors within the near-surface region via the blending function $f$, the curvature calculated from the unit normals can give rise to large spurious values from the complicated $\nabla \cdot \hat{\mathbf{n}}_a^{(p)}$. 

\begin{figure}
    \centering
    \includegraphics[width=0.5\textwidth]{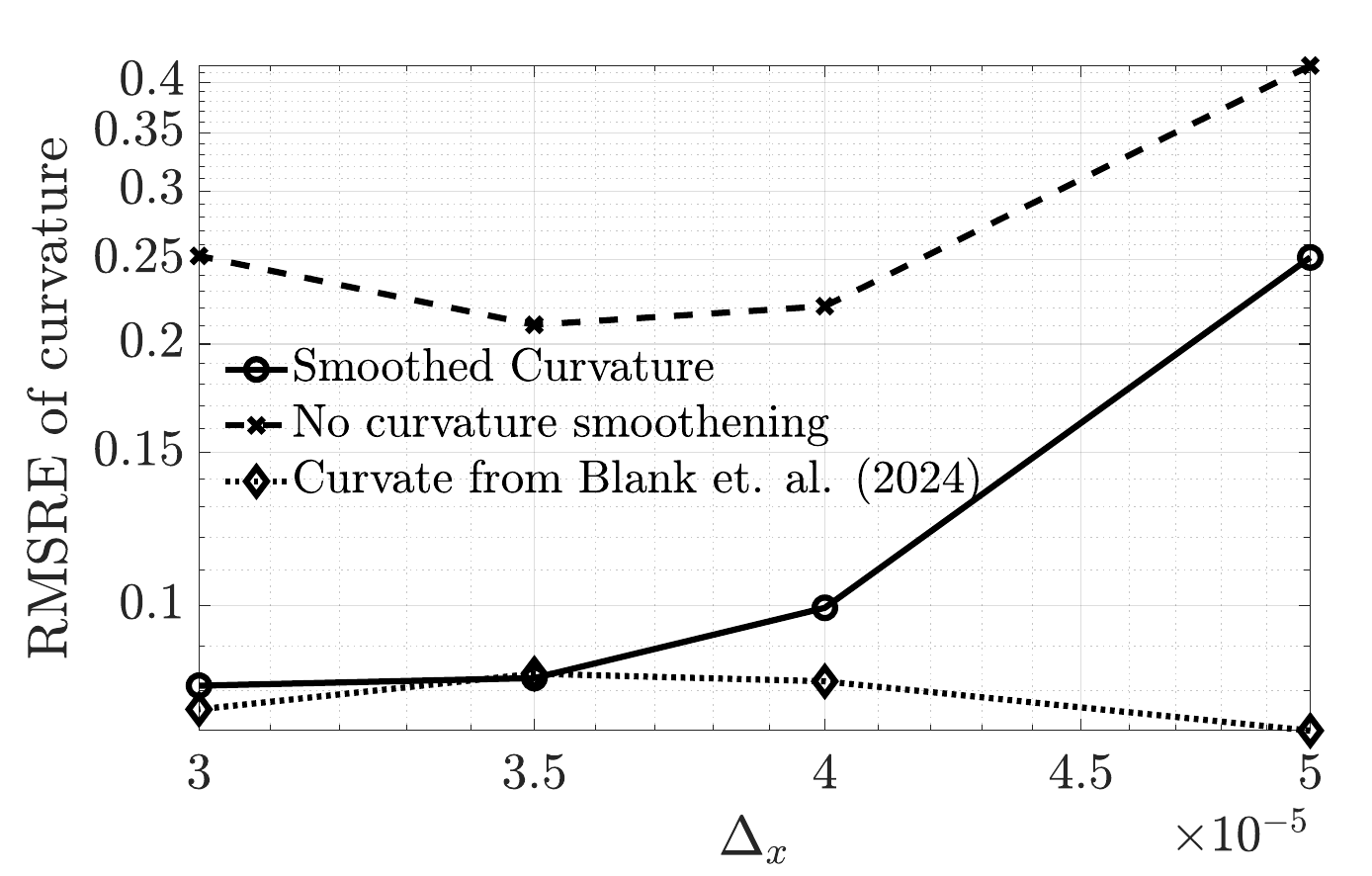}
  \caption{Root mean square relative error of the smoothed curvature in the proposed pinning scheme is compared with the root mean square relative error of the curvature $\kappa_a=\nabla \cdot \hat{\mathbf{n}}_a^{(p)}$, and the curvature obtained from \citet{blank2024surface} for different resolutions.}
  \label{fig:RRMSE_kappa}
\end{figure}

The curvature can be written as
\begin{align}
    \kappa &= \nabla \cdot \frac{f(z_a-Z_s) \hat{\mathbf{n}}_a^{(p)} + (1-f(z_a-Z_s)) \mathbf{\hat{n}}_a^*}{|f(z_a-Z_s) \hat{\mathbf{n}}_a^{(p)} + (1-f(z_a-Z_s)) \mathbf{\hat{n}}_a^*|}.
    \label{eq:kappa_analysis}
\end{align}

To mitigate instabilities due to the spurious values in curvature, we smooth the curvature within the near-surface region only as follows:
\begin{equation} \displaystyle
    \tilde{\kappa}_a = \frac{\sum\limits_{b \in \Omega_{csf}} \frac{m_b}{\rho_b} \kappa_{b}W_{ab}}{\sum\limits_{b \in \Omega_{csf}} \frac{m_b}{\rho_b}W_{ab}}\qquad\forall \, a \in\Omega_{ns},
    \label{eq:smoothKappaCorrection} 
\end{equation}
where $\kappa_b$ is calculated using Eq. \ref{eq:curvature_corrected}. The effect of smoothing the curvature is shown in Fig. \ref{fig:RRMSE_kappa}, where the root mean square relative error of curvature is calculated for the curvature obtained at the contact line of a hemispherical droplet held on a substrate, considering the same radius of the pinning curve with the theoretical curvature ($1/R$). The error in curvature is calculated for a sessile droplet, initialized with a hemispherical shape, in the absence of gravity. The root mean square relative error is calculated for the smoothed curvature, the curvature without smoothing for the pinning scheme, and the curvature calculated for the same droplet with equilibrium contact angle of $90^o$ \citet{blank2024surface}, for different resolutions, after droplet equilibrates at $0.04$ s. It is observed that the curvature obtained is consistent when smoothed.
A pseudo-code for the implementation of the pinning scheme is presented in \ref{app:pseudocode}. The model is implemented only at the near substrate region which requires three additional interaction steps(such as normal computation) in comparison with the wetting force model by \citet{blank2024surface}. Because of the local nature of this treatment, the runtime performance of the whole solver is hardly affected ($<5\% $ increase in runtime).

\subsection{Particle shifting}
In ISPH, the particles tend to move along streamlines, causing anisotropy, results in clustering of particles. Anisotropic spacing results in errors in SPH approximation and leads to wrong identification of interior particles belonging proximity to free surface. We apply the particle motion correction formulation discussed for the finite volume particle method \cite{nestor2008moving}, to avoid local stretching and compression in the particle spacing. The particle shifting method is based on identifying the anisotropic spacing in the liquid region and defining a shifting vector pointing the particle to recover isotropy. The interior liquid particles in our simulation are shifted at every time-step according to:
\begin{equation}
\delta \mathbf{r}_a = C \alpha \mathbf{R}_a,
\label{eq:position_shift}
\end{equation}
here, $\delta \mathbf{r}_a$ is the shift in position by particle $a$, $C$ is a constant set to $0.01$ for the static substrate cases and $0.05$ otherwise. For the test cases where the substrate is in motion, the shifting magnitude $\alpha$ is set as $U_{\max}dt$, where $U_{\max}$ is the maximum velocity of the substrate. For all the other cases $\alpha$ is set as the capillary wavelength ($\sqrt{\sigma/\rho g}$). $\mathbf{R}_a$ is the shifting vector for particle $a$, given as:
\begin{equation}
\mathbf{R}_a = \sum_{b} \frac{\overline{r}_a ^2}{r_{ab}^2} \mathbf{n}_{ab},
\label{eq:shifting_vector}
\end{equation}
where,
\begin{equation}
    \mathbf{n}_{ab} = \frac{\mathbf{r}_b-\mathbf{r}_a}{|\mathbf{r}_b-\mathbf{r}_a|},
    \label{eq:unit_dist_vect}
\end{equation}
and $\overline{r}_a$ is the average spacing of all the neighbours of $a$, given by
\begin{equation}
\overline{r}_a = \frac{1}{N_a} \sum_b^{N_a} r_{ab}.
\label{eq:avg_particle_spacing} 
\end{equation} 
Here,  $r_{ab}$ is the distance between the particle $a$ and its neighbour $b$
and $N_a$ is the cardinal number of $\mathcal{N}_a$.

At the shifted position, any field variable (say $\phi_a$) associated with the particle is interpolated based on the Taylor series as: 
\begin{equation}
    \phi_{a'}=\phi_a + \delta r_{aa'} \cdot (\nabla \phi)_{a} + O(\delta r_{aa'}^2).    
\end{equation}
However, this treatment is avoided for the free surface particles, as the shifting vector at the free surface region is ill-defined.
\section{\label{sec:res} Validation and results}
In this section, we discuss different static and dynamic test cases involving the pinned contact line. In the first test case, the spreading and pinning of sessile and pendant droplets is simulated, in which the circular pinning curve is specified. In the second test case, we compare the profile of a static droplet in equilibrium in the presence of gravity with the solution of the Young-Laplace equation. In the third test case, we consider a complex star-shaped pinning curve onto which a droplet, initialized with a circular contact line, gets pinned.
Next, we validate our simulation against experimental results for droplets pinned on an oscillating substrate.  
We also validate the simulation of axial extension of a pinned liquid bridge and its breakup with experimental data. 
In all the test cases, except those involving pinch off of the liquid domain, we ensure that the minimum radius of curvature in the domain is greater than five times the initial particle spacing. Also, we found that water like viscosity requires much higher spatial resolution and therefore we choose test cases with higher viscosity up to 10 times that of water.   

\subsection{Spreading and pinning of contact line of sessile and pendant droplets}
\begin{figure}[h!]
    \centering
    \begin{subfigure}[b]{0.24\textwidth}
        \centering
        \includegraphics[width=\textwidth]{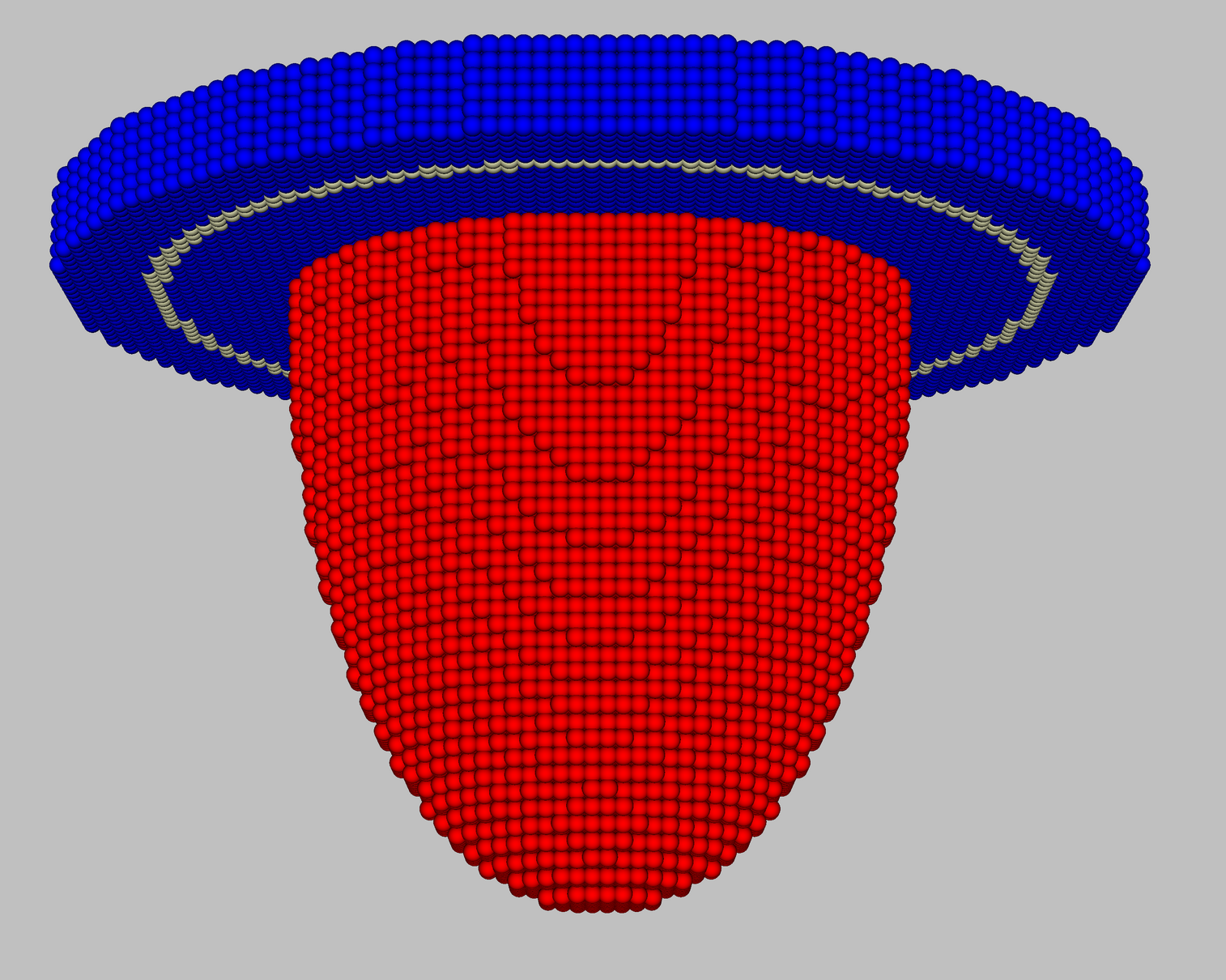}
        \caption{$t^*=0.0$}
        \label{fig:pendant1}
    \end{subfigure}
    \begin{subfigure}[b]{0.24\textwidth}
        \centering
        \includegraphics[width=\textwidth]{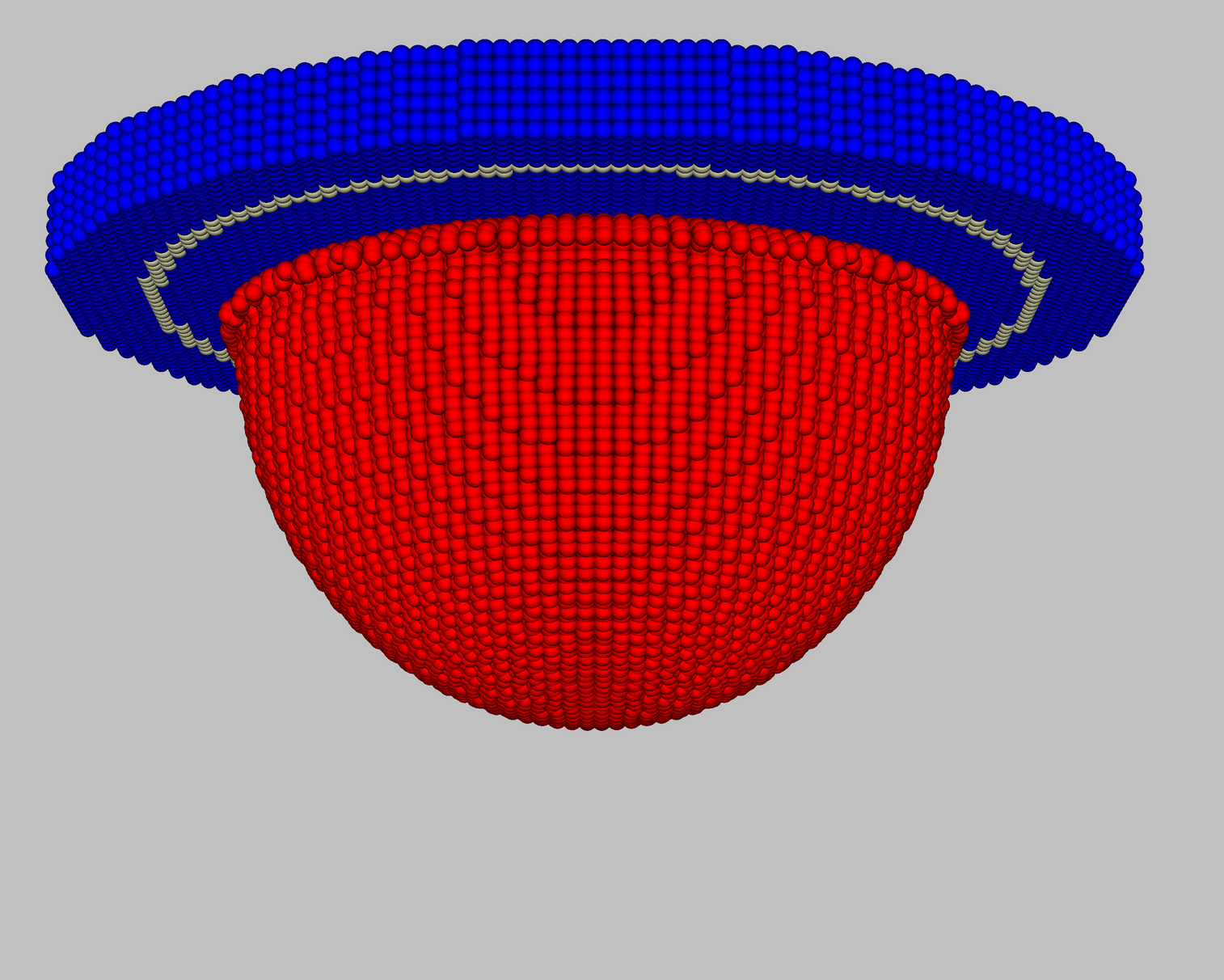}
        \caption{$t^*=0.64$}
        \label{fig:pendant2}
    \end{subfigure}
    \begin{subfigure}[b]{0.24\textwidth}
        \centering
        \includegraphics[width=\textwidth]{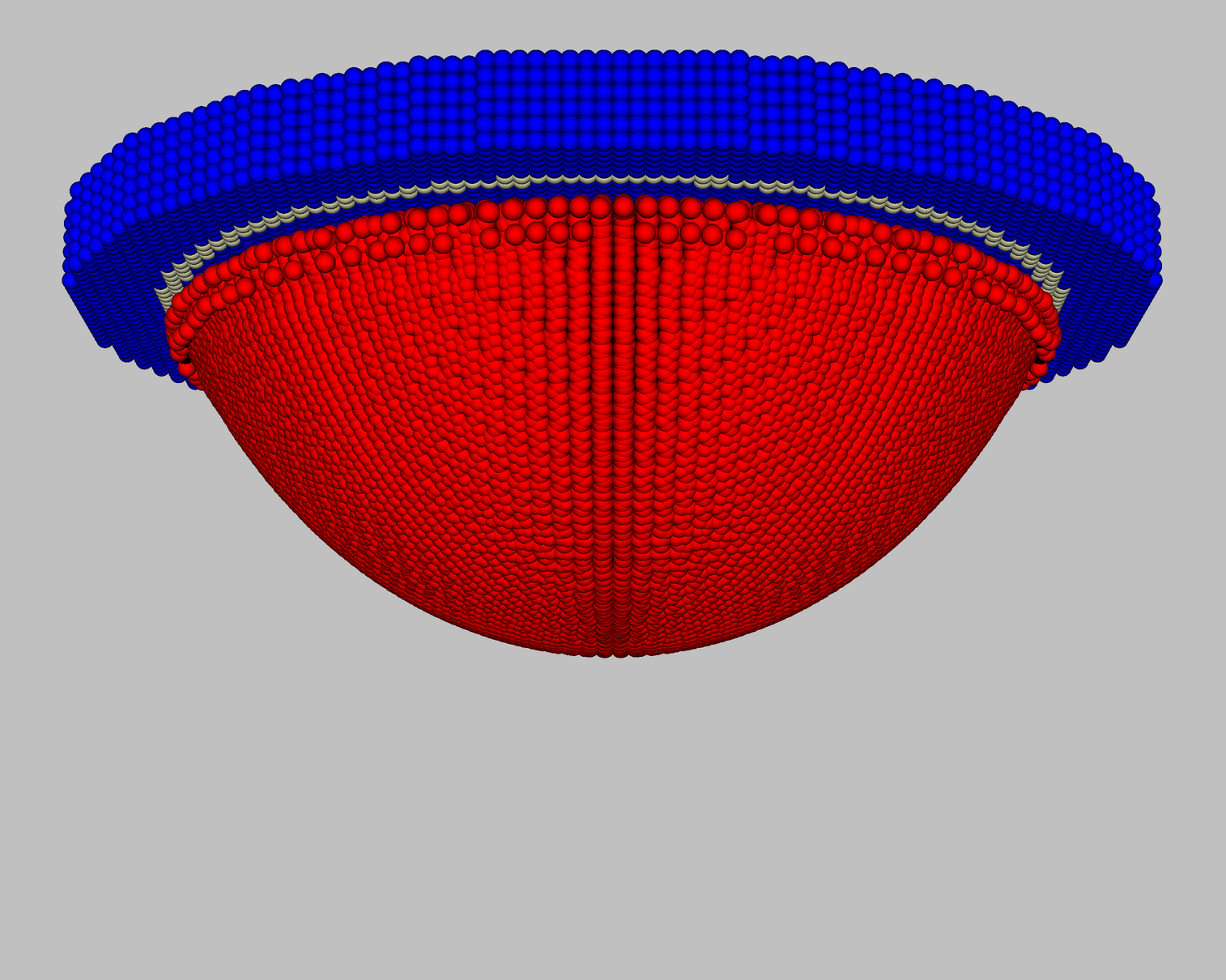}
        \caption{$t^*=3.2$}
        \label{fig:pendant3}
    \end{subfigure}
    \begin{subfigure}[b]{0.24\textwidth}
        \centering
        \includegraphics[width=\textwidth]{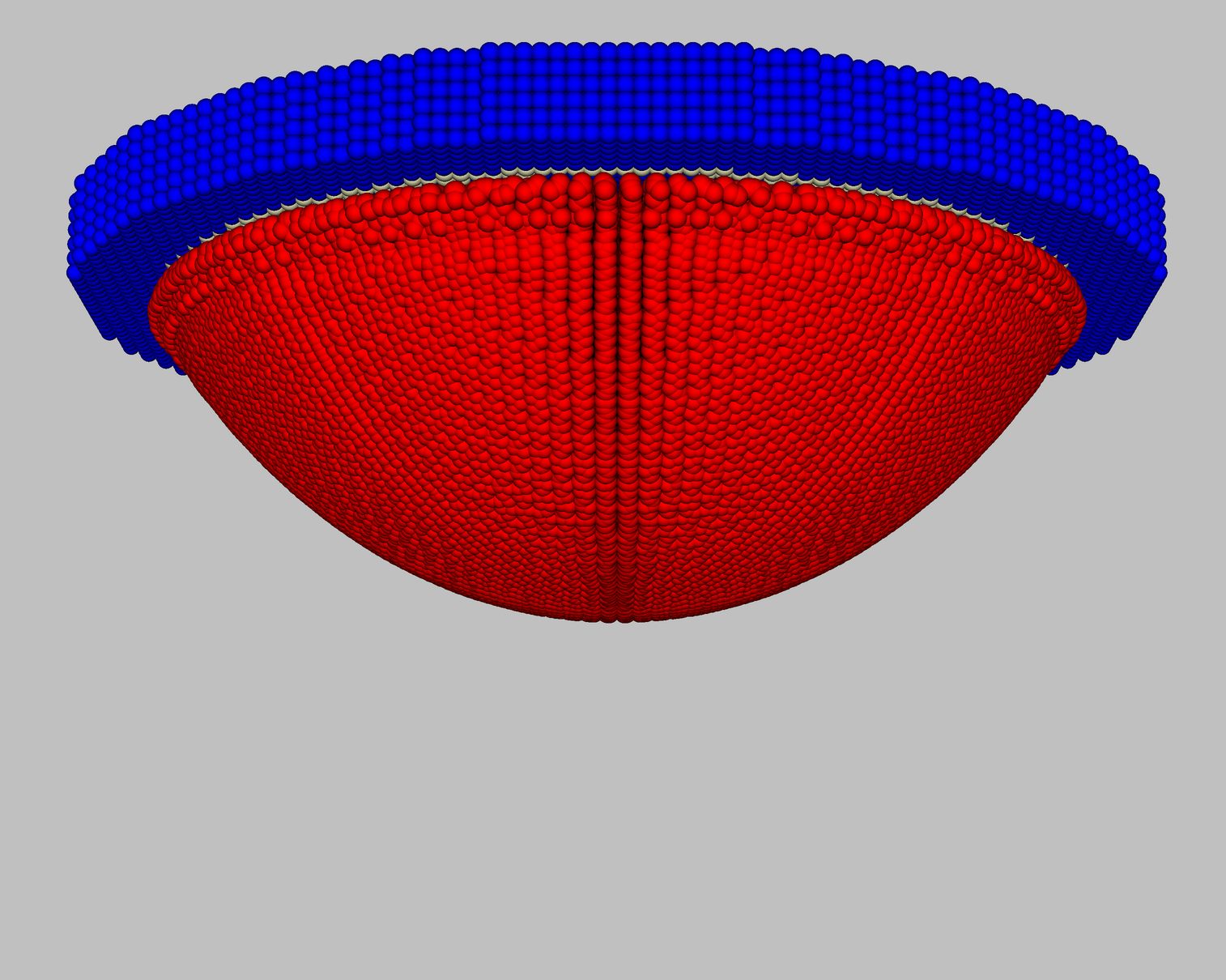}
        \caption{$t^*=6.0$}
        \label{fig:pendant4}
    \end{subfigure}
    \caption{Pinning of a spreading pendant droplet. The red colored particles represent the liquid and the blue colored particles are substrate particles. The yellow circle on the substrate represents the pinning curve.}
    \label{fig:pendant_snap}
\end{figure}
\begin{figure}[h!]
    \centering
    \begin{subfigure}[b]{0.24\textwidth}
        \centering
        \includegraphics[width=\textwidth]{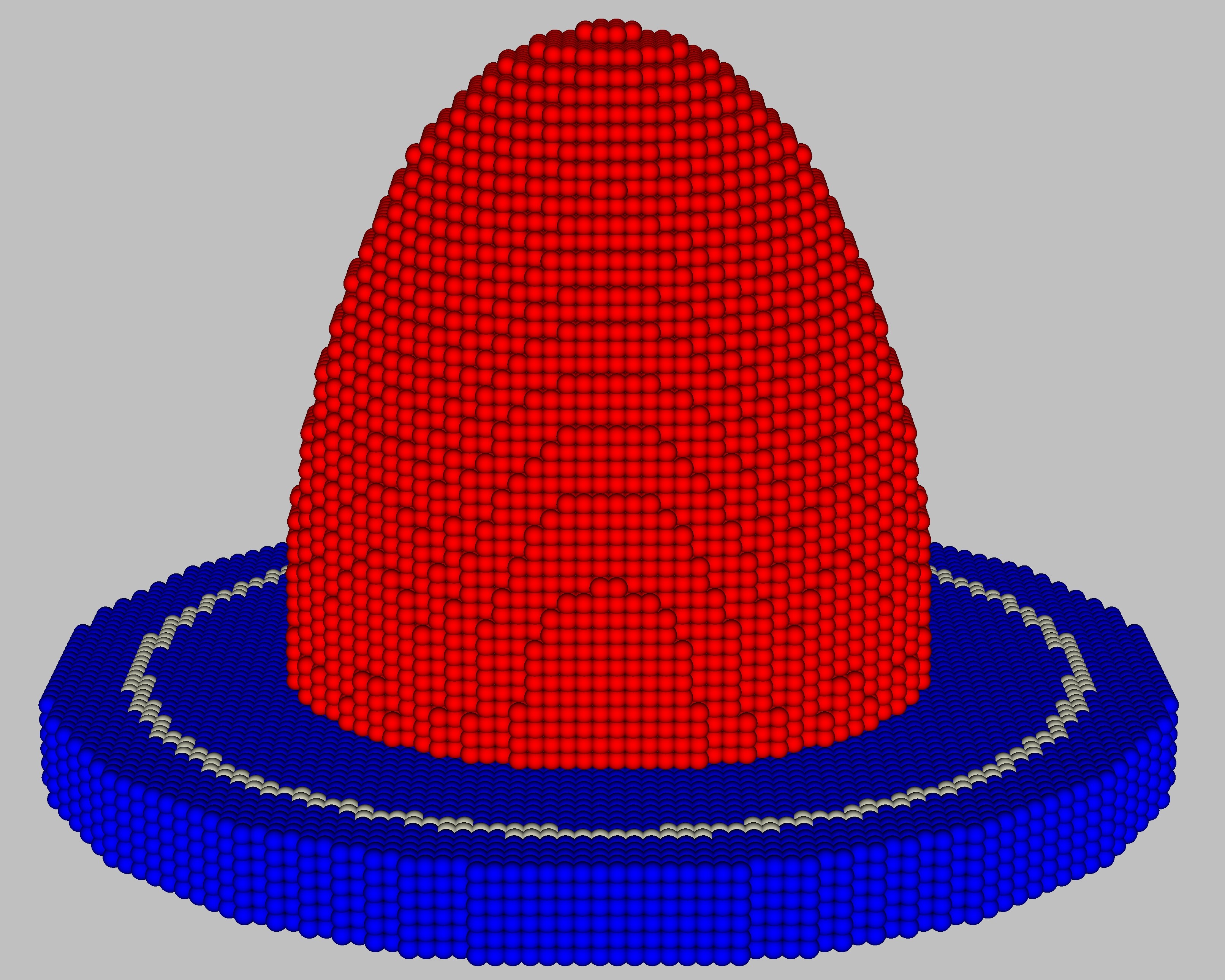}
        \caption{$t^*=0.0$}
        \label{fig:sessile1}
    \end{subfigure}
    \begin{subfigure}[b]{0.24\textwidth}
        \centering
        \includegraphics[width=\textwidth]{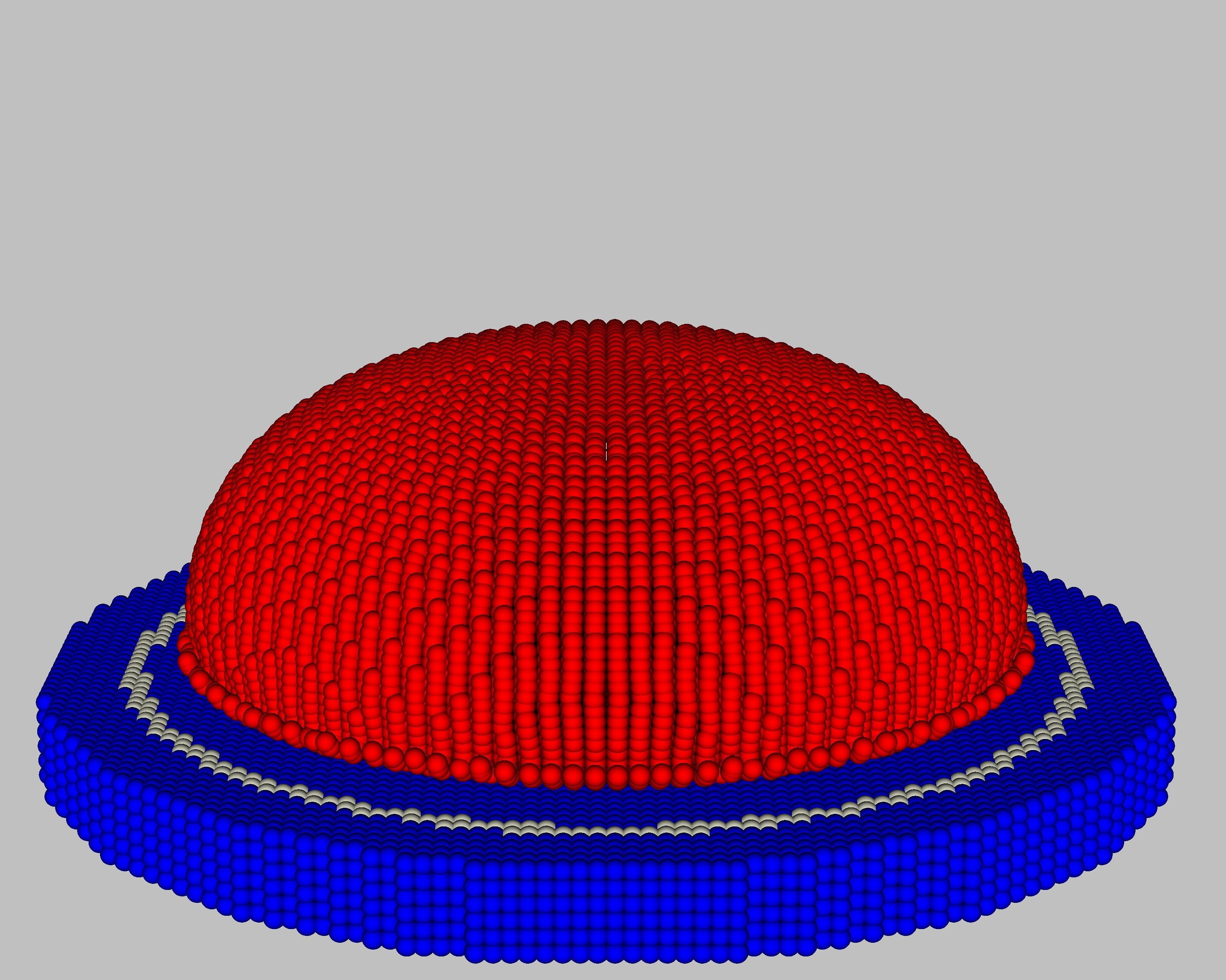}
        \caption{$t^*=0.64$}
        \label{fig:sessile2}
    \end{subfigure}
    \begin{subfigure}[b]{0.24\textwidth}
        \centering
        \includegraphics[width=\textwidth]{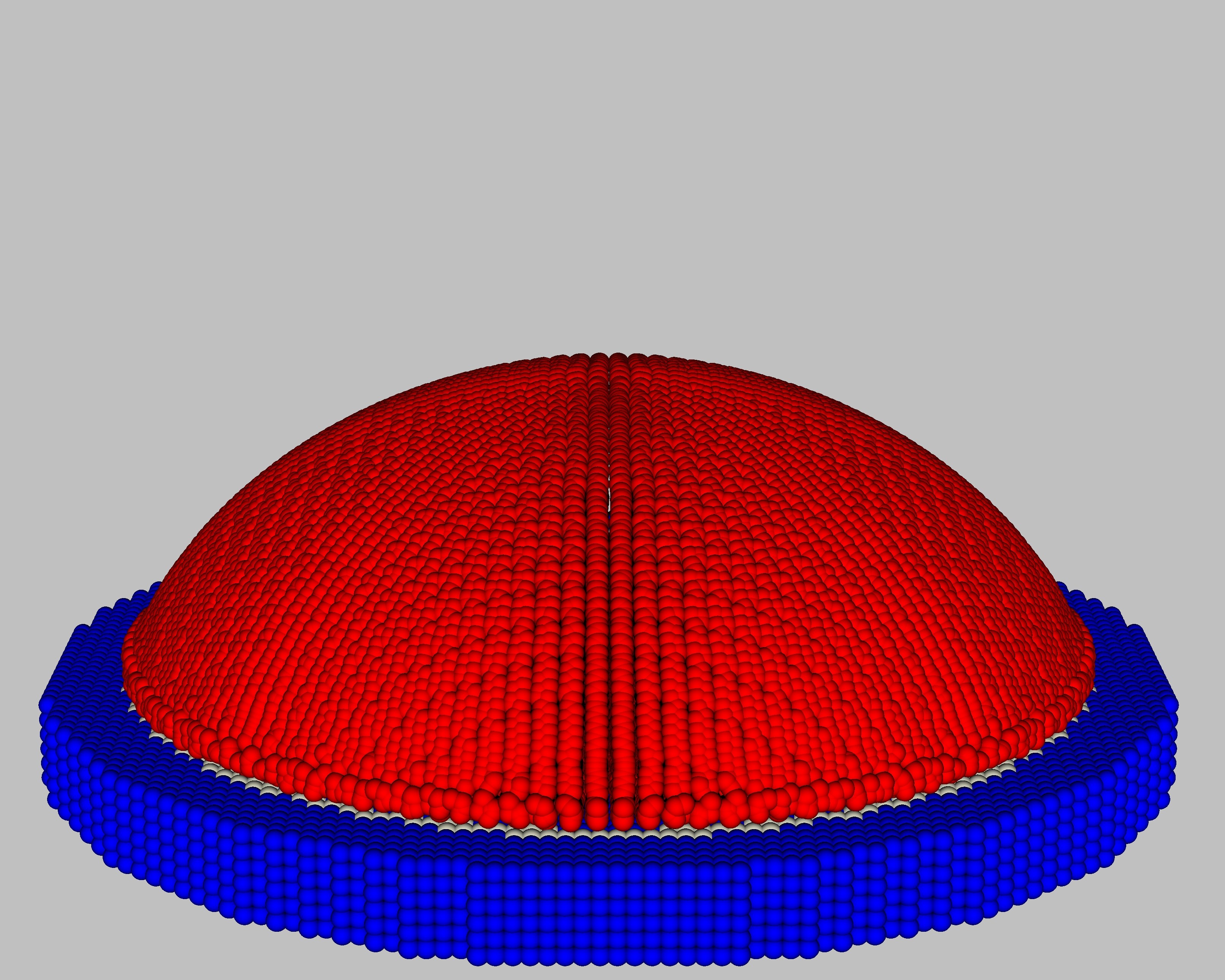}
        \caption{$t^*=3.2$}
        \label{fig:sessile3}
    \end{subfigure}
    \begin{subfigure}[b]{0.24\textwidth}
        \centering
        \includegraphics[width=\textwidth]{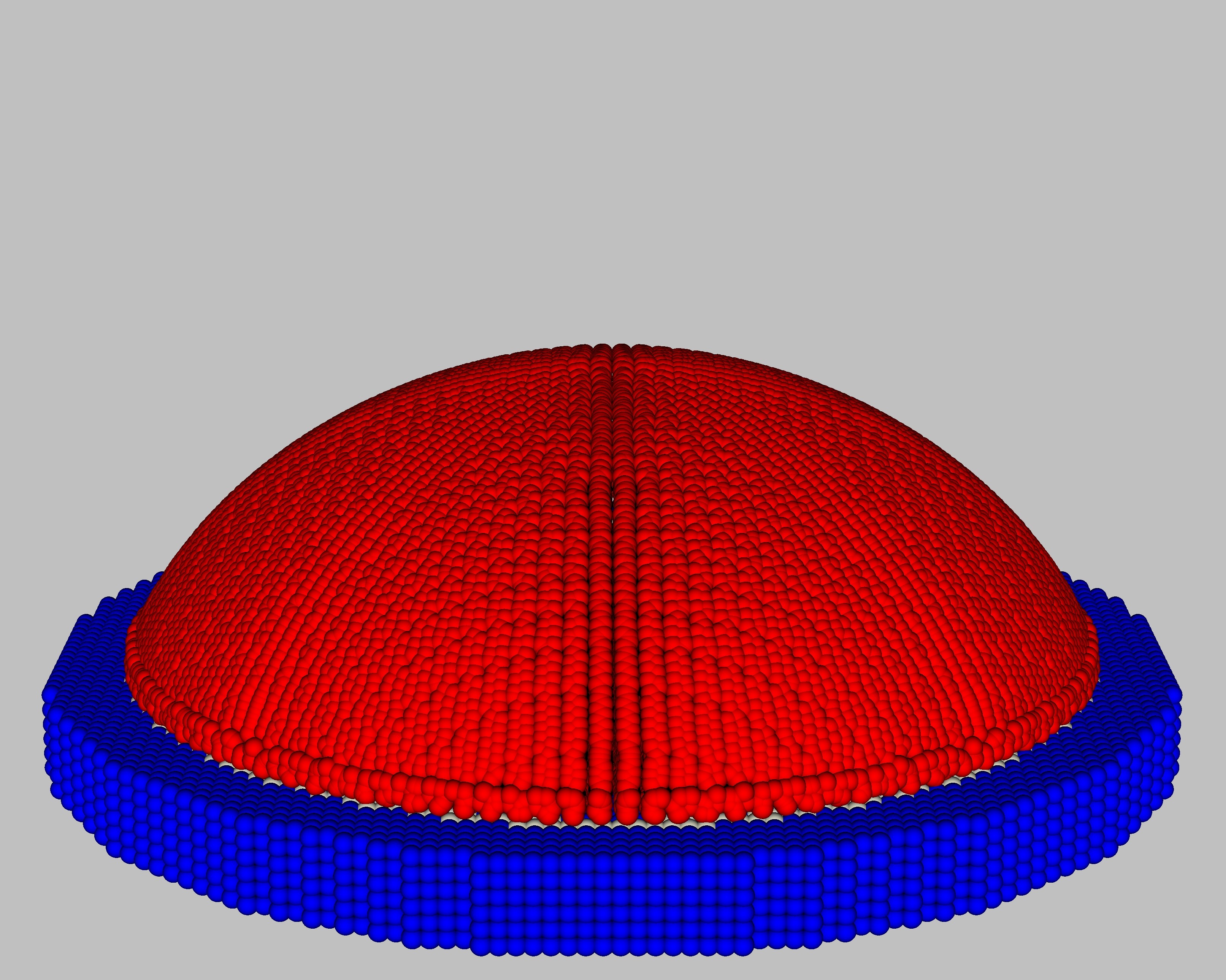}
        \caption{$t^*=6.0$}
        \label{fig:sessile4}
    \end{subfigure}
    \caption{Pinning of a spreading sessile droplet. The other details identical to Fig. \ref{fig:pendant_snap} }
    \label{fig:sessile_snap}
\end{figure}
\begin{figure}[h!]
\centering
\includegraphics[width=0.75\textwidth]{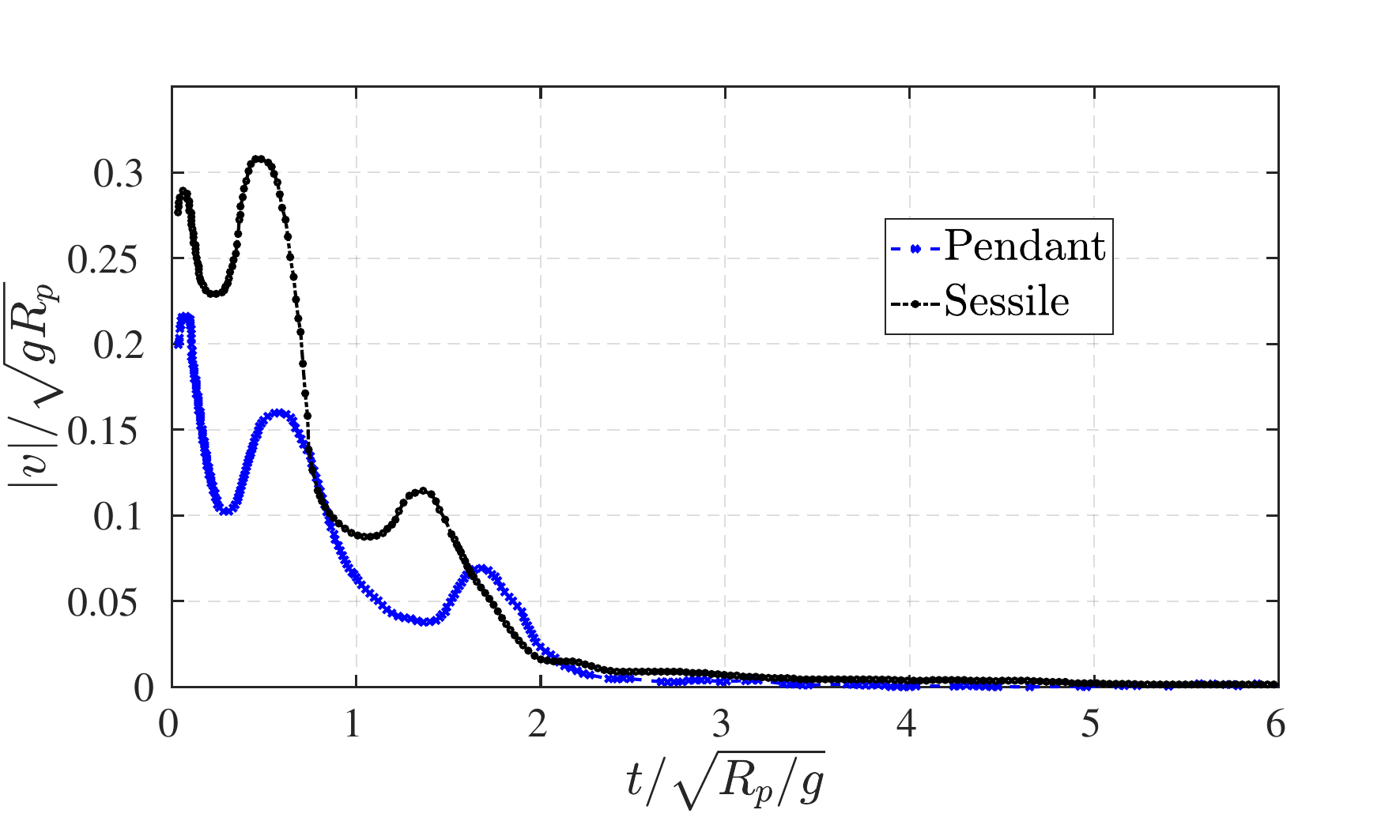}
\caption{Plot of the dimensionless contact line velocity magnitude($|v|/\sqrt{g R_p}$) against dimensionless time($t/\sqrt{R_p/g}$) for spreading and pinning sessile and pendant droplets.}
\label{fig:cl_vel_plot}
\end{figure}
The objective of this test case is to check whether the solver is able to capture the evolution of sessile and pendant droplets on a substrate in which a predefined circular pinning curve is specified. A static contact angle of $30^o$ is imposed away from the pinning curve, and, initially, the contact line and the pinning curve do not coincide. The droplets are initialized as a semi-prolate ellipsoid with a semi-major axis of $3.2$ $\mathrm{mm}$ in z-direction (normal to the substrate) and semi-minor axes (along the substrate) of $1.6$ $\mathrm{mm}$, which also corresponds to the initial contact line radius. The simulations are carried out for a liquid of viscosity($\mu$) $0.09\text{ } \mathrm{Pa\,s}$, density($\rho$) $1222$ $\mathrm{kg/m^3}$, and surface tension coefficient($\sigma$) $0.063$ $\mathrm{N/m}$. The Bond number ($\rho g {R_{eq}^2}/\sigma$) here is $0.48$, indicating that the spreading of the pendant droplet will not be affected significantly by gravity\cite{gennes2004capillarity}. Fig. \ref{fig:pendant_snap} and \ref{fig:sessile_snap} show the time snapshot of the simulation of the pendant and sessile droplet, respectively. An initial semi-prolate ellipsoid shape is considered to provide greater inertial force, making the three-phase contact line approach the pinning curve with a higher velocity than that with a hemisphere. Fig. \ref{fig:cl_vel_plot} shows the time series of the magnitude of the velocity of the contact line, in which we see evidence of damped oscillation of the droplets as they spread. The pinning model becomes relevant when any particle $a\in \Omega_{ns}$ is in the range of $2h$ from the pinning curve. As the contact line reaches the pinning curve, the magnitude of the contact line velocity approaches zero, and does not show any variation after contact line pinning. The contact line velocity of the sessile droplet is observed to be higher compared to the pendant drop, since gravity acts in the same direction of the wetting force in the case of the sessile droplet and against the wetting force in the pendant case. From this test case, it is evident that the contact line motion can be pinned to a circular pinning curve, even when the droplet is not wetting the pinning curve initially.

\subsection{Equilibrium shape of a pinned droplet}

In this section, we demonstrate the shape evolution of a pinned droplet in the presence of surface tension forces and gravity. The equilibrium shape achieved by the droplet is compared with  the experimental data given by  \citet{zhang2018975_eq_droplet_exp} and the theoretical solution of the Young-Laplace equation (derived in \ref{appendix:YL_equation}) for a pinned sessile droplet under the influence of gravity. Here, the pinning curve on the substrate is a circle of radius $R_p$. Four different cases, corresponding to varying equilibrium contact angles and Bond numbers, are simulated (the parameters are listed in Table \ref{table:BoCA}), with non-dimensional viscosity ($\mu \sqrt{gR_p}/\sigma$) equal to $0.025$. In the SPH simulations, the initial liquid volume and pinning radius $R_p$ are fixed according to the desired contact angle and Bond number. The droplets are initialized with cylindrical volumes, for which the cylinder radius is $R_p$.

\begin{figure}[!ht]
    \centering
    \begin{subfigure}[b]{0.24\textwidth}
        \centering
        \includegraphics[width=\textwidth]{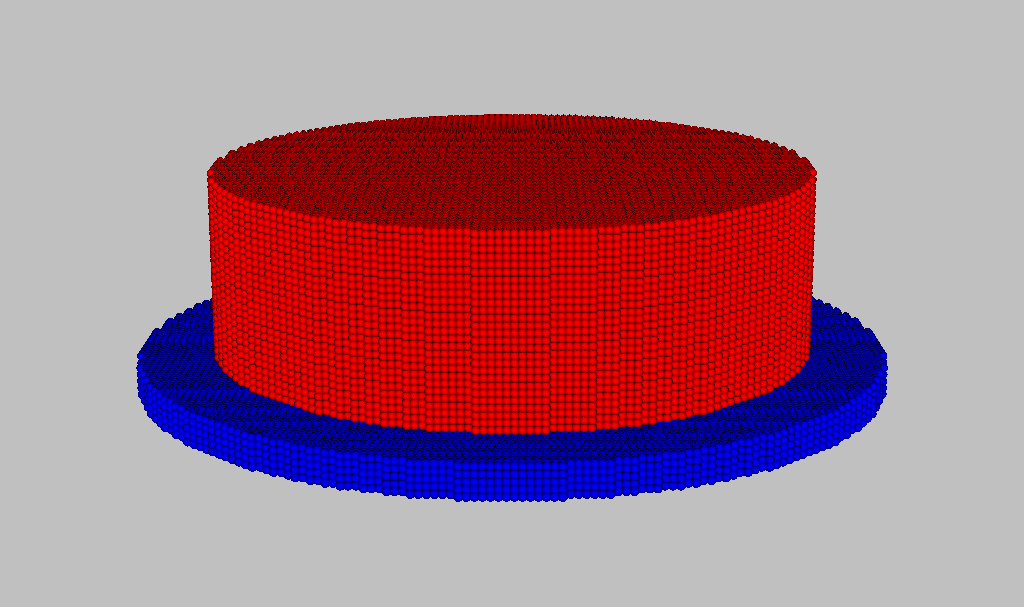}
        \caption{$t^*=0.0$}
        \label{fig:ca78_1}
    \end{subfigure}
    \begin{subfigure}[b]{0.24\textwidth}
        \centering
        \includegraphics[width=\textwidth]{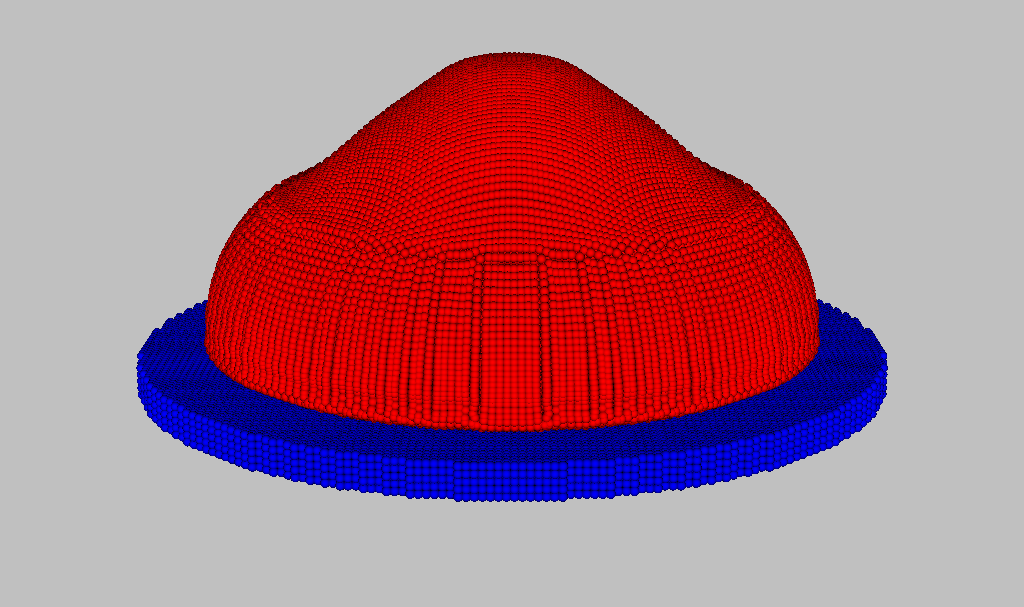}
        \caption{$t^*=0.34$}
        \label{fig:ca78_2}
    \end{subfigure}
    \begin{subfigure}[b]{0.24\textwidth}
        \centering
        \includegraphics[width=\textwidth]{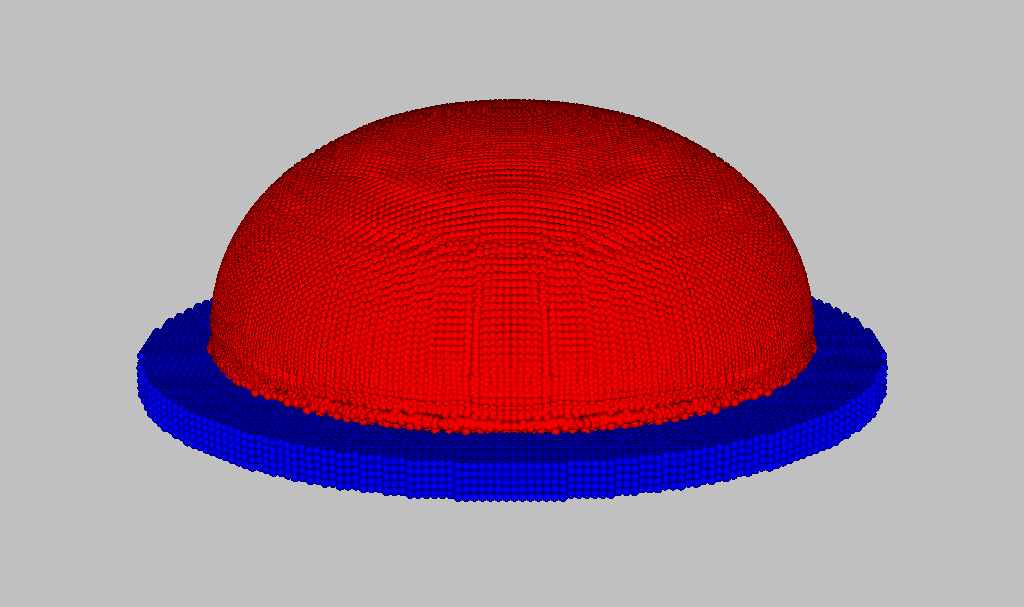}
        \caption{$t^*=1.8$}
        \label{fig:ca78_3}
    \end{subfigure}
    \begin{subfigure}[b]{0.24\textwidth}
        \centering
        \includegraphics[width=\textwidth]{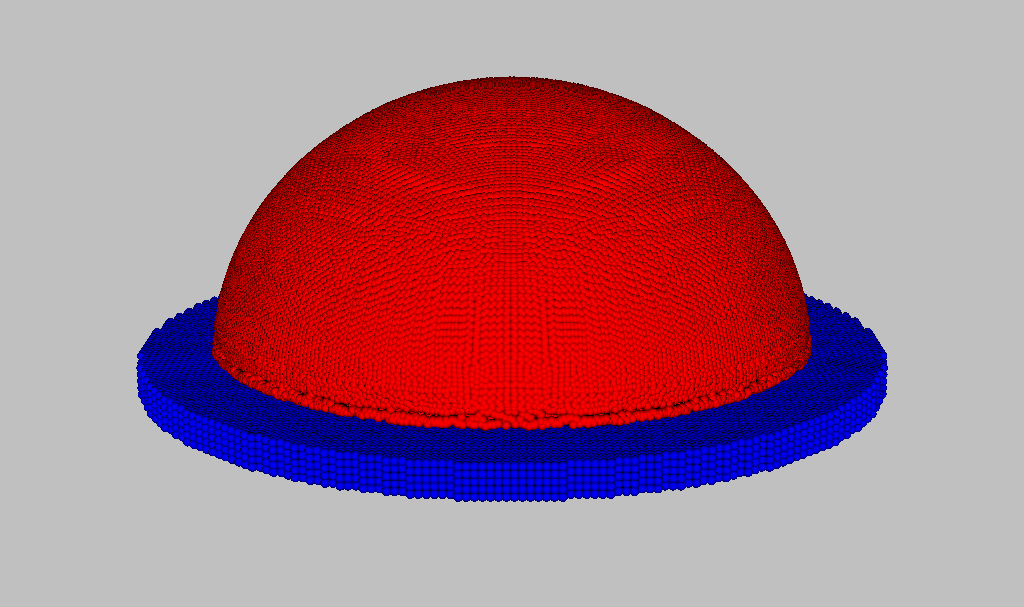}
        \caption{$t^*=10.5$}
        \label{fig:ca78_4}
    \end{subfigure}
    \caption{Evolution of a sessile droplet with its three-phase contact line pinned, initialized as a cylindrical volume. The Bond number of the droplet is $0.74$.}
    \label{fig:ca78_eq}
\end{figure}

\begin{figure}[!ht]
    \centering
    \begin{subfigure}[b]{0.24\textwidth}
        \centering
        \includegraphics[width=\textwidth]{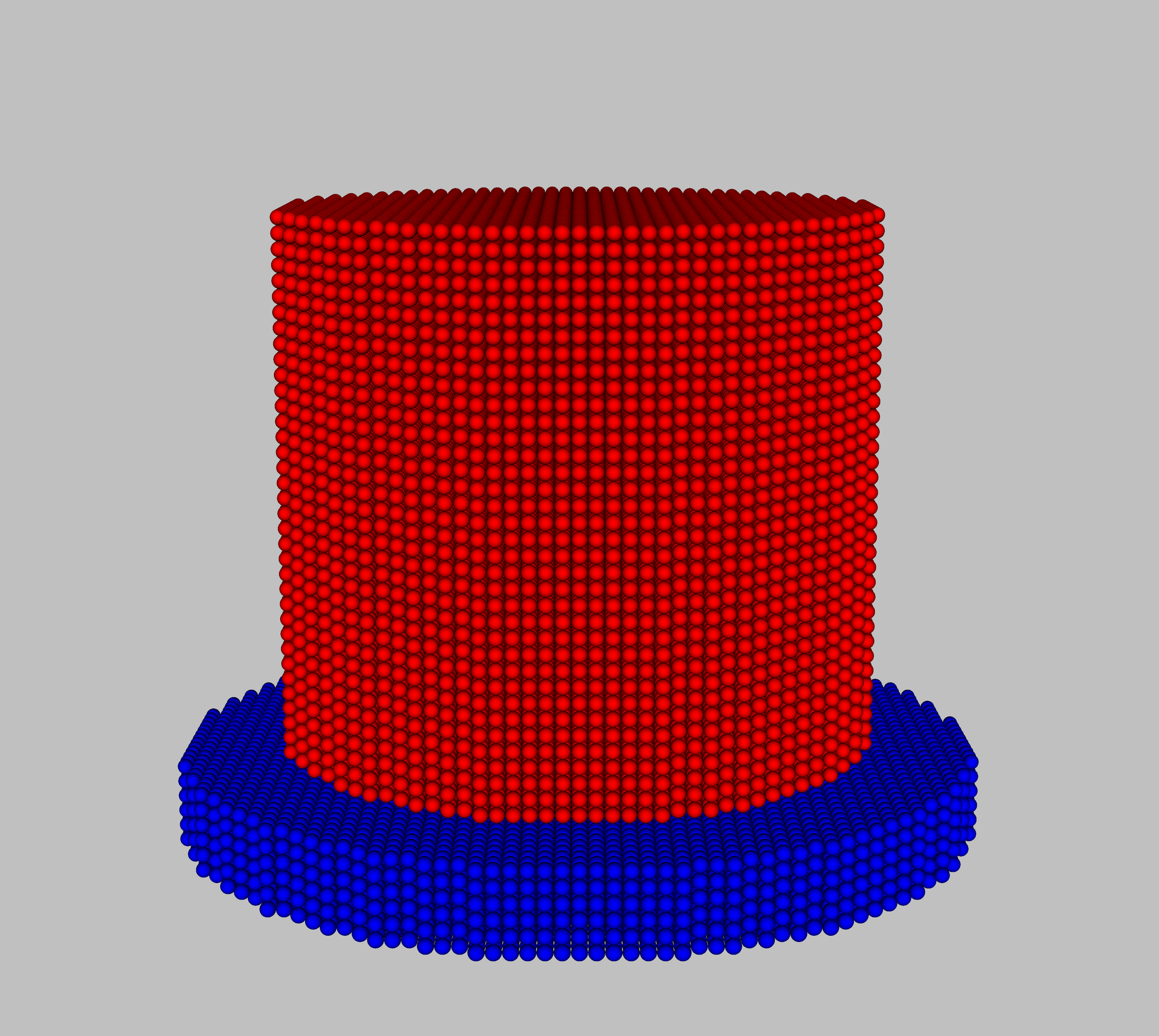}
        \caption{$t^*=0.0$}
        \label{fig:a3}
    \end{subfigure}
    \begin{subfigure}[b]{0.24\textwidth}
        \centering
        \includegraphics[width=\textwidth]{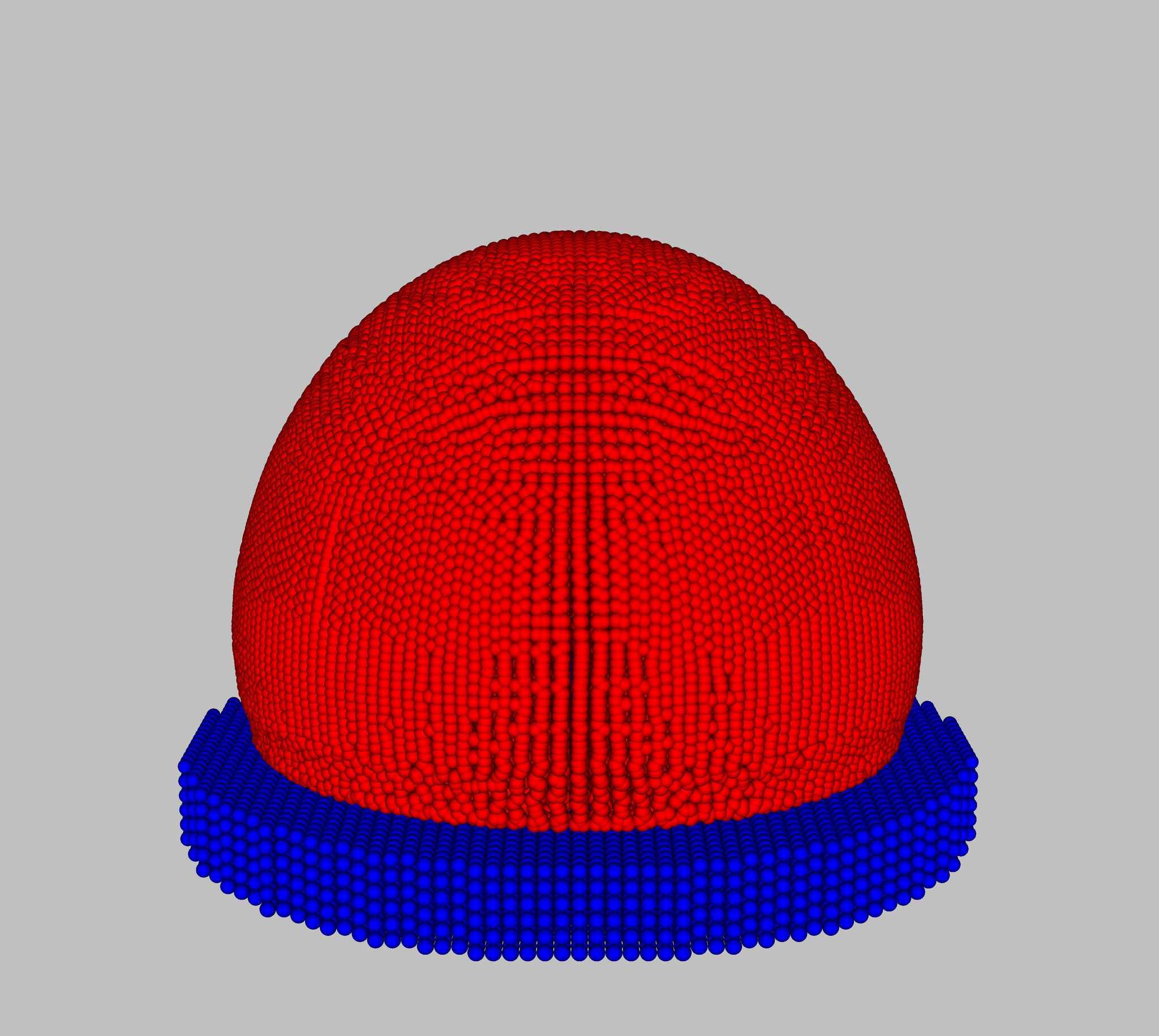}
        \caption{$t^*=0.34$}
        \label{fig:b3}
    \end{subfigure}
    \begin{subfigure}[b]{0.24\textwidth}
        \centering
        \includegraphics[width=\textwidth]{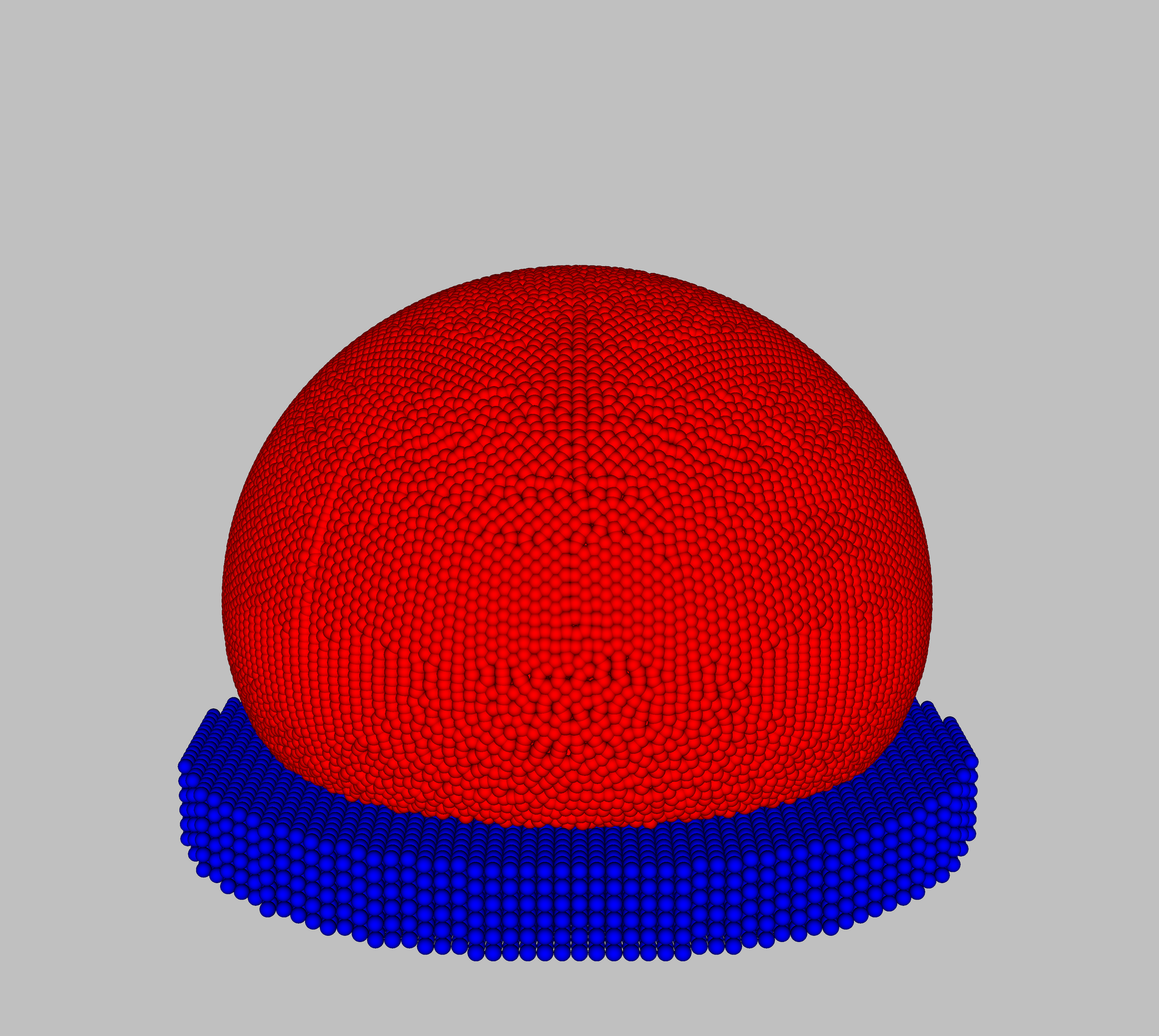}
        \caption{$t^*=1.8$}
        \label{fig:a4}
    \end{subfigure}
    \begin{subfigure}[b]{0.24\textwidth}
        \centering
        \includegraphics[width=\textwidth]{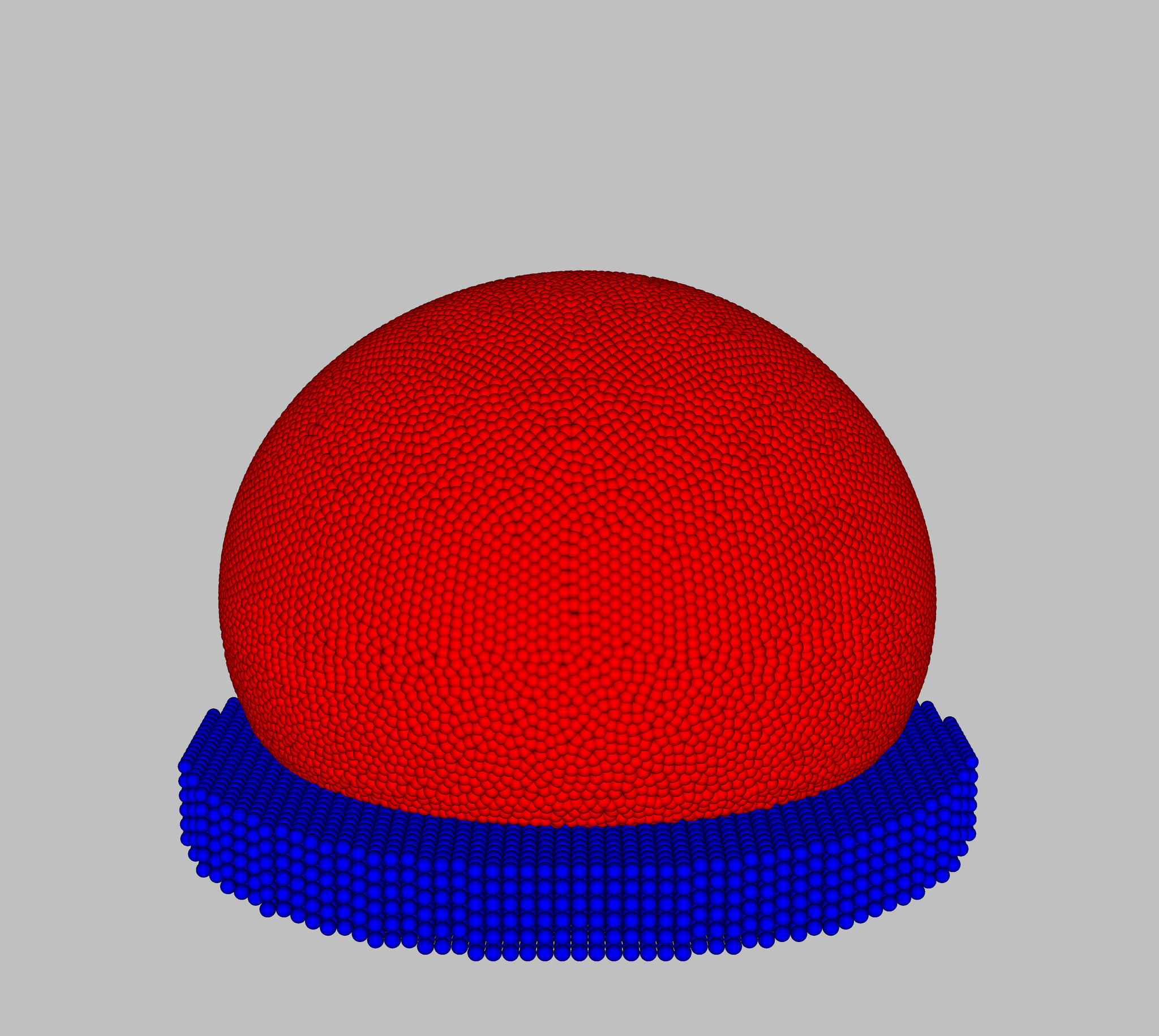}
        \caption{$t^*=10.5$}
        \label{fig:b4}
    \end{subfigure}
    \caption{Evolution of a sessile droplet with its three-phase contact line pinned, initialized as a  cylindrical volume.  The Bond number of the droplet is $0.34$.}
    \label{fig:ca125_eq}
\end{figure}

The simulations are performed with $1.48\times 10^5$ liquid particles for all cases. Figs. \ref{fig:ca78_eq} and \ref{fig:ca125_eq} represent the time snapshots of the shapes of sessile droplets evolving to their respective equilibrium states. The initial cylindrical shape of the droplet results in large curvature and therefore a large surface tension force density at the top edge, which in turn leads to interfacial oscillations of large magnitude. The viscous forces damp these oscillations, causing the droplets to eventually reach their respective equilibrium contact angles. 

\begin{table}[]
\begin{tabularx}{\textwidth}{cwc{1in}wc{1.5in}X}
\toprule
Case & Contact Angle & \begin{tabular}[c]{@{}c@{}}Bond Number \\ ($Bo=\rho g {{R_p}^2}/\sigma$)\end{tabular} & Comparison \\ \midrule
C1 & $125^o$ & 0.34 & Simulation with theory and experiment \\
C2 & $78^o$ & 0.74 & Simulation with theory and experiment \\
C3 & $45^o$ & 1.0 & Simulation with theory \\
C4 & $150^o$ & 0.25 & Simulation with theory \\ \bottomrule
\end{tabularx}
\caption{Contact angle and Bond number for different cases of comparison of the equilibrium droplet shape obtained from simulation with theory and experiment.}
\label{table:BoCA}
\end{table}

\begin{figure}[h!]
\centering
\includegraphics[width=0.95\textwidth]{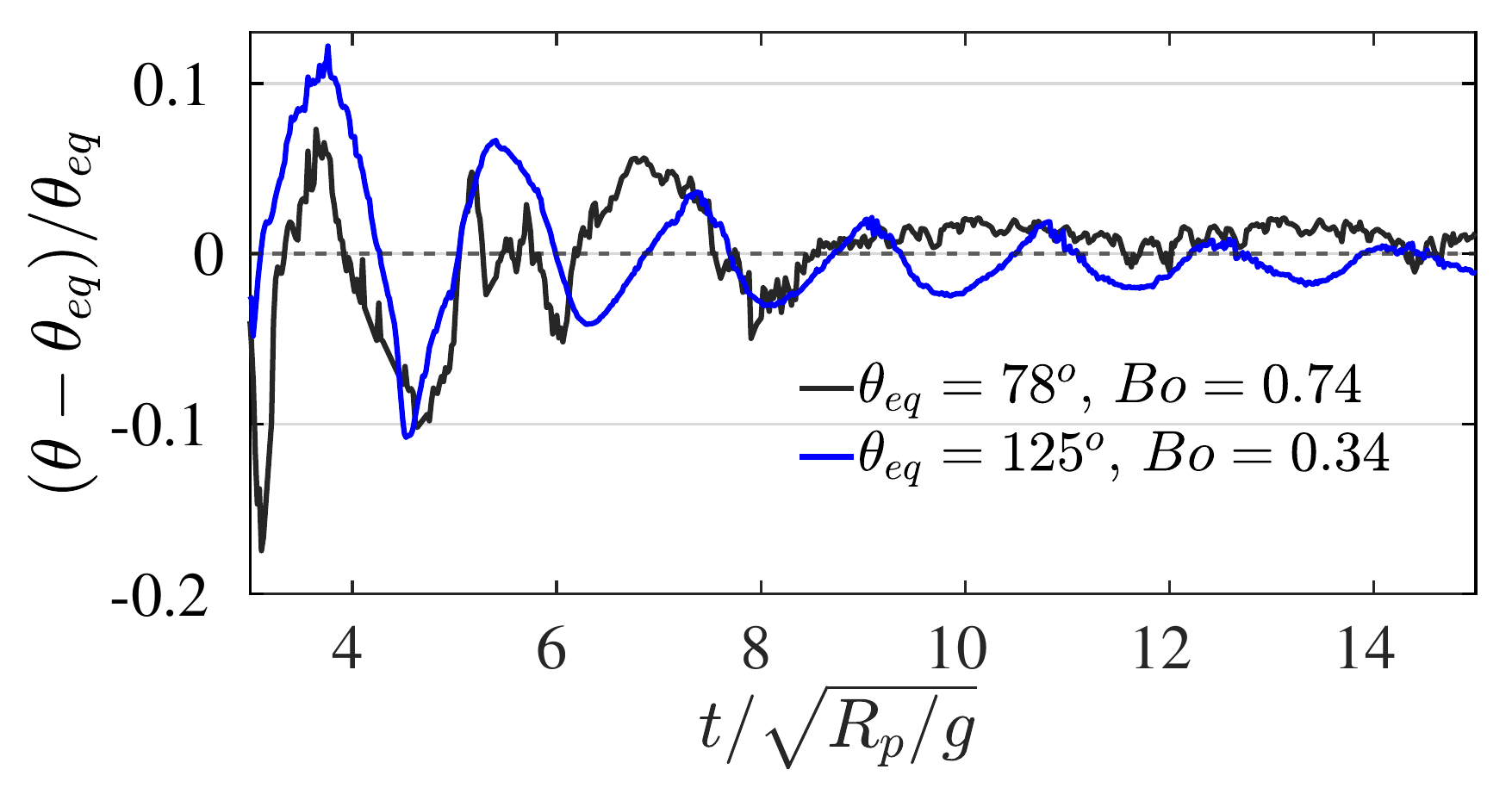}
\caption{Relative variation of the evolving contact angle($\left(\theta-\theta_{eq}\right)/\theta_{eq}$) with non-dimensional time($t/\sqrt{R_p/g}$). Here, $\theta$ is the instantaneous contact angle and $\theta_{eq}$ is the equilibrium contact angle. The equilibrium contact angle for the droplets of Bond number $0.74$ and $0.34$ are $78^o$ and $125^o$, respectively.}
\label{fig:static_CA_osc_ND}
\end{figure}

\begin{figure}[h!]
\centering
 \begin{subfigure}[b]{0.485\textwidth}
        \centering
    \includegraphics[width=\textwidth]{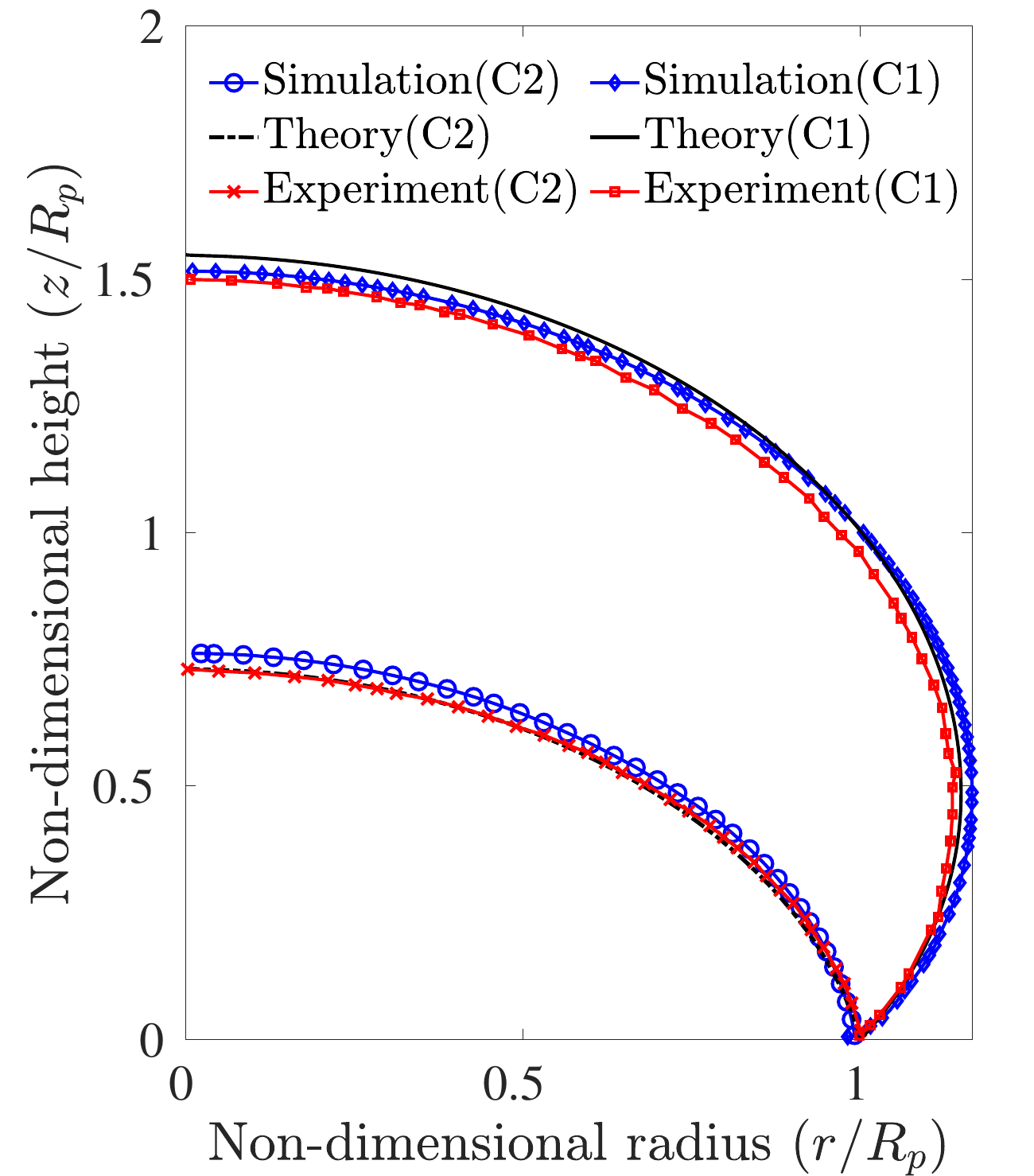}
    \caption{}
    \label{fig:sessile_equilibrium}
\end{subfigure}
 \begin{subfigure}[b]{0.49\textwidth}
        \centering
    \includegraphics[width=\textwidth]{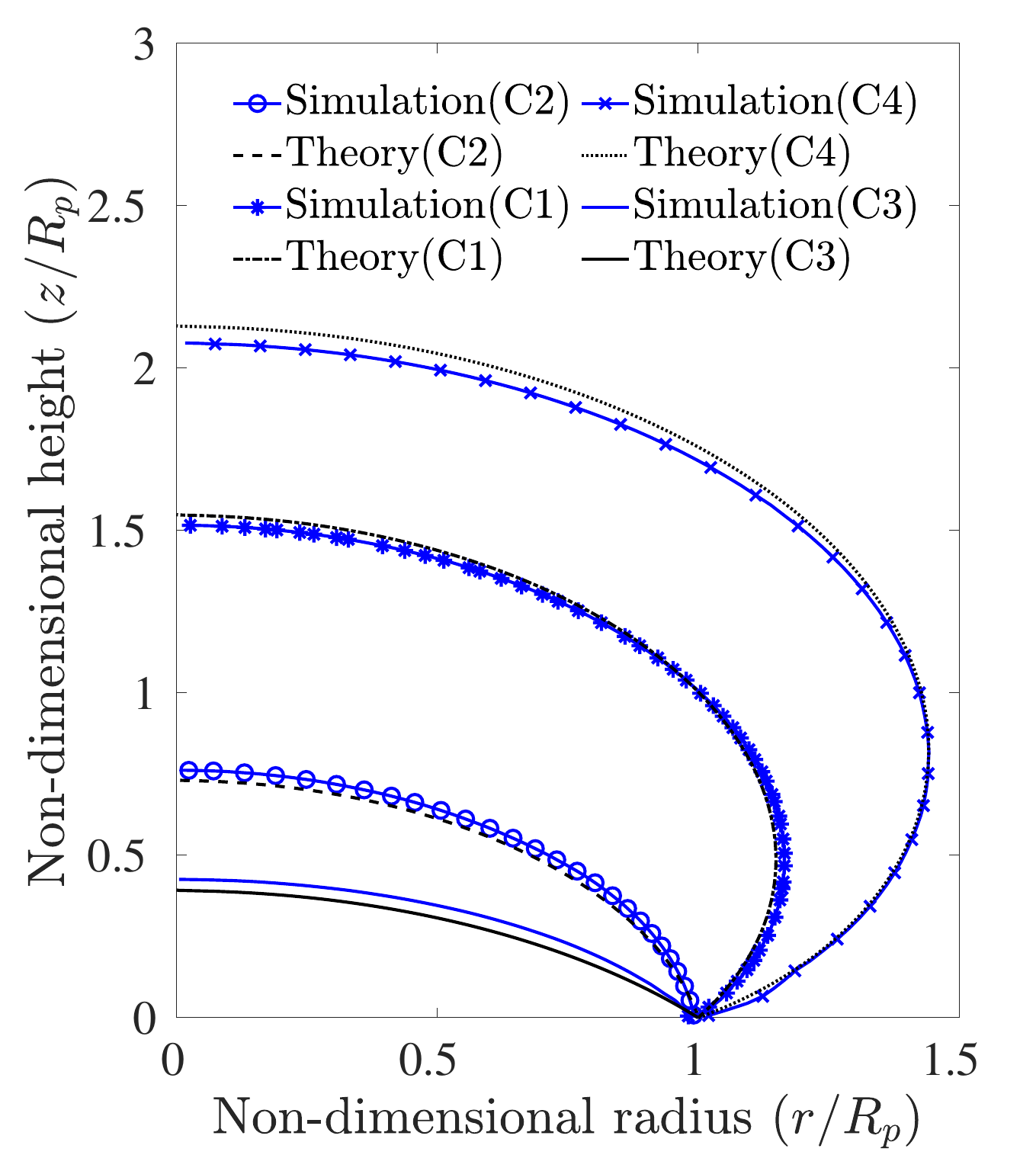}
    \caption{}
    \label{fig:sessile_equilibrium_tVs}
\end{subfigure}
\caption{(a) A comparison of the profile of a static droplet (for cases C1,C2, Table. \ref{table:BoCA}) obtained from numerical simulation with the corresponding profile obtained from theory and experiment \cite{zhang2018975_eq_droplet_exp}. (b) A comparison of profile for droplets obtained from simulation with theory for different Bond numbers and contact angles, for all cases in Table. \ref{table:BoCA}.}
    \label{fig:sessile_equilibrium_plot}
\end{figure}
We estimate the contact angle at each instance in time by evaluating the slope of a second order polynomial fitted to the location of the interfacial points $\mathbf{r}_{pa}$ near the substrate. Fig. \ref{fig:static_CA_osc_ND} shows a gradual reduction of the contact angle with time, until it tends to an equilibrium shape. The contact angle reached at $t/\sqrt{R_p/g}=15$ is within 5\% of the target equilibrium contact angle $\theta_{eq}$. The equilibrium shapes obtained from the simulation are compared with the shapes obtained from theory and the experiment shown in Fig. \ref{fig:sessile_equilibrium}. Further, for different combinations of Bond numbers and contact angles (shown in Table \ref{table:BoCA}), we obtain the profile and the volume of the droplet by solving the second-order ODE. 
A comparison of the equilibrium shape of different droplets obtained from the simulations is in good agreement with the theoretical droplet shape as shown in Fig. \ref{fig:sessile_equilibrium_tVs}. The error between simulation and theory is higher for cases C1 and C2, for which the droplet volume is large, and therefore the effective grid resolution is lower. From the comparison, it is evident that the pinning model is able to satisfactorily reproduce the equilibrium shape of the droplet as observed from the theory and experiment. 
\begin{figure}[h!]
    \centering
    \begin{subfigure}[b]{0.49\textwidth}
        \centering
        \includegraphics[width=\textwidth]{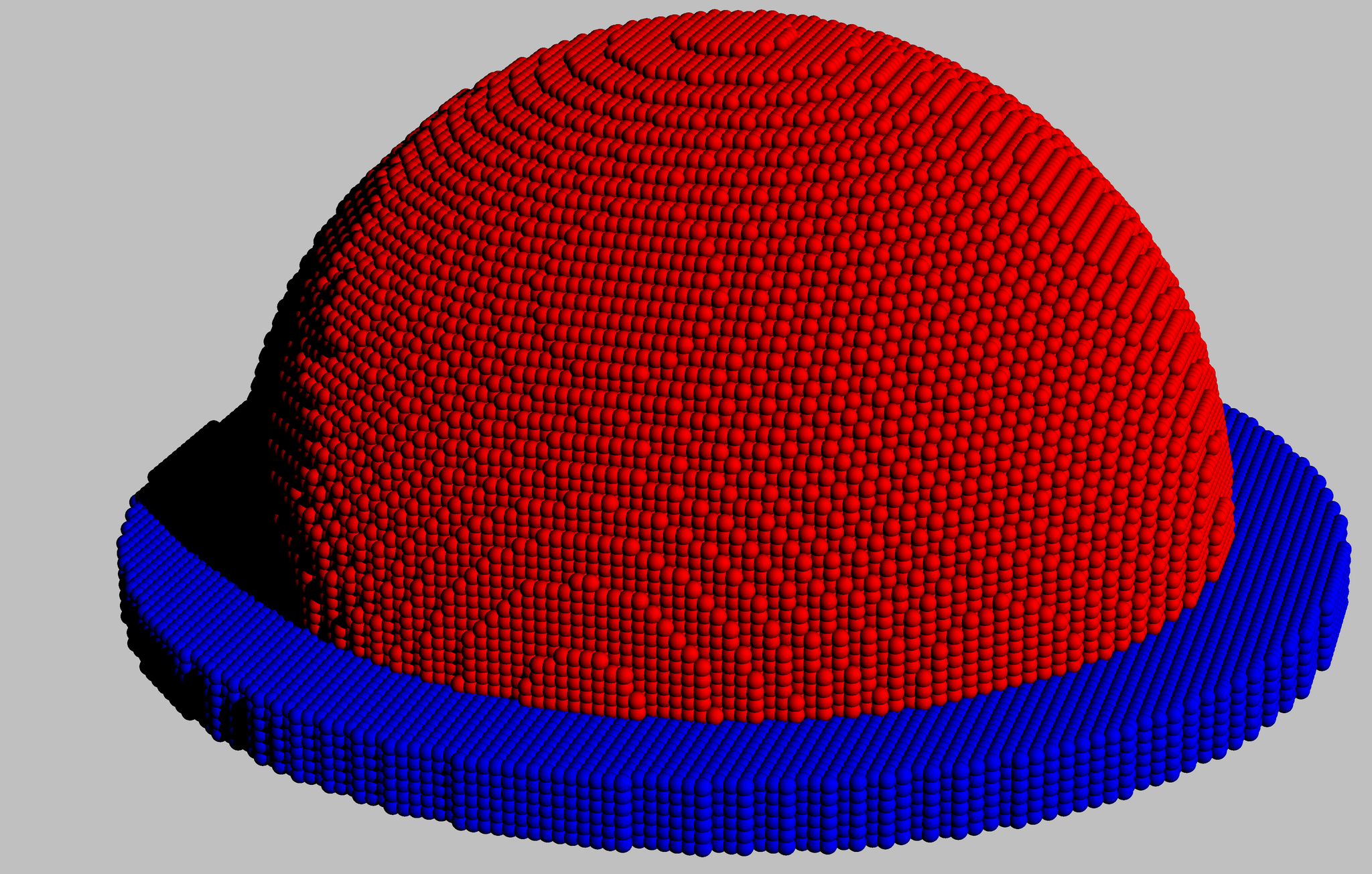}
        \caption{$t^* = 0.0$}
        \label{fig:pin2star_1}
    \end{subfigure}
    \begin{subfigure}[b]{0.49\textwidth}
        \centering
        \includegraphics[width=\textwidth]{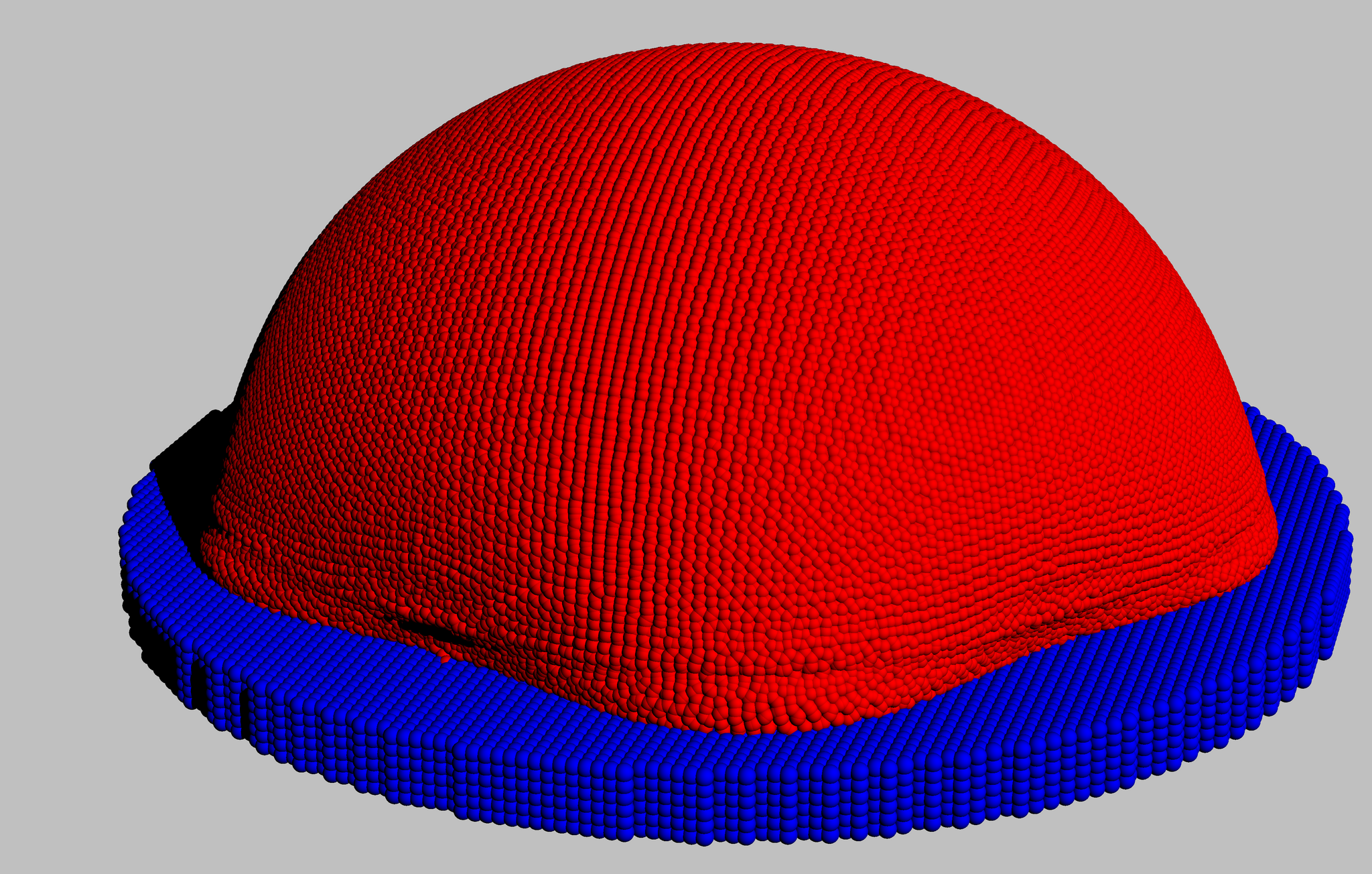}
        \caption{$t^* = 0.34$}
        \label{fig:pin2star_2}
    \end{subfigure}
    \begin{subfigure}[b]{0.49\textwidth}
        \centering
        \includegraphics[width=\textwidth]{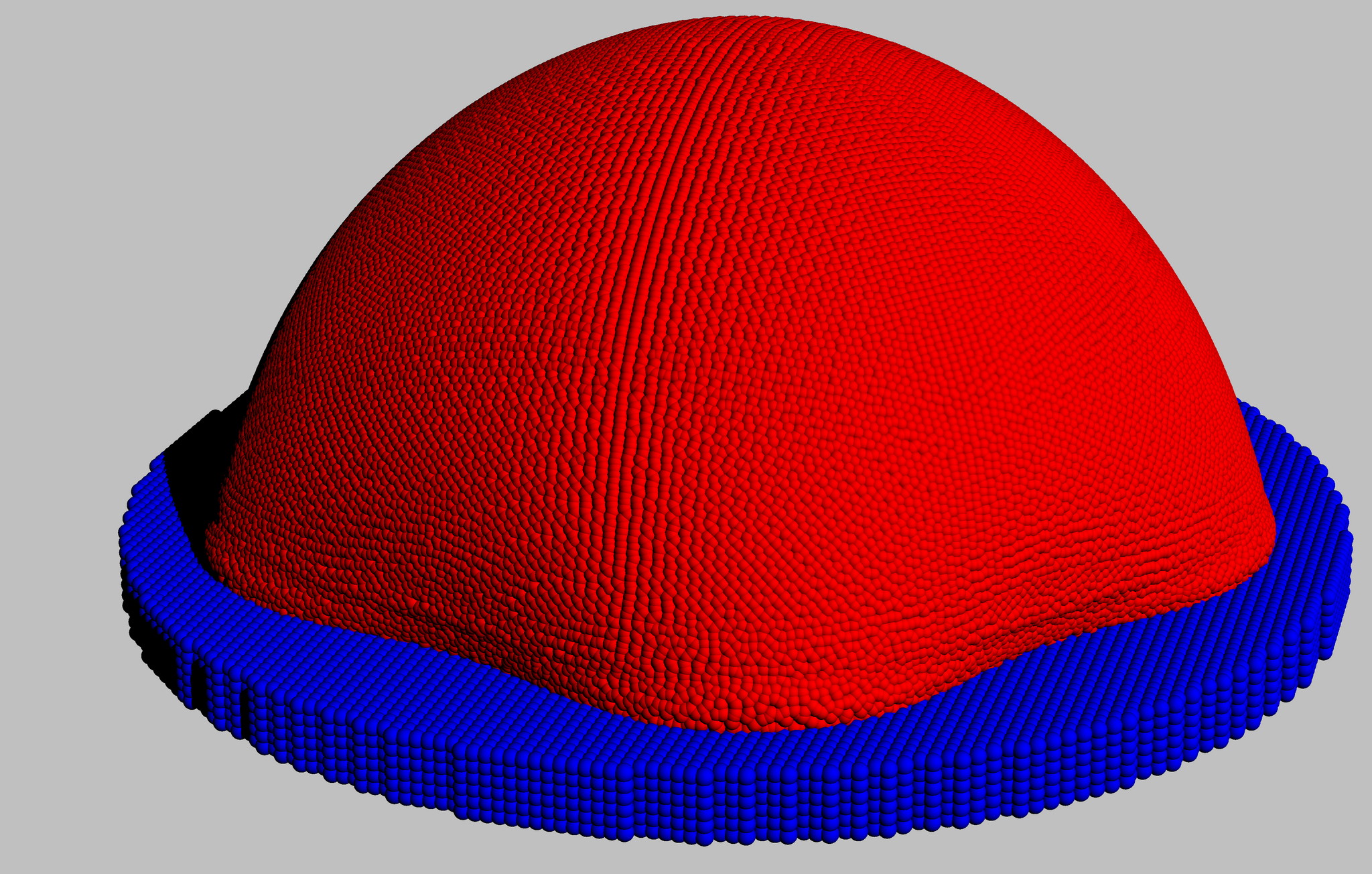}
        \caption{$t^* = 3.4$}
        \label{fig:pin2star_3}
    \end{subfigure}
    \begin{subfigure}[b]{0.49\textwidth}
        \centering
        \includegraphics[width=\textwidth]{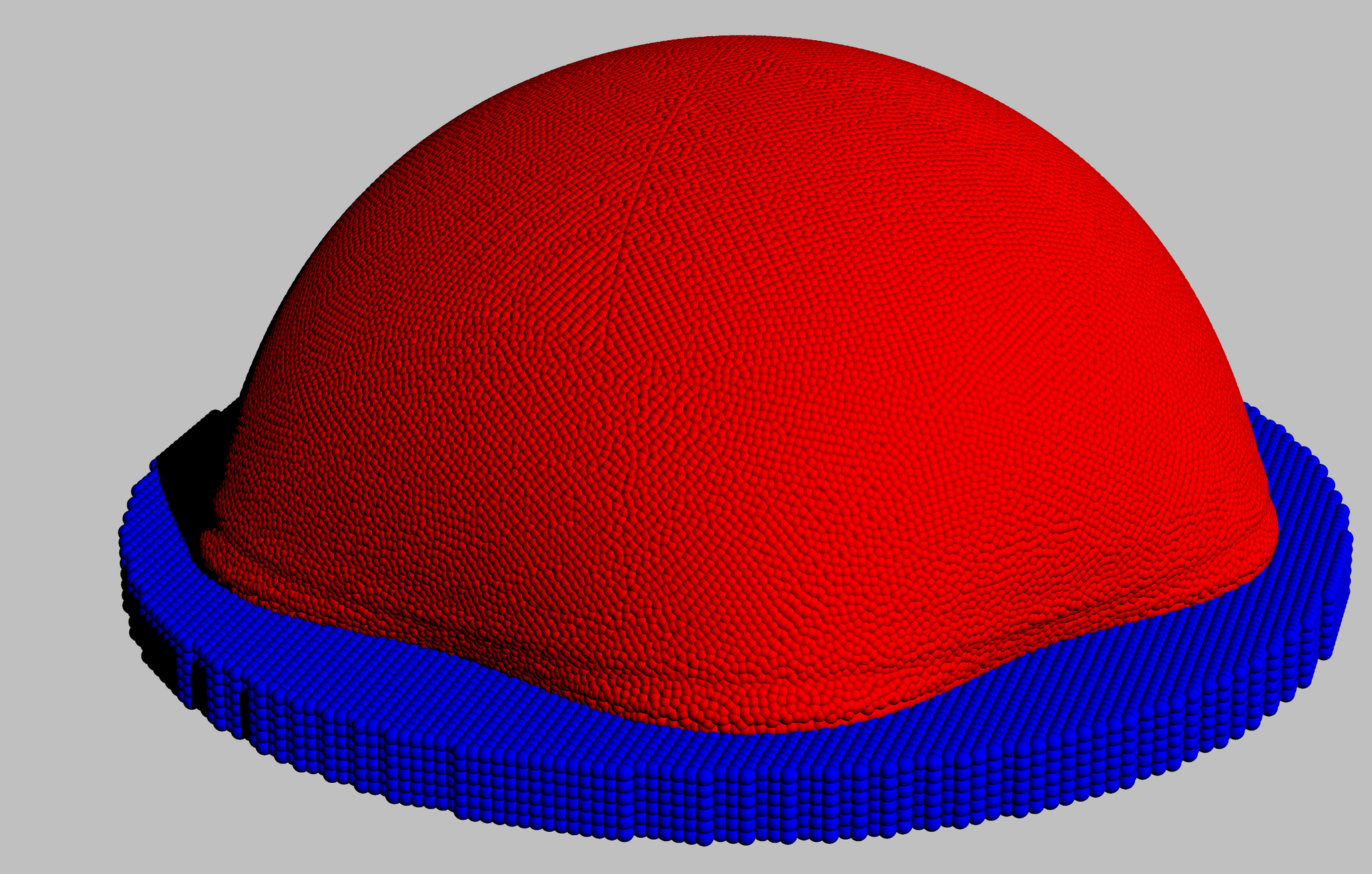}
        \caption{$t^* = 16.0$}
        \label{fig:pin2star_4}
    \end{subfigure}
    \caption{Shape evolution of an initial hemispherical droplet get its contact line pinned to a star-shaped pinning curve.}
    \label{fig:star}
\end{figure}
\begin{figure}[h!]
    \centering
    \begin{subfigure}[t]{0.35\textwidth}
        \centering
                 \includegraphics[width=\textwidth]{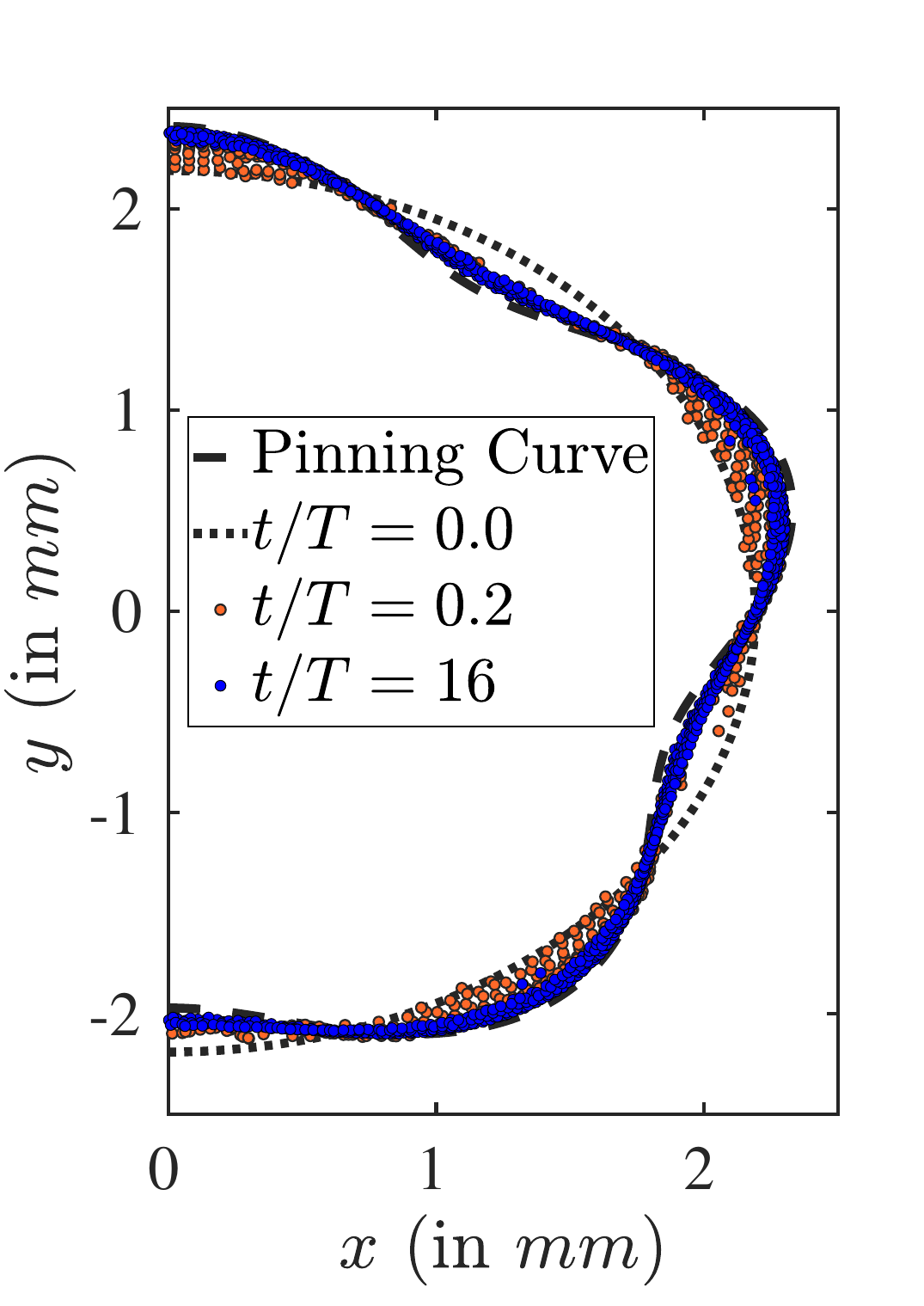}
        \caption{}
        \label{fig:pin2star_cl_2}
    \end{subfigure}
    \begin{subfigure}[t]{0.62 \textwidth}
        \centering
        \includegraphics[width=\textwidth]{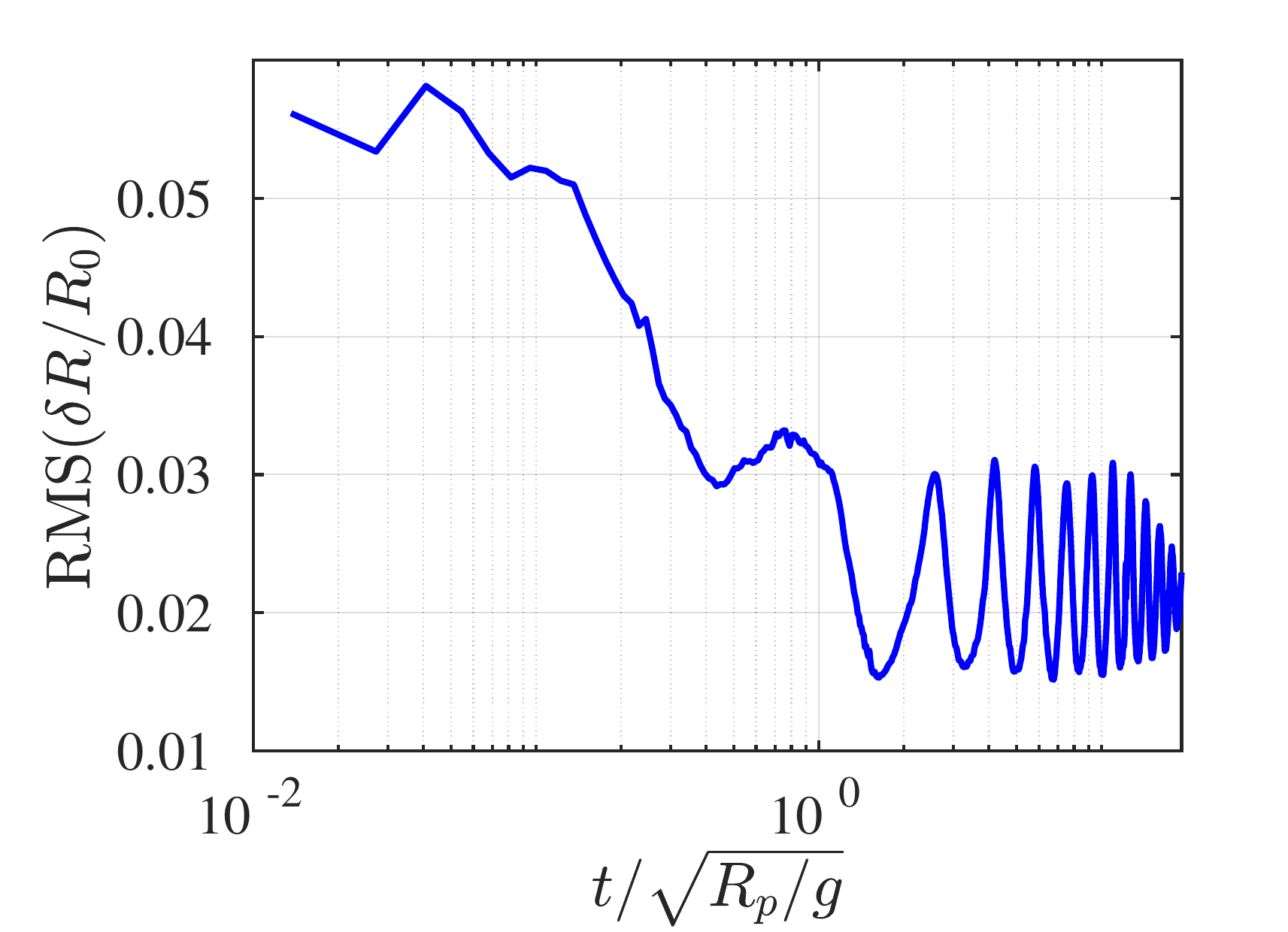}
        \caption{}
        \label{fig:star_rms_plot}
    \end{subfigure}
     \caption{(a) Evolution of  shape of three-phase contact line in presence of a star shaped pinning curve. (b) Evolution of non-dimensional Root Mean Square (RMS) error $\delta R$ between three-phase contact line position and pinning curve with time.}
     \label{fig:star_cl}
\end{figure}

\subsection{Pinning of hemispherical droplet by star-shaped pinning curve}
The test cases discussed so far are simulated for circular pinning curves.
In this test case, for varying curvature, we consider the evolution of a hemispherical droplet with an initial circular contact line towards a star-shaped pinning curve parameterized by: 
\begin{eqnarray}
    X(\theta)&=& R_0 (1+0.1\text{ sin}(5\theta))\text{ cos}(\theta)\\
    Y(\theta)&=& R_0 (1+0.1\text{ sin}(5\theta))\text{ sin}(\theta) \\
    Z(\theta)&=&0
    \label{eq:star}
\end{eqnarray}
Here, $R_0$ is the average radius of the pinning curve. 
To simulate the test case, a droplet with viscosity($\mu$) $\SI{0.01}{Pa.s}$, density($\rho$) $\SI{1000}{Kg/m^3}$ and surface tension coefficient($\sigma$) $\SI{0.063}{N/m}$ is initialized with a hemispherical shape with contact-line radius $R_0=\SI{2.2}{mm}$.
The droplet is discretized using $1.87\times10^5$ particles. Fig. \ref{fig:star} represents the shape evolution of the droplet with time. It can be observed that the droplet tries to maintain a minimum surface area in the free surface region even though the contact line takes the shape of a wavy circular pattern. The contact line (Fig. \ref{fig:pin2star_cl_2})
evolves from an initial circular shape to the wavy pinning curve at large time. Thus, parts of the contact line lying outside the area enclosed by the pinning curve dewet the substrate, whereas parts of contact line lying inside this area tend to wet the substrate further. The evolution of the normalized root mean square error between the contact line and the pinning curve over time (Fig. \ref{fig:star_rms_plot}), shows that after the initial transient regime ($t\approx \sqrt{R_p/g}$) the error remains below 3\%. Thus, the solver is able to successfully capture the transient pinning of contact line to a wavy pinning curve.
\subsection{Pinned droplet on a vertical cylinder}
\begin{figure}[!ht]
    \centering
    \begin{subfigure}[b]{0.4\textwidth}
        \centering
        \includegraphics[width=\textwidth]{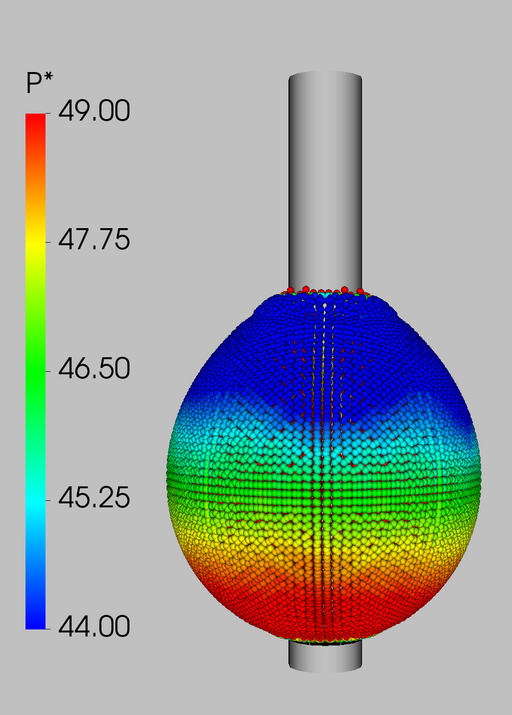}
        \caption{$t^* = 12.8$}
        \label{fig:cylinder1}
    \end{subfigure}
    \begin{subfigure}[b]{0.3\textwidth}
        \centering
        \includegraphics[width=\textwidth]{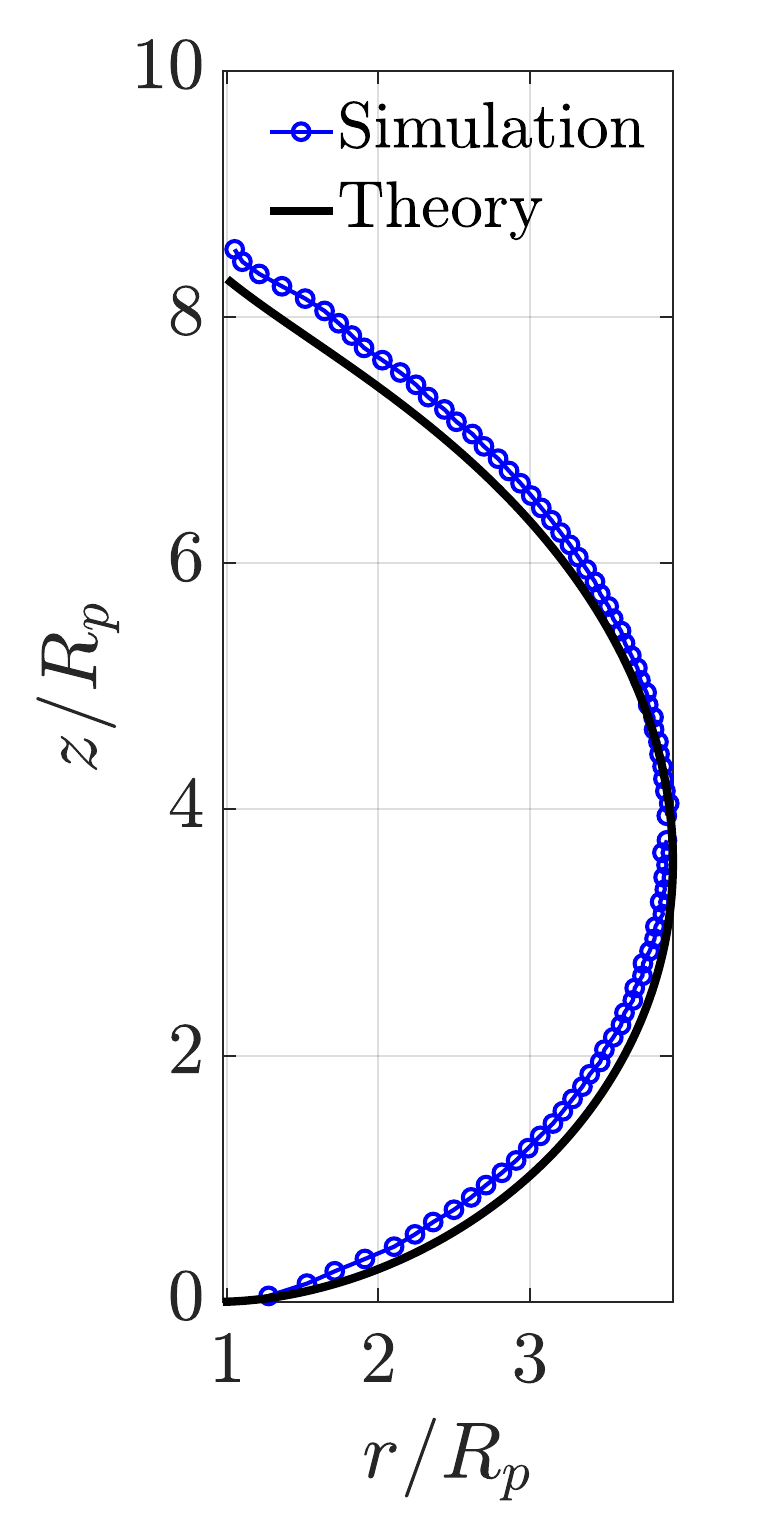}
        \caption{}
        \label{fig:cylinder2}
    \end{subfigure}
    \caption{Comparison of the evolved droplet shape held on a cylinder with the corresponding profile obtained from theory at dimensionless time($t^*=t/\sqrt{g/R_c}$) of $12.8$. (a) Dimensionless pressure($P^* = p/(\rho g R_p)$) contour of the droplet held over a cylinder. (b) Comparison of the droplet profile obtained in simulation with the profile obtained by solving the ODE.}
    \label{fig:cylinder}
\end{figure}
In all the test cases above, the pinning treatment of the contact line is implemented on a flat surface. To show the method's generalization to 3D curves, here we present a test case of a droplet spread around a vertical cylinder, pinned along a circumferential line. The droplet forms two contact lines one at top and one at bottom of the droplet. The top contact line is assigned a wetting contact angle\cite{blank2024surface} ($\theta_{eq} = 45^o$), while the bottom contact line is pinned along a circular curve using the proposed pinning treatment. 

The Bond number($\rho g (2 \pi R_c h)/\sigma$) is $0.36$, where $R_c$ is the radius of the cylinder and $h$ is the initial wetted height of the cylinder. The droplet is initialized as a volume of revolution of an ellipse of volume $\SI{6.26}{mm^3}$, and the pinning line as a circle with radius($R_p =R_c-0.5\Delta_x$). The test case has been simulated with $31,400$ liquid particles. It is to be noted that the blending function used here is $f(\eta) = f(r_a-R_c)$, where $r_a$ is the radius of the particle $a$ and $R_c$ is the radius of the cylinder.

The dimensionless pressure contour of the droplet at equilibrium is shown in Fig. \ref{fig:cylinder1}. We compare the droplet profile obtained from the simulation with the profile obtained from theory by solving the ordinary differential equation derived from the Young-Laplace equation(discussed in  \ref{appendix:YL_equation_cylinder}) for an axisymmetric vertical droplet under the influence of gravity as shown in Fig. \ref{fig:cylinder2}. The comparison shows good agreement between the simulation and theory, demonstrating that the implemented pinning treatment effectively constrains the contact line motion on a three-dimensional curved surface.
 
\subsection{Pinned droplet on an oscillating substrate}
\begin{figure}[!ht]
    \centering
    \begin{subfigure}[b]{0.32\textwidth}
        \centering
        \includegraphics[width=\textwidth]{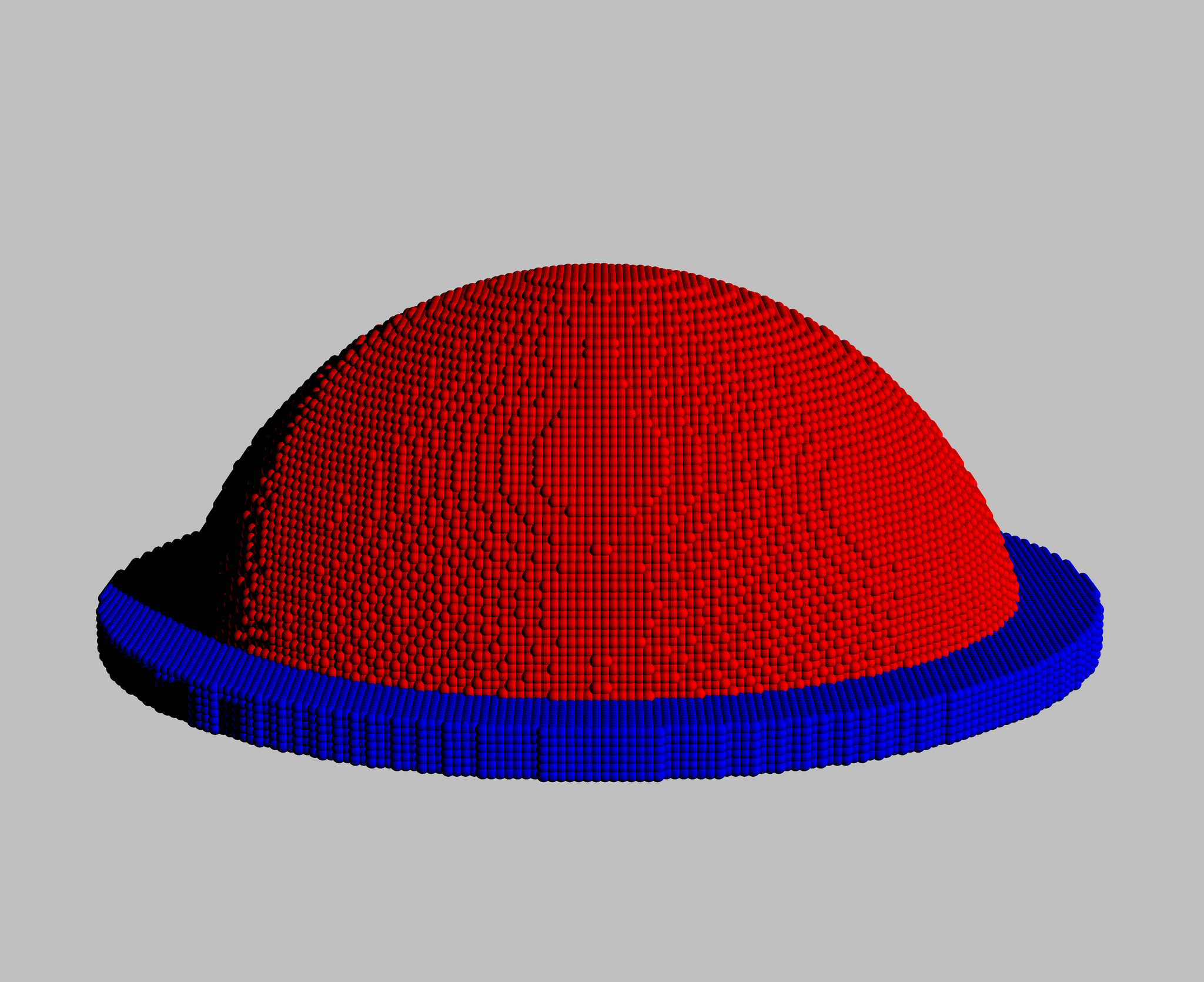}
        \caption{$t = 0.0$ $s$}
        \label{fig:osc_1}
    \end{subfigure}
    \begin{subfigure}[b]{0.32\textwidth}
        \centering
        \includegraphics[width=\textwidth]{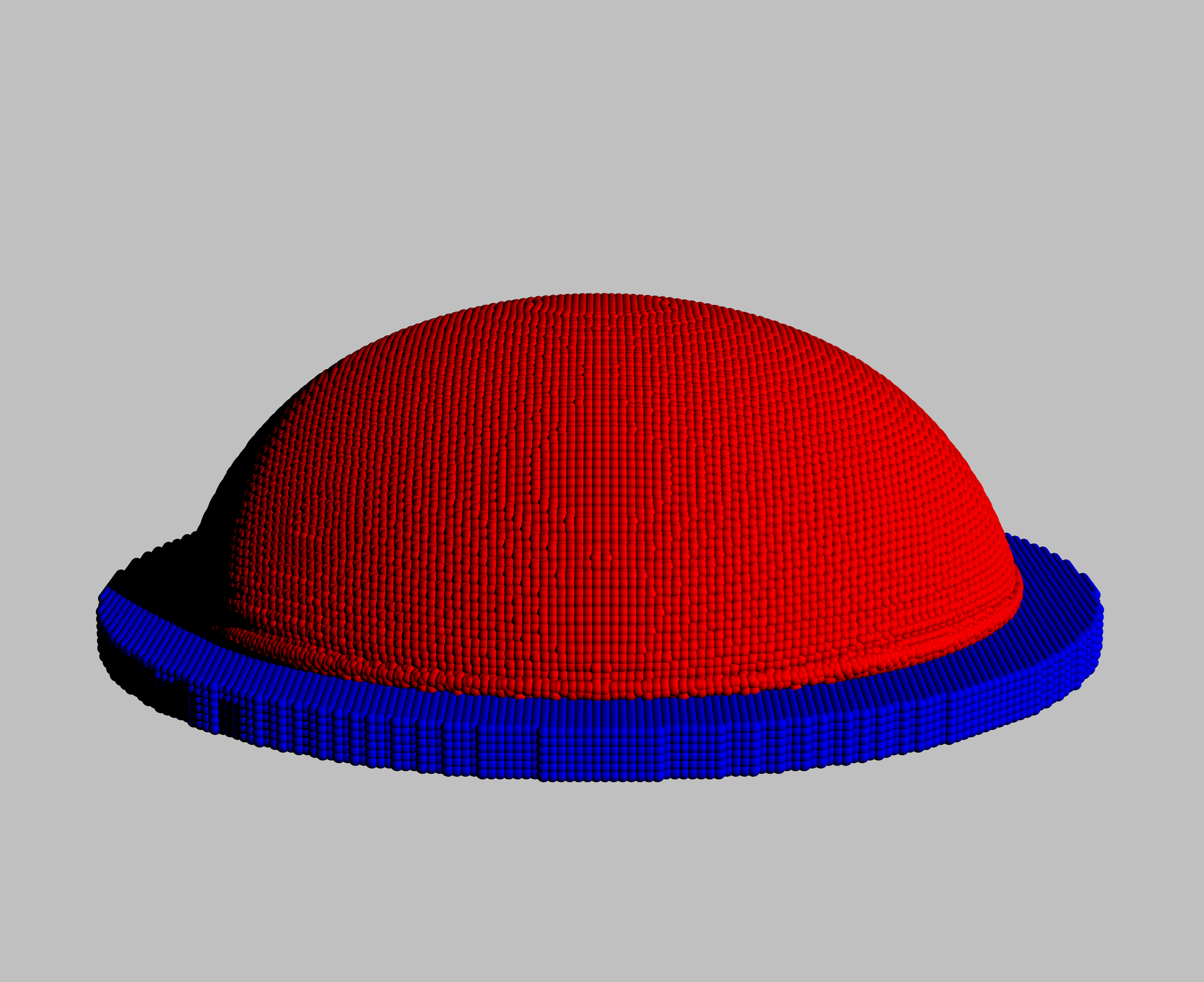}
        \caption{$t^* = 0.0$}
        \label{fig:osc_2}
    \end{subfigure}
    \begin{subfigure}[b]{0.32\textwidth}
        \centering
        \includegraphics[width=\textwidth]{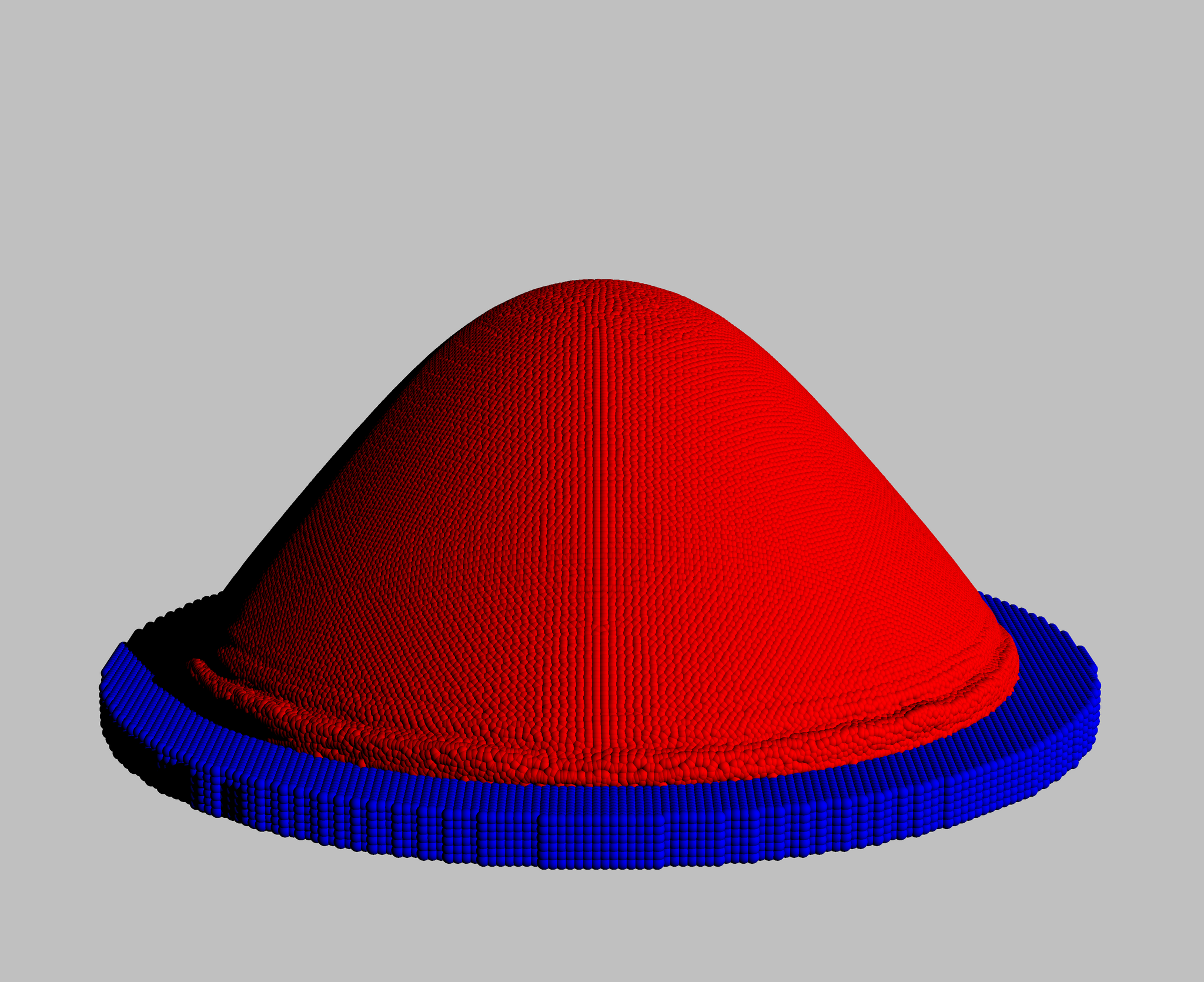}
        \caption{$t^* = 1/4$}
        \label{fig:osc_3}
    \end{subfigure}
    \begin{subfigure}[b]{0.32\textwidth}
        \centering
        \includegraphics[width=\textwidth]{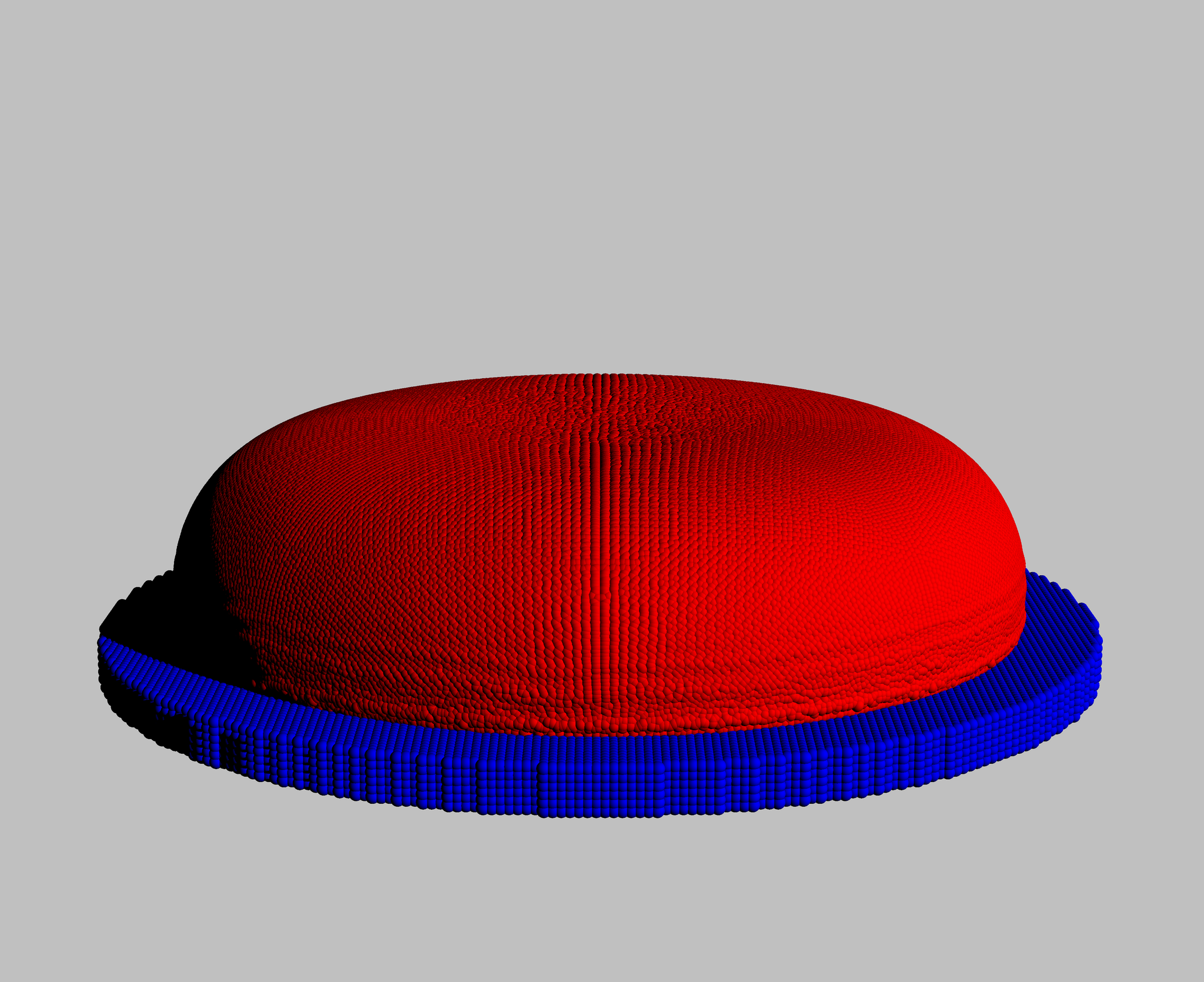}
        \caption{$t^* = 1/2$}
        \label{fig:osc_4}
    \end{subfigure}
    \begin{subfigure}[b]{0.32\textwidth}
        \centering
        \includegraphics[width=\textwidth]{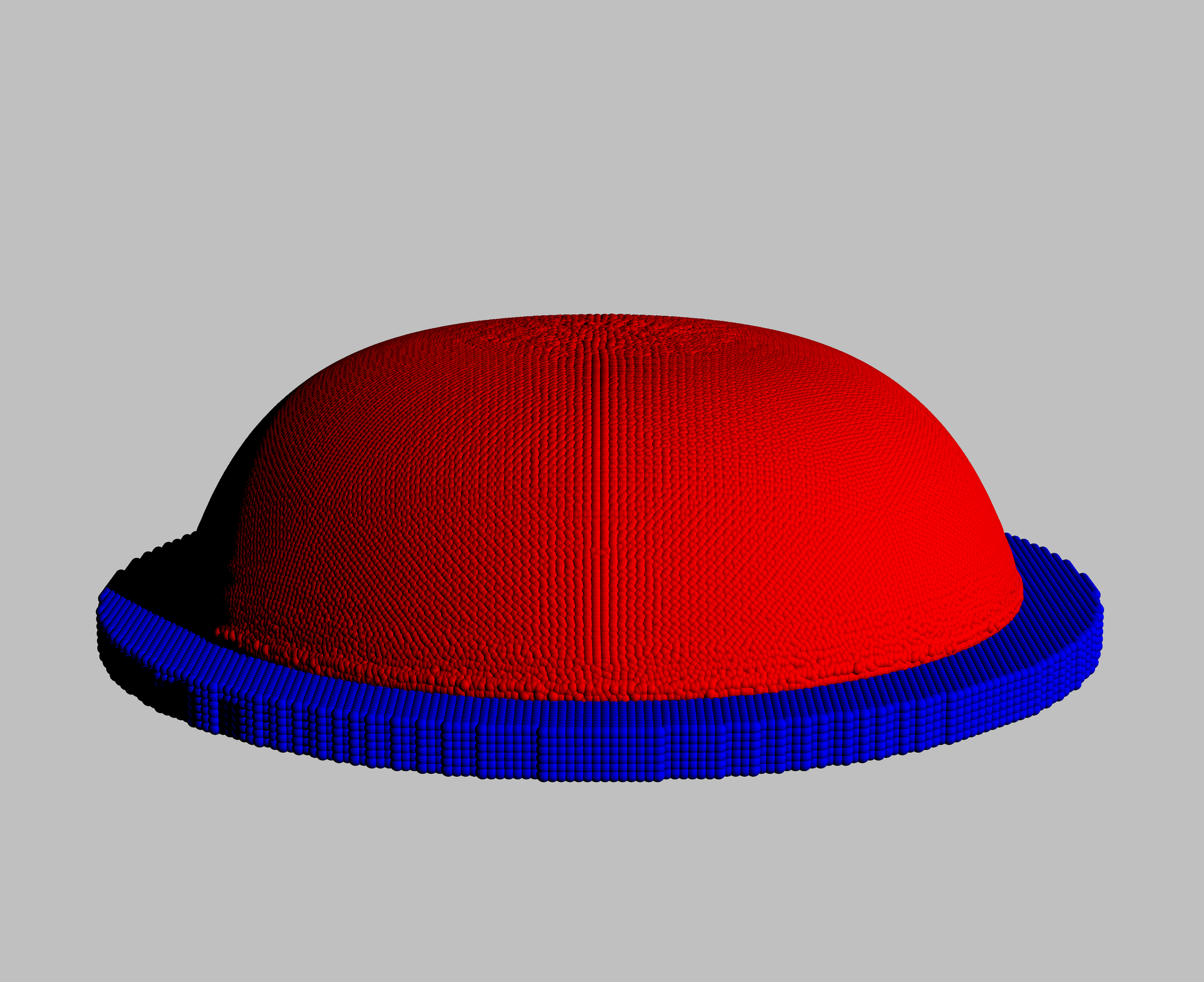}
        \caption{$t^* = 3/4$}
        \label{fig:osc_5}
    \end{subfigure}
    \begin{subfigure}[b]{0.32\textwidth}
        \centering
        \includegraphics[width=\textwidth]{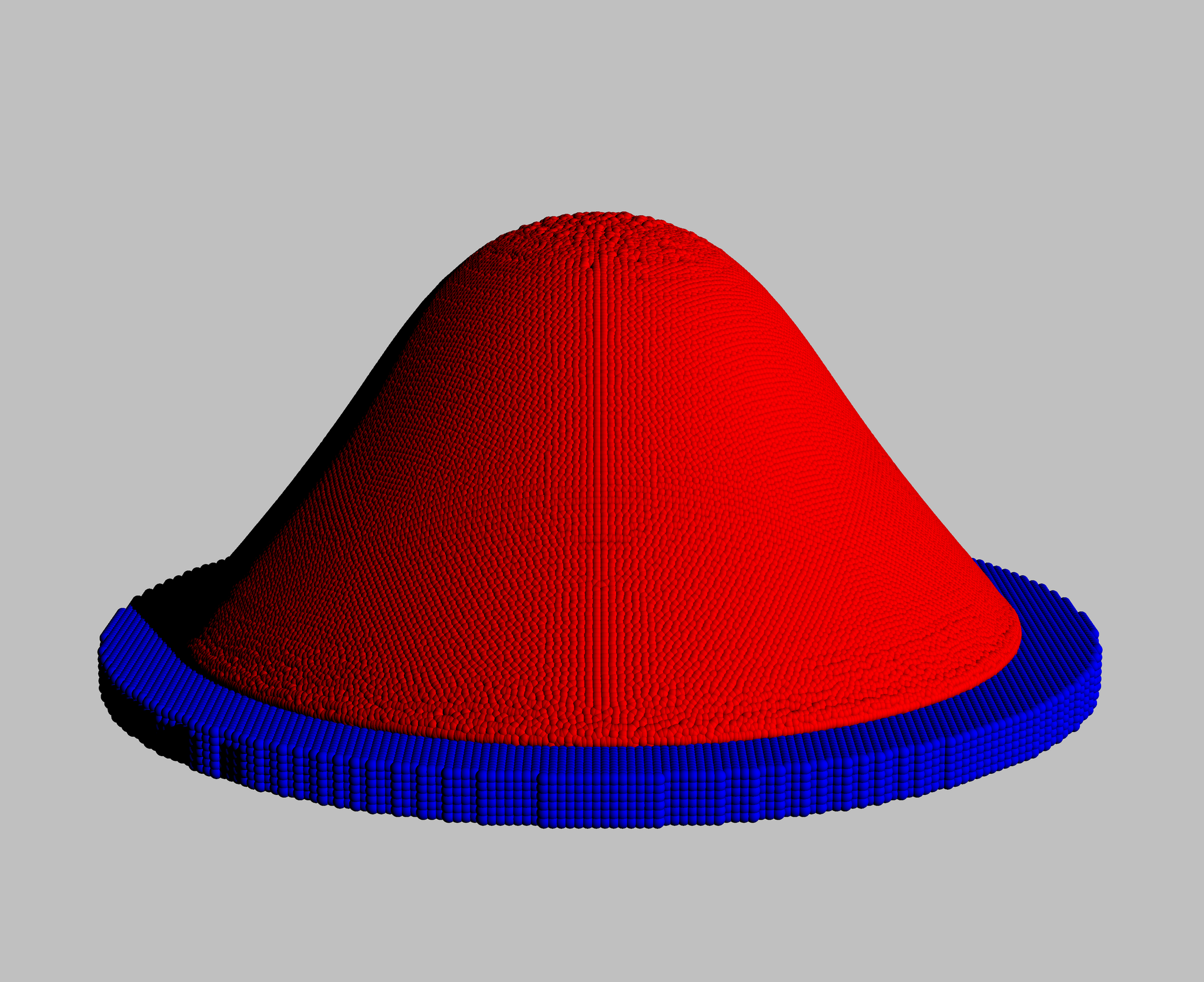}
        \caption{$t^* = 1$}
        \label{fig:osc_6}
    \end{subfigure}
    \caption{Shape evolution of a pinned sessile droplet during cyclic oscillation of the substrate. (a) Initialized state. (b) Shape at equilibrium ($t=0.005$ $sec$), before start of oscillation. Shapshots of droplet shape after reaching periodic state at (c) $t^* = 1/4$, (d) $t^* = 1/2$, (e) $t^* = 3/4$, and (f) $t^* = 1$. Here, $t^* = t/T$ wrapped around $[0,1]$, $t$ is the physical time, and $T$ is the oscillation time period of the substrate.}
    \label{fig:osc_sub}
\end{figure}

\begin{figure}[h!]
\centering
\includegraphics[width=0.8\textwidth]{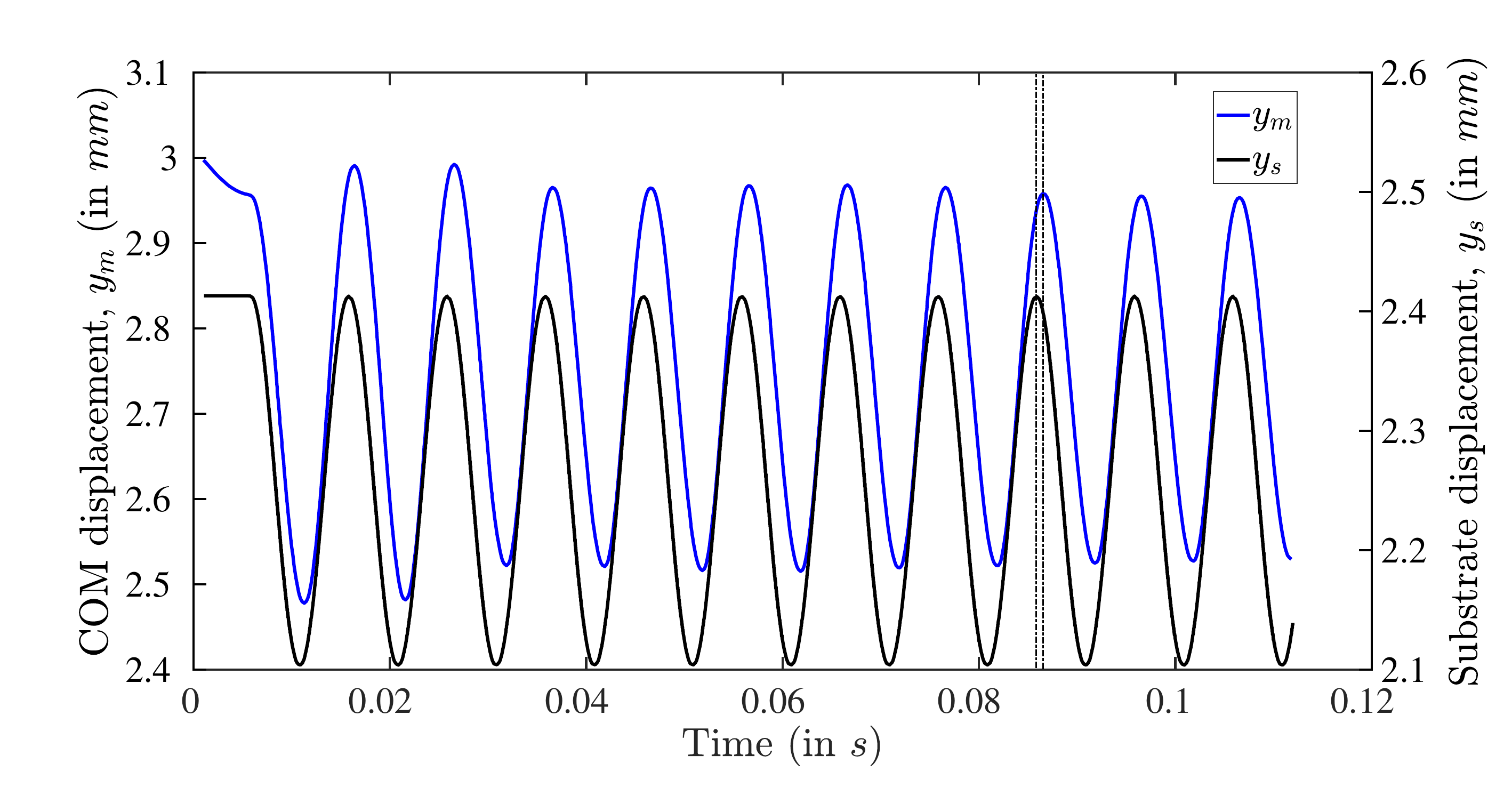}
\caption{Variation of position of the centre of mass (COM) of the droplet ($y_m$) and position of the oscillating substrate ($y_s$) with respect to time. The dashed vertical lines represent the location of peaks of the two curves, to highlight the small phase-shift observed in the SPH simulations.}
\label{fig:COM_270K}
\end{figure}

\begin{figure}[h!]
\begin{subfigure}[b]{0.59\textwidth}
        \includegraphics[width=\textwidth]{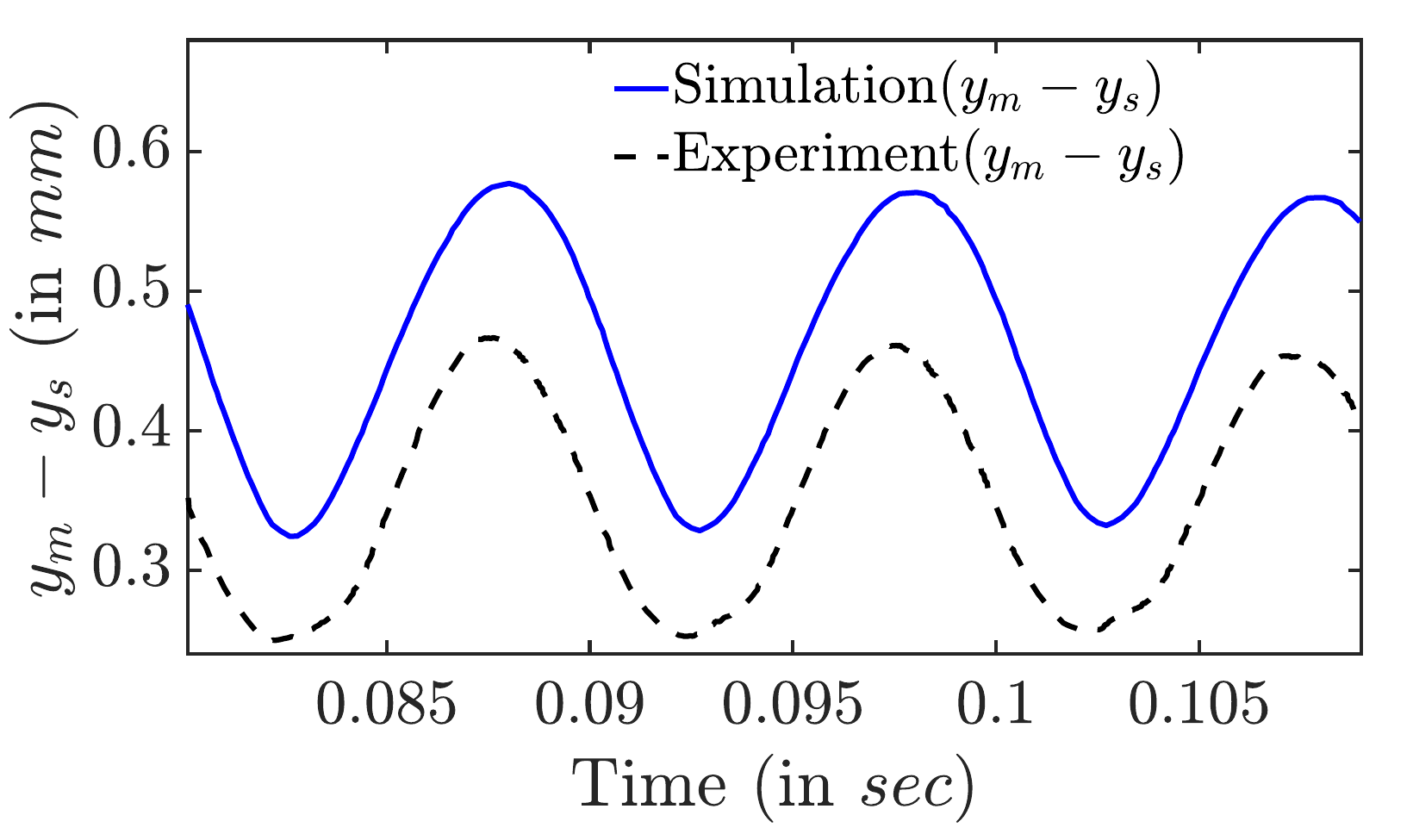}
        \caption{}
        \label{fig:deepu_comp_osc}
    \end{subfigure}
    \begin{subfigure}[b]{0.345\textwidth}
        \centering
        \includegraphics[width=\textwidth]{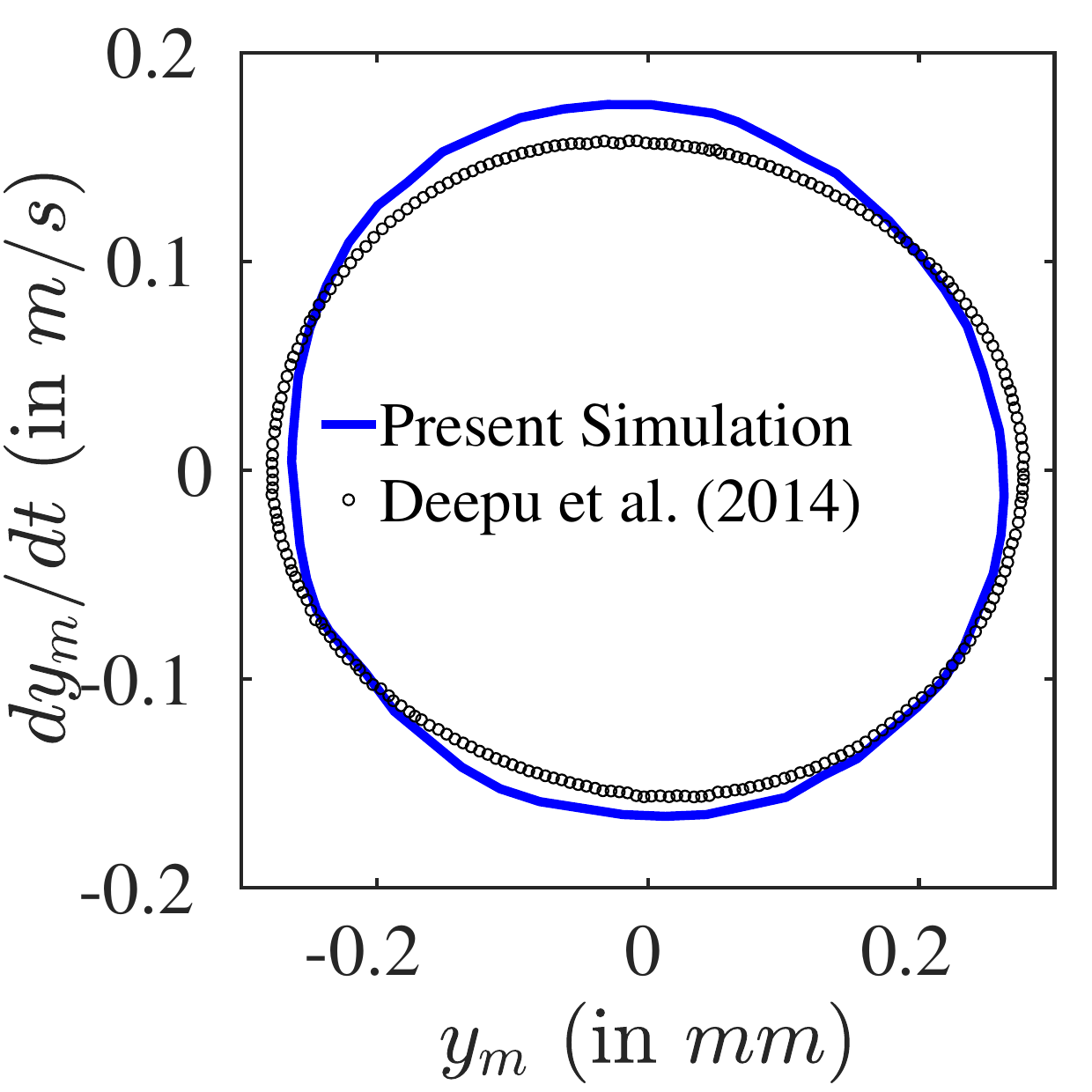}
        \caption{}
        \label{fig:phase_comp}
    \end{subfigure}
    \caption{(a) Comparison of time series of displacement of the center of mass (COM) of the droplet with respect to the substrate $\Delta y_m(t)=y_m(t)-y_s(t)$ between simulations and experiments \cite{deepu2014oscillation}. (b) Comparison of the phase trajectory $\dot{y}_m$-vs-$y_m'$ of the center of mass of the droplet between simulation and experiments \cite{deepu2014oscillation}.}
    \label{deepu_comp}
\end{figure}
\begin{figure}[h!]
\centering
\includegraphics[width=0.7\textwidth]{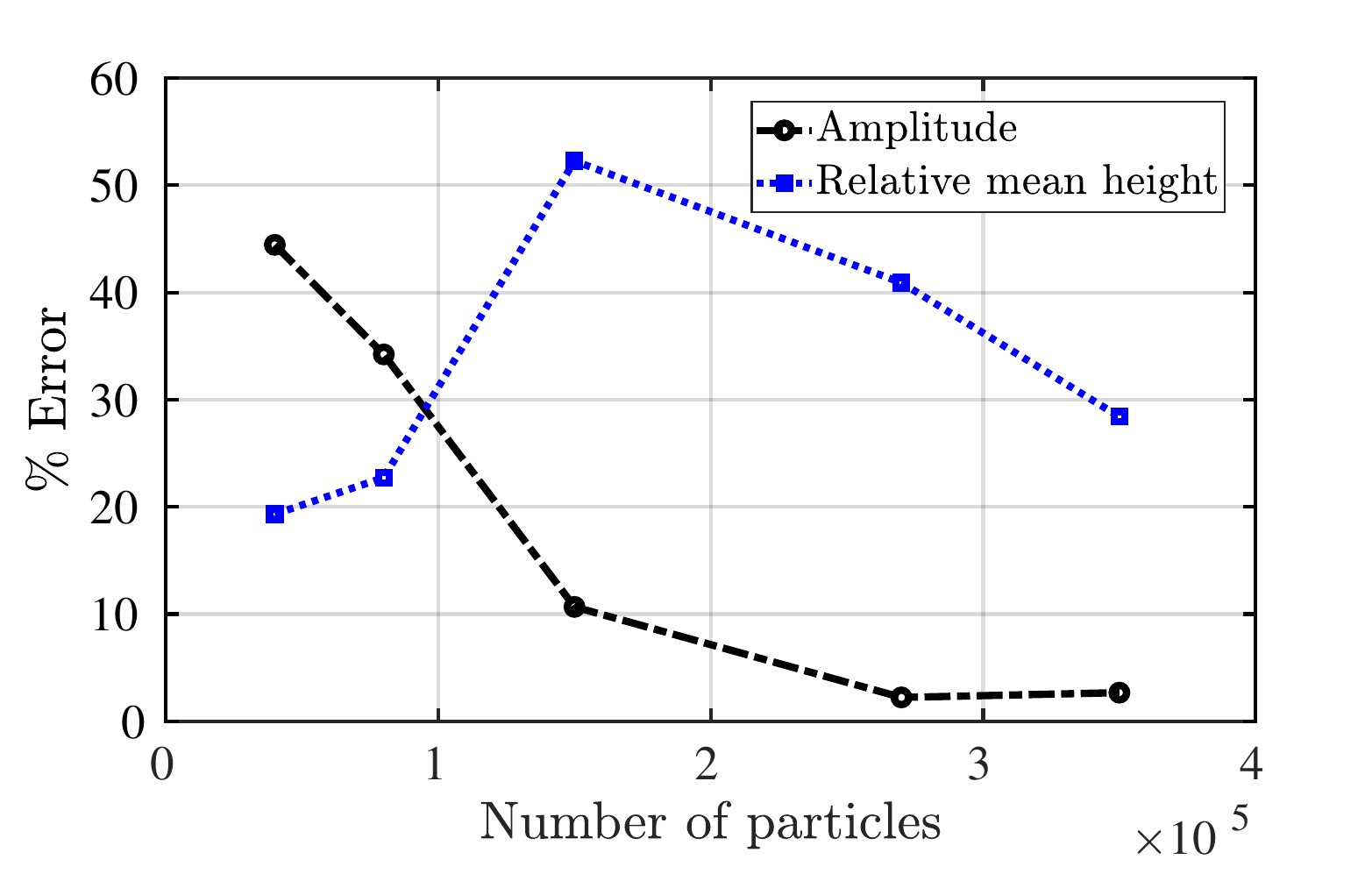}
\caption{Percentage relative error in amplitude and mean value of $\Delta y=y_m-y_s$ between simulation and experimental data \cite{deepu2014oscillation}, plotted with respect to number of SPH particles.}
\label{fig:error_osc}
\end{figure}

In this section, we further validate the proposed method with a dynamic test case of a pinned sessile droplet subjected to a vibrating substrate against experimental data \cite{deepu2014oscillation}. We reproduce the conditions of the experiment in the SPH simulation by considering the liquid droplet as mixture of $60 \%$ glycerol and $40\%$ water, in which the viscosity($\mu$) of liquid is $\SI{0.0128}{Pa.s}$, density($\rho$) is $\SI{1169.6}{Kg/m^3}$, and the surface tension coefficient($\sigma$) is $\SI{0.0667}{N/m}$. The liquid droplet is initialised as a spherical cap with a volume of $\SI{6.67}{\mu l}$. The Bond number of the sessile droplet is $0.49$, which is same as the experimental data. 
In the simulation, the substrate is kept fixed until $t=\SI{0.005}{s}$, at which point the liquid droplet reaches equilibrium. 
Fig. \ref{fig:osc_sub} shows the snapshots of the shape evolution of the sessile droplet, simulated with $3.5 \times 10^5$ particles,  as the substrate oscillates. The substrate is  oscillated vertically with a frequency ($\omega / 2 \pi$) of $\SI{100}{Hz}$; its vertical position is given as $y_s(t)=y_0+\delta y_s\cos(\omega t) $, where $y_0$ is a fixed reference height and $\delta y_s$ is the oscillation amplitude.  
Fig. \ref{fig:COM_270K} shows the variation of the substrate height $y_s$ and the height of center of mass (COM) of the sessile droplet $y_m$ with respect to time in a parallel coordinate plot. The initial horizontal line in the $y_s$-vs-$t$ curve represents the equilibration period during which no substrate motion is imposed, and the center of mass of the droplet reduces with time.  At a large value of time, the height of the COM of droplet reaches a periodic state, although a small phase shift remains between $y_m(t)$ and $y_s(t)$ (indicated by dashed vertical reference lines in Fig. \ref{fig:COM_270K}).
Fig. \ref{fig:deepu_comp_osc} compares the time series of the displacement of droplet COM with respect to the substrate position $\Delta y_m(t)=y_m(t)-y_s(t)$. The amplitude of the oscillation  of the droplet COM with the substrate position is in good agreement with the experiment reported. There is a large difference in the mean relative height of the COM $\overline{\Delta y_m}$ predicted by the simulation in comparison with experiments; here, $\bar{(\cdot)}$ indicates cycle averaged quantities. On the other hand, the amplitude of oscillations predicted by the SPH simulations match the experimental values reasonably well. 
The variation of time derivative of the COM displacement ($\dot y_m$) with the fluctuation in COM position $y_m'(t)=y_m(t)-\overline{y}_m$, is shown for both SPH simulations and experimental data in Fig. \ref{fig:phase_comp}. {The time range considered for the phase plot is from $\SI{0.0164}{s}$ to $\SI{0.0267}{s}$}. The phase plot is a closed circular curve for  both the simulation and the experiment, indicating that the oscillation of COM reaches a periodic and approximately sinusoidal state. The range of $\dot{y}_m$ and $y_m'(t)$ from the two data sets also compare reasonably well. A grid convergence study is performed for which the number of particles are varied from $8.0\times10^4$ to $3.5\times 10^5$. The absolute percentage error in the amplitude and the mean height with respect to the number of particles are shown in Fig. \ref{fig:error_osc}. 
With increasing number of liquid particles, the magnitude of error between predicted amplitude of $\Delta y_m(t)$ and the experimental value reduces drastically; the percentage error remains below $2.7\%$ for simulations having  $2.7\times10^5$ or higher number of SPH particles. 
The absolute percentage error in the mean height $\overline{y}_m$ reduces relatively slowly, reaching values below $30\%$ for simulation with $3.5\times10^5$ liquid particles. Results from this test case demonstrate that the pinning algorithm allows for simulation of pinned sessile droplets in the presence of substrate oscillations. While a reasonably good match is observed for the amplitude of oscillation of $\Delta y_m$ between experiments and simulation, higher grid resolution is required for matching the mean value of $\Delta y_m$ from experiments.  

\subsection{Axial extension of a liquid bridge with pinned contact line}

\begin{figure}[h!]
    \centering
    \begin{subfigure}[b]{0.32\textwidth}
        \centering
        \includegraphics[width=\textwidth]{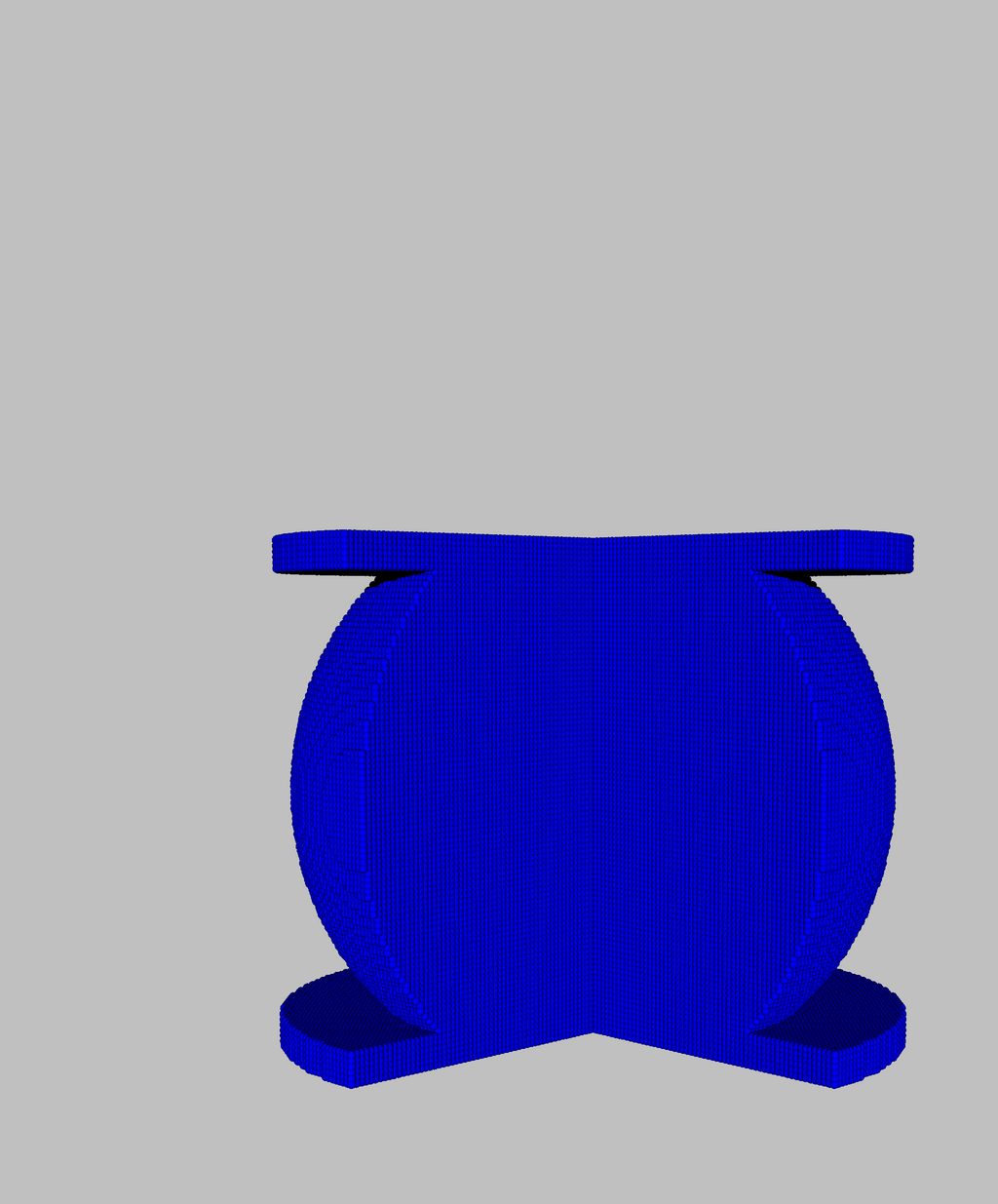}
        \caption{$t/T = 0.0$}
        \label{fig:lb1}
    \end{subfigure}
    \begin{subfigure}[b]{0.32\textwidth}
        \centering
        \includegraphics[width=\textwidth]{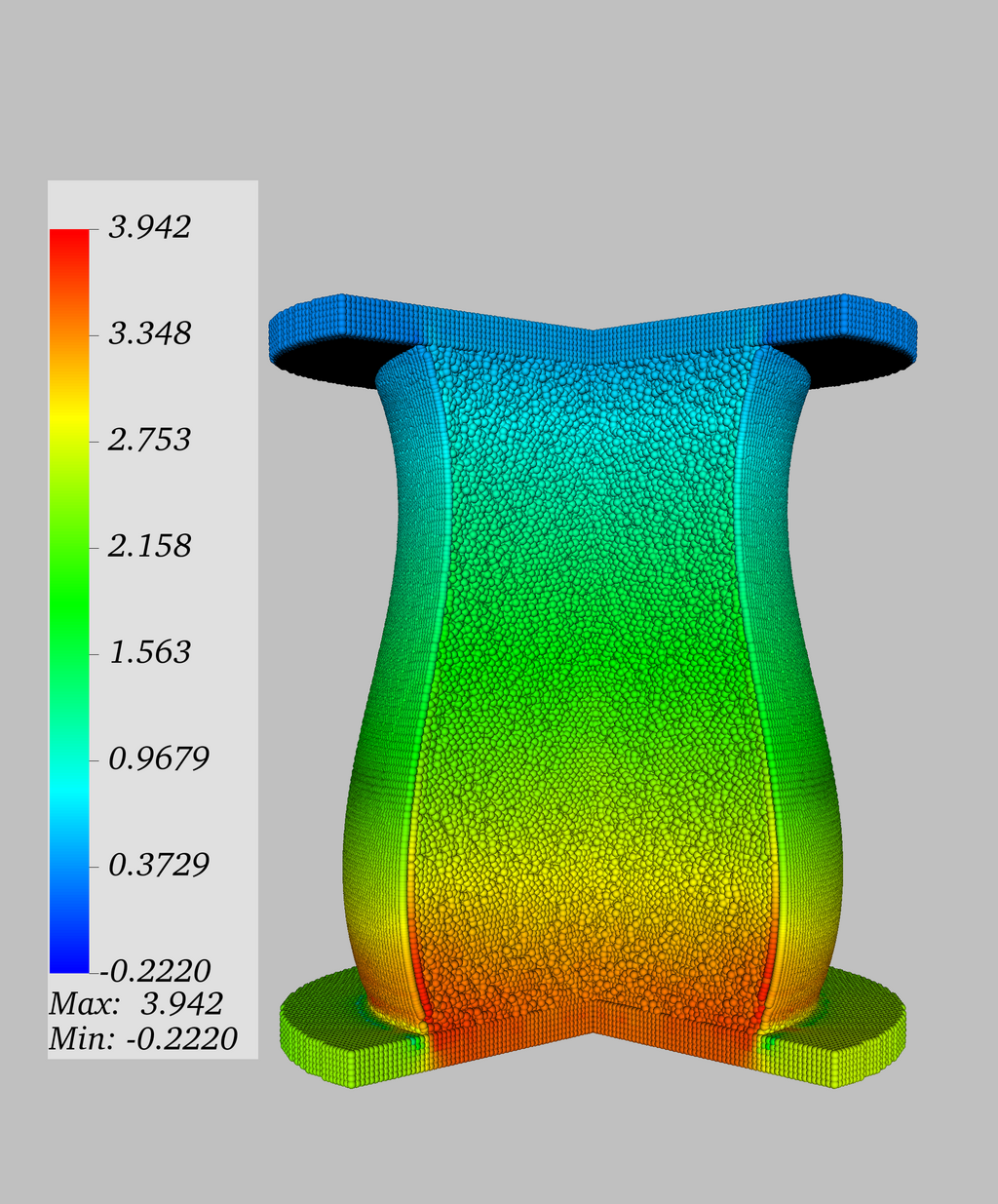}
        \caption{$t/T = 20.75$}
        \label{fig:lb2}
    \end{subfigure}
    \begin{subfigure}[b]{0.32\textwidth}
        \centering
        \includegraphics[width=\textwidth]{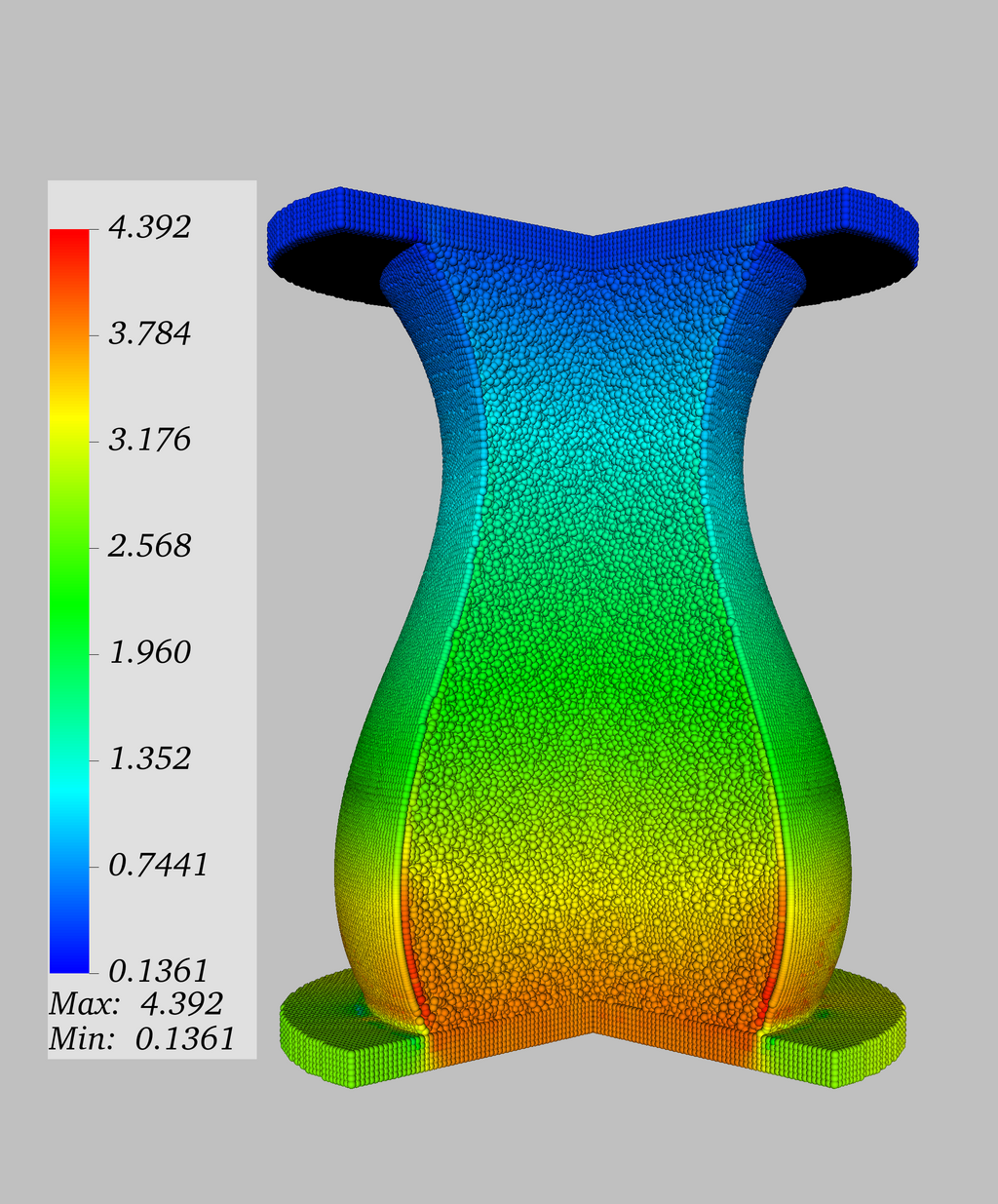}
        \caption{$t/T = 30.0$}
        \label{fig:lb3}
    \end{subfigure}
    \begin{subfigure}[b]{0.32\textwidth}
        \centering
        \includegraphics[width=\textwidth]{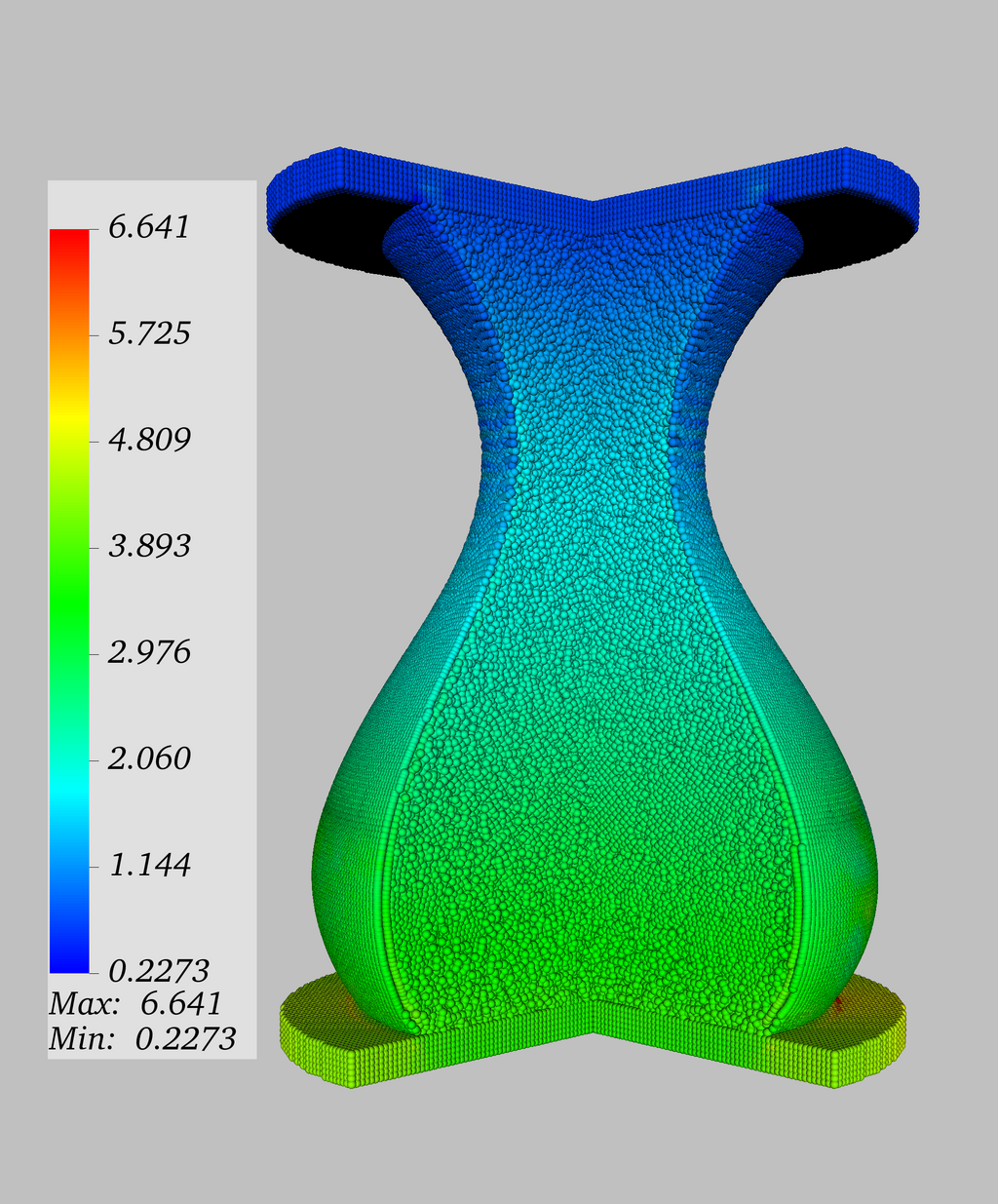}
        \caption{$t/T = 33.41$}
        \label{fig:lb4}
    \end{subfigure}
    \begin{subfigure}[b]{0.32\textwidth}
        \centering
        \includegraphics[width=\textwidth]{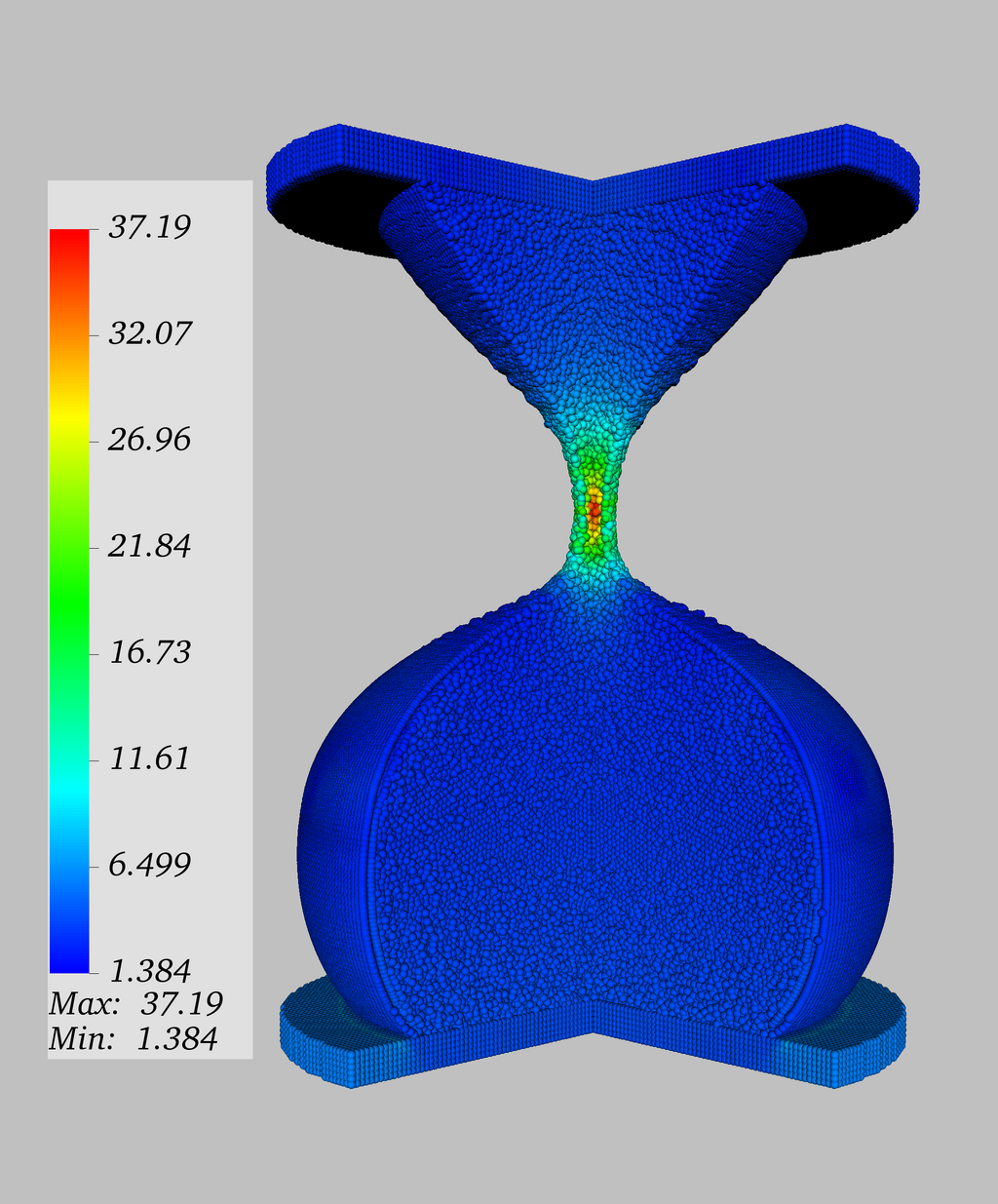}
        \caption{$t/T = 35.4$}
        \label{fig:lb5}
    \end{subfigure}
    \begin{subfigure}[b]{0.32\textwidth}
        \centering
        \includegraphics[width=\textwidth]{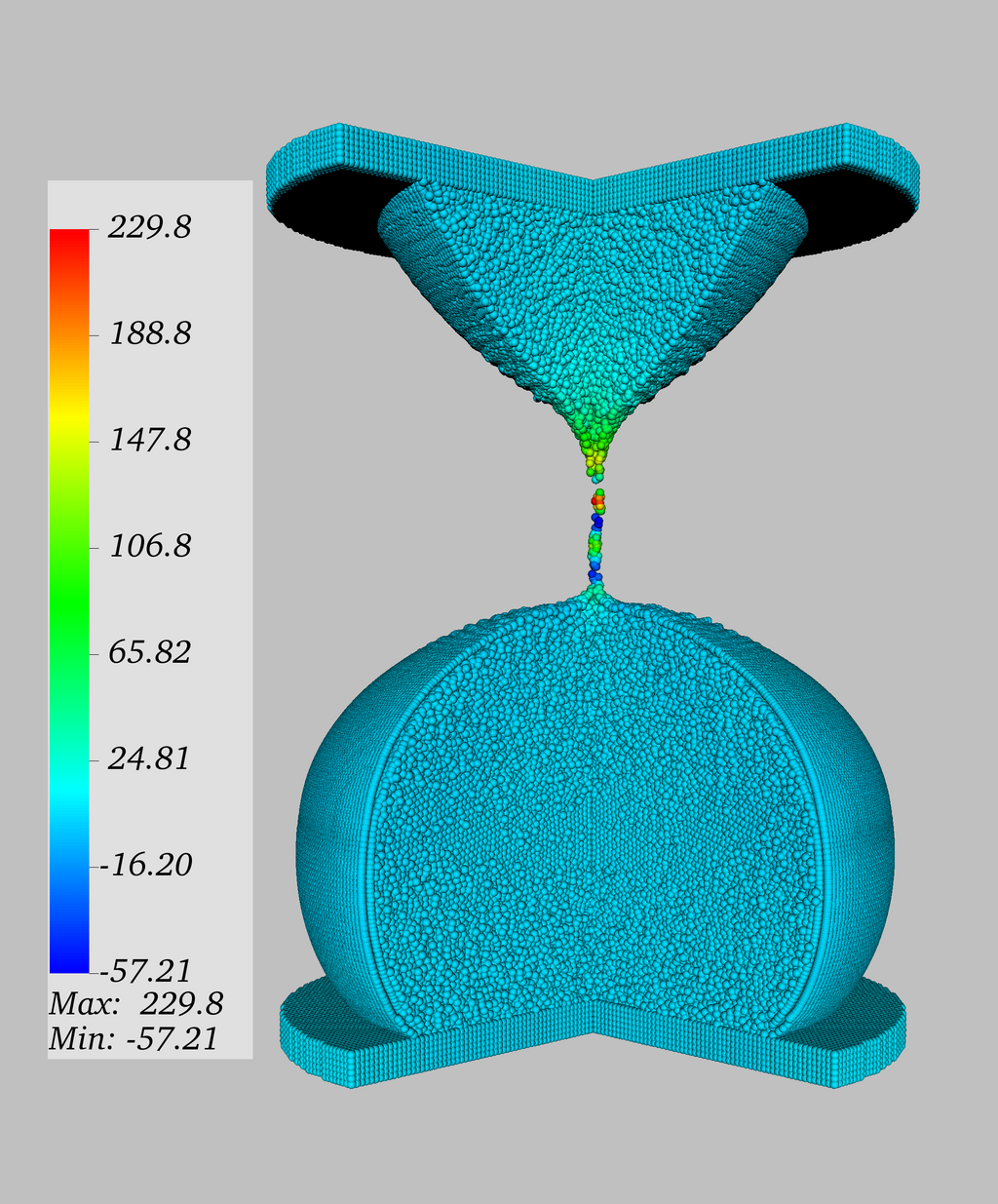}
        \caption{$t/T = 35.49$}
        \label{fig:lb6}
    \end{subfigure}
    \caption{Evolution of stretched liquid bridge, showing rendering of SPH particles and iso-contours of dimensionless pressure ($p R_p/(\sigma)$) along cross sectional planes at different dimensionless time instances ($t/T$, where $T=\sqrt{R_p/g}$).}
    \label{fig:pinned_lb}
\end{figure}
\begin{figure}[h!]
    \centering
    \begin{subfigure}[b]{0.49\textwidth}
        \centering
            \includegraphics[width=\textwidth]{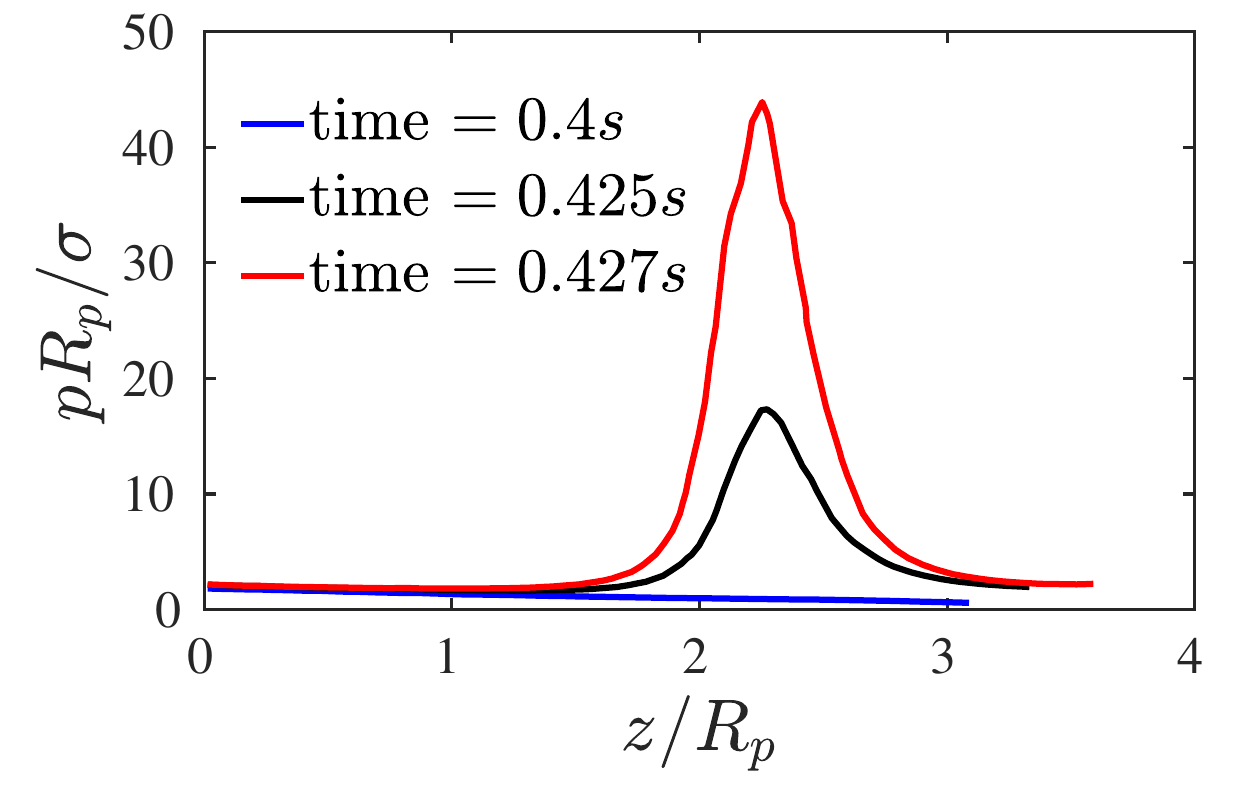}
        \caption{}
        \label{fig:lb_cl_p}
    \end{subfigure}
    \begin{subfigure}[b]{0.49\textwidth}
        \centering
        \includegraphics[width=\textwidth]{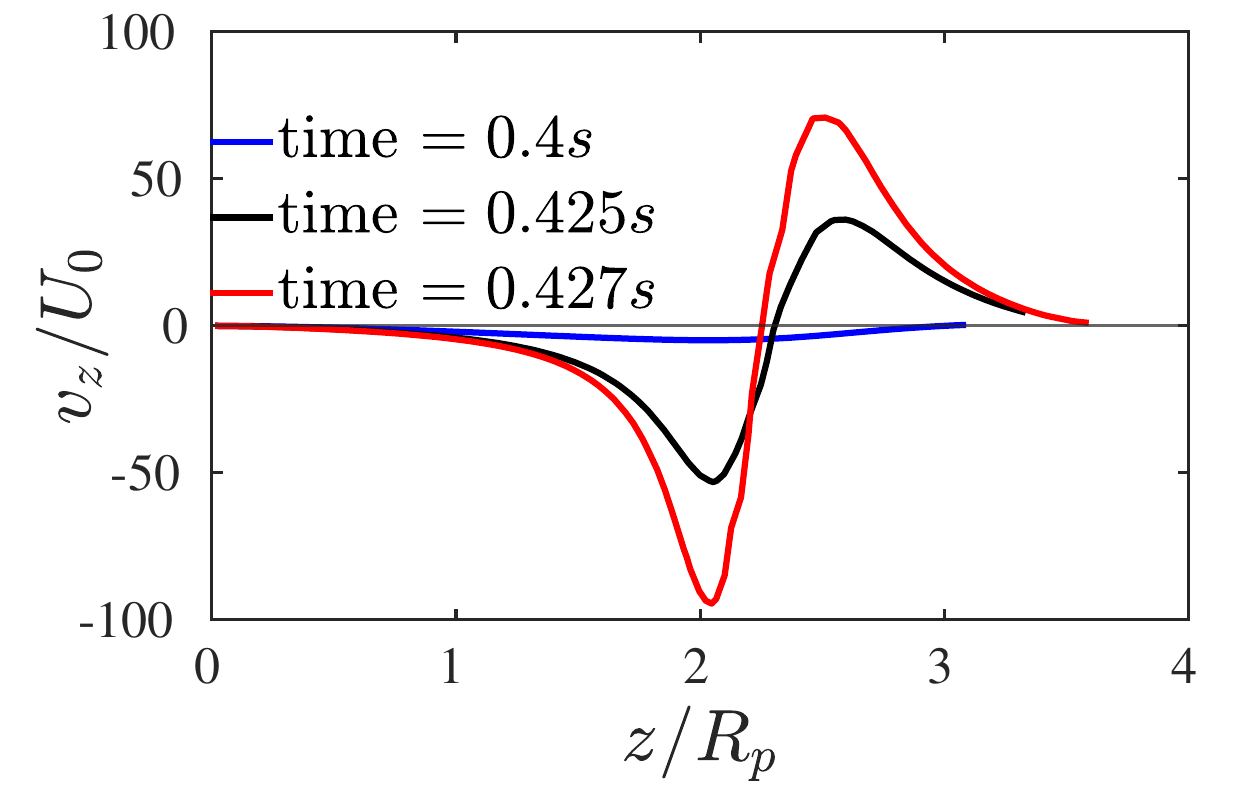}
        \caption{}
        \label{fig:lb_cl_vz}
    \end{subfigure}
    \caption{(a) Variation of non-dimensional pressure ($p R_p/\sigma$) along the non-dimensional axial length($z/R_p$) of the liquid bridge at different time instances. (b) Variation of non-dimensional axial velocity ($v_z/U_0$) variation along the non-dimensional axial length of the liquid bridge at different time instances.}
    \label{fig:cl_liq_br}
\end{figure}
\begin{figure}[h!]
    \centering
    \begin{subfigure}[b]{0.24\textwidth}
        \centering
        \includegraphics[width=\textwidth]{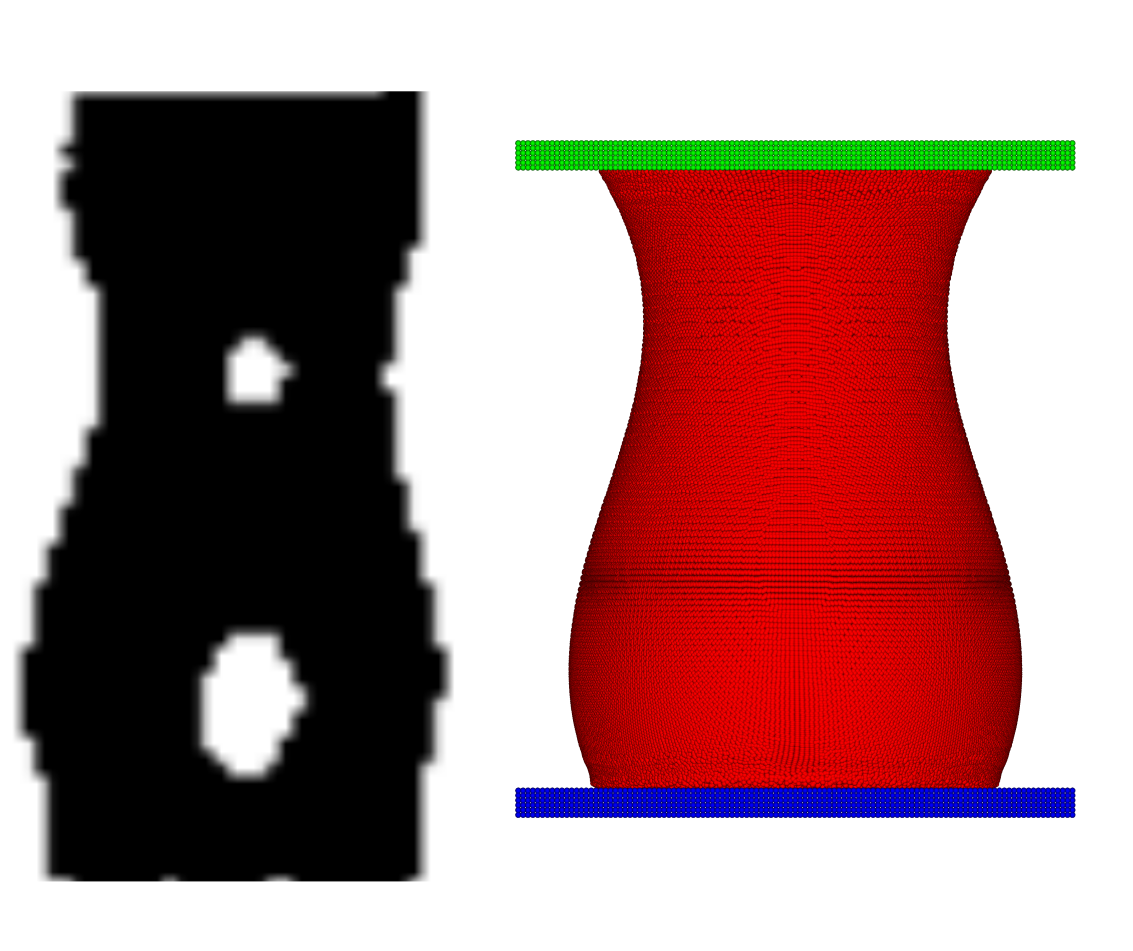}
        \caption{$r_{min}/R_p = 0.74$}
        \label{fig:zhang_profile_comparison1}
    \end{subfigure}
    \begin{subfigure}[b]{0.24\textwidth}
        \centering
        \includegraphics[width=\textwidth]{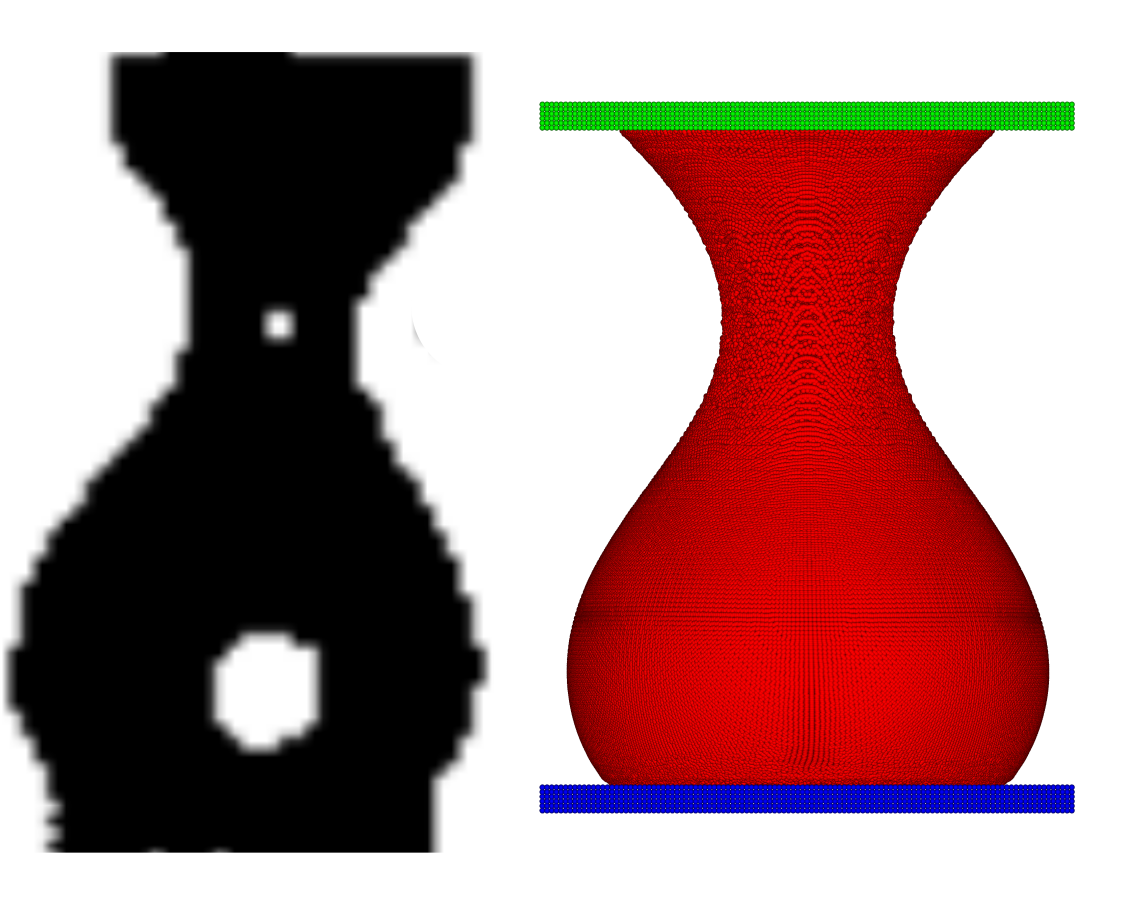}
        \caption{$r_{min}/R_p = 0.42$}
        \label{fig:zhang_profile_comparison2}
    \end{subfigure}
        \begin{subfigure}[b]{0.24\textwidth}
        \centering
        \includegraphics[width=\textwidth]{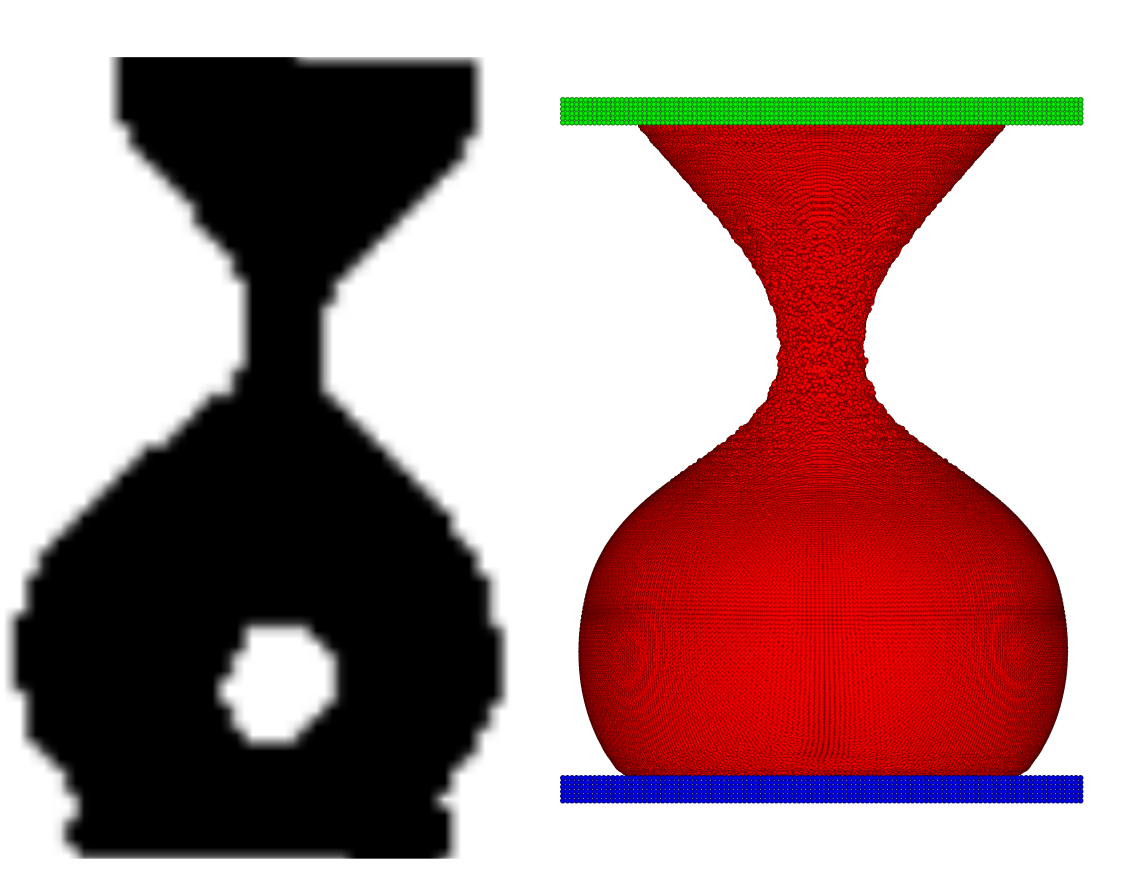}
        \caption{$r_{min}/R_p = 0.3$}
        \label{fig:zhang_profile_comparison3}
    \end{subfigure}
        \begin{subfigure}[b]{0.24\textwidth}
        \centering
        \includegraphics[width=\textwidth]{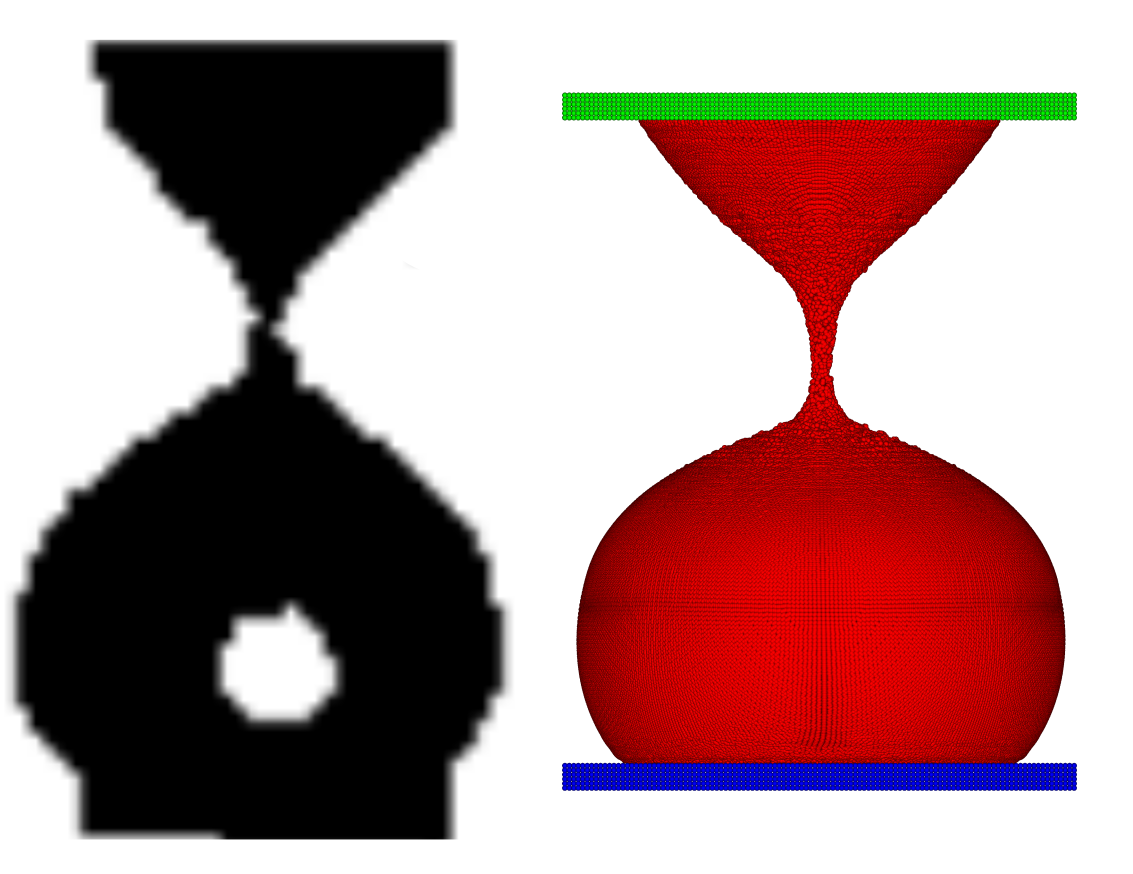}
        \caption{$r_{min}/R_p = 0.2$}
        \label{fig:zhang_profile_comparison4}
    \end{subfigure}
    \caption{Side by side comparison of the evolved surface from numerical simulation with images from the experiment by \citet{zhang1996nonlinear}. The comparison is made for images corresponding to identical non-dimensional minimum radius ($r_{min}/R_p$) of the liquid bridge. The images taken from the experiment here are screenshots from the article, and the aspect ratio of the images are preserved. The vertical location of top and bottom substrates in the images from from experiments and simulations are matched.}
    \label{fig:zhang_profile_comparison}
\end{figure}
\begin{figure}[h!]
        \centering
        \includegraphics[width=0.6\textwidth]{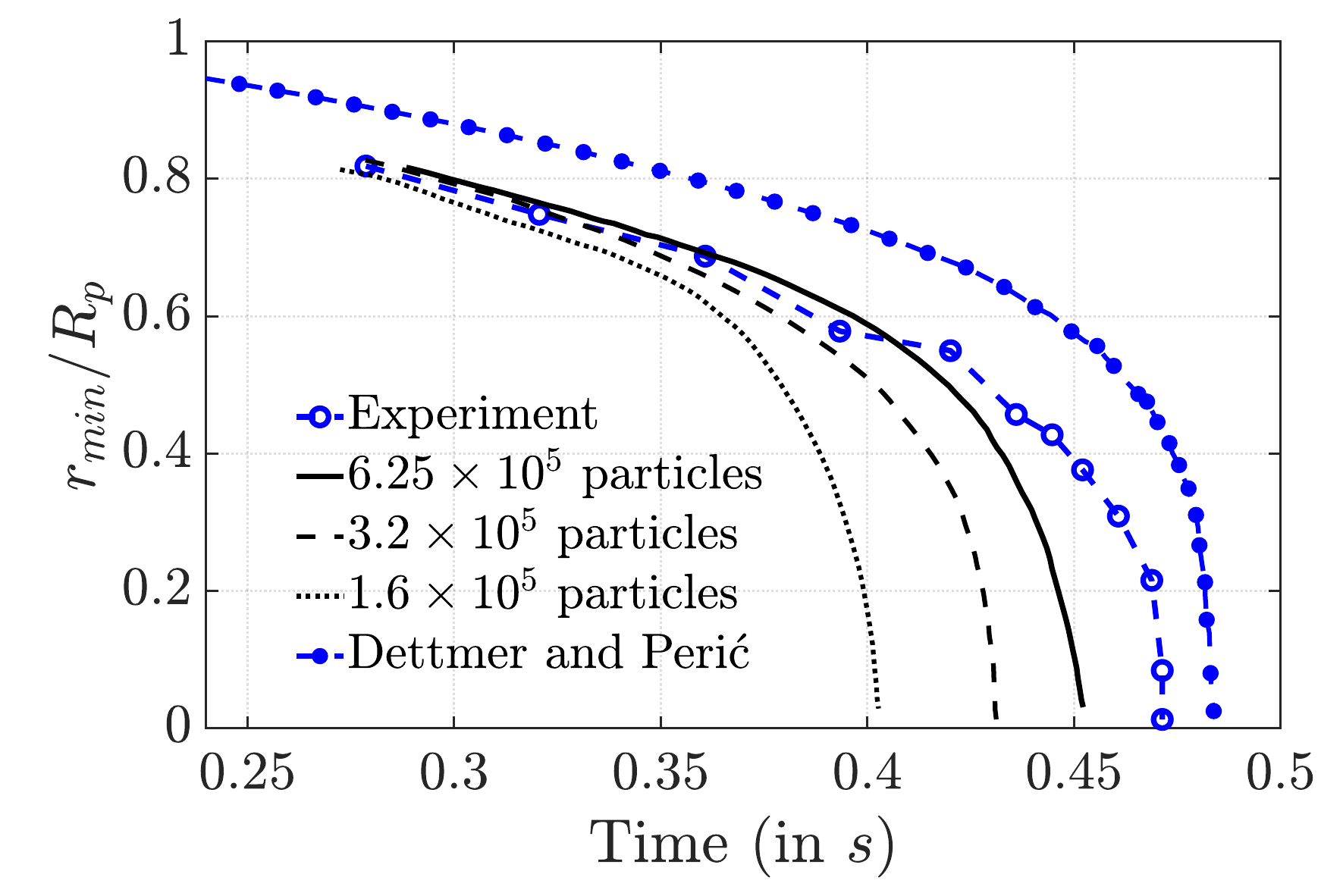}
    \caption{A comparison of the evolution of dimensionless minimum radius($r_{min}/R_p$) obtained from the numerical simulation with experiment\cite{zhang1996nonlinear} and finite element axisymmetric solution \cite{dettmer2006computational} obtained with $1350$ finite elements.The comparison is shown for increasing number of SPH particles from $1.6\times 10^5$ liquid particles to $6.25\times 10^5$ liquid particles.}
    \label{fig:liq_min_rad_comp}
\end{figure}
Liquid bridges have been extensively studied experimentally for Newtonian fluids \cite{zhang1996nonlinear, padday1997, qian2011} and viscoelastic fluids for multiple decades and have been numerically simulated \cite{zinelis2024transition} in two dimensions to study the stretching and break-up behaviours with the axisymmetric assumption. In these studies the major focus has been on the stretching region where a thread is formed. Several experimental studies report liquid bridges bounded on the substrate by deliberate pinning of the contact line to a surface discontinuity \cite{zhang1996nonlinear, dodds2012dynamics}. The stretching motion may break the symmetry \cite{dodds2012dynamics} of the liquid bridge warranting a three dimensional numerical simulation.
In this test case, we compare our results with the experiment by \citet{zhang1996nonlinear} on axial stretching of liquid bridge between two flat substrates being pulled apart at a constant velocity. The three-phase contact line of the liquid bridge is pinned on the substrates along  circular pinning curves.
To reproduce the experimental conditions \cite{zhang1996nonlinear}, we consider the liquid to be $85\%$ glycerol and water solution with viscosity $\mu=\SI{0.1129}{Pa s}$, surface tension coefficient $\sigma=\SI{0.066}{N/m}$ and density $\rho=\SI{1223}{kg/m^3}$. The liquid bridge is initialized as a truncated sphere held between two substrates separated by a distance $h_0$, such that its volume is $\SI{0.04}{cm^3}$. The radius of pinning curve on both the substrates is $R_p=\SI{0.16}{cm}$ and the initial slenderness ratio is $h_0/R_p=2$. The Bond number $Bo\,(\rho g R_p^2 / \sigma)$ is 0.46, the Capillary number  $Ca\, (\mu U_0/\sigma)$ is 0.01 and the Reynolds Number $Re\,(\rho U_0 R_p/ \mu)$ is 0.104. The bridge is stretched by moving the top substrate vertically upwards with a constant velocity of $U_0=\SI{0.6}{cm/s}$, while keeping the bottom plate fixed. 
The evolution of the liquid bridge is simulated with $6.25\times 10^5$ SPH particles. 

Figure \ref{fig:pinned_lb} shows snapshots of the liquid bridge at different time instances. Due to the presence of gravity in the negative $z$ direction, the liquid bridge displays a bulge in the lower half of the domain and a neck in the upper half of the domain. The minimum radius of the necking region reduces with time, until a thin liquid thread connects the sessile liquid mass at bottom with the pendant liquid mass at the top substrates. The non-dimensional pressure contour at the center-plane of the liquid bridge (contours in Fig. \ref{fig:pinned_lb}) indicates that the pressure varies primarily along the axial direction at such low Capillary numbers. The axial variation of centre-line dimensionless pressure ($p R_p/\sigma$) and dimensionless axial velocity ($v_z/U_0$) is plotted at different time instances in Fig. \ref{fig:cl_liq_br}, starting from $t=0.4$ s. It is clear from the dimensionless pressure plot (Fig. \ref{fig:lb_cl_p}) that with increasing time, the pressure in the neck region increases, primarily due to thinning of the neck. This leads to flushing of liquid from the neck region towards the two ends of the liquid bridge, as shown in Fig. \ref{fig:lb_cl_vz}, where the axial velocity becomes negative below the neck and positive above the neck. This positive feedback between the large local pressure at the neck and the efflux of liquid towards the end of the liquid bridge leads to further thinning and eventual breakage of the neck region. 

To validate the shape of the liquid bridge, we show a side-by-side comparison of the shape of the liquid bridge predicted by the SPH simulations with the images of experiments performed by \citet{zhang1996nonlinear} (see Fig. \ref{fig:zhang_profile_comparison}).
It can be observed that the shapes of the pendant and sessile masses, and the neck region in the simulation agree reasonably well with the experiment  until $r_{min}/R_p\geq 0.3$ (Fig. \ref{fig:zhang_profile_comparison}(a)--(c)). For lower values of $r_{min}$ (say $r_{min}/R_p=0.2$ shown in Fig. \ref{fig:zhang_profile_comparison}(d)), when a thin thread is formed, we observe differences between the experiment and the simulation due to insufficient resolution of the SPH domain in the necking region.

A grid convergence test is performed for this case, in which the number of particles are varied from $1.6\times 10^5$ to $6.25\times 10^5$. Fig. \ref{fig:liq_min_rad_comp} shows the variation of $r_{min}/R_p$ with time for different grid resolutions. We observe excellent grid convergence up to $t=0.4$ s, for which the neck region is well resolved by the SPH particles.  Beyond this time instance, the grid convergence is poor, which is not surprising given the small radius of the liquid bridge thread in this regime. The simulation with $6.25\times 10^5$ particles shows the best agreement with the experimental measurements for the time evolution of $r_{min}$. 
Since pinning of contact line on plane substrates have not been attempted using Eulerian mesh based methods to the best of our knowledge, a comparison with simulations using axisymmetric finite element method by \citet{dettmer2006computational} is presented in Fig. \ref{fig:liq_min_rad_comp}. It is evident that the proposed scheme predicts the minimum radius evolution during the initial and intermediate part of the stretching better than that predicted using the axisymmetric finite element method. 
Near the pinch off, the prediction is evidently resolution dependent. Nevertheless, the near-pinch off regime can be modeled using a universal scaling law, as explained in \cite{eggers1993universal}.
Results from this test case demonstrate that the pinning algorithm is able to capture the essential flow physics of a stretched liquid bridge with pinned contact line, while at the same time reproducing the experimental measurements by \citet{zhang1996nonlinear} reasonably well.

\section{\label{sec:concl} Conclusion}

In this work, we propose a model for pinning the three-phase contact line of a liquid drop in the presence of a substrate. The Continuum Surface Model (CSF) is modified for implementing the scheme. The proposed approach identifies a ``near-substrate" region in the vicinity of the pinning curve where the standard CSF model is modified. The surface tension force for interfacial particles in the near-substrate region is directed along a unit normal that is perpendicular to the position vector of the interfacial particle with respect to the pinning curve. The unit normal prescribed at these interfacial particles is then extrapolated to all the neighboring particles in the near-substrate region. The unit normal vector of the surface tension force is smoothly blended between the near-substrate and CSF region. 

Several test cases are carried out to highlight the robustness and versatility of the technique. In all the test cases, the contact angle remains in the limit of $30^o<\theta<155^o$, which is in the limit discussed in the methodology. A limitation of the model is to capture the contact line on super-hydrophilic or super-hydrophobic substrates. No studies have been conducted with respect to Marangoni flows and evaporation at the interface.
The pinning scheme is able to successfully simulate the spreading of the three-phase contact line towards circular and wavy pinning curves. The contact angle predicted for pinned droplets in equilibrium is consistent with the droplet volume and radius of the pinning curve. The scheme also functions well during simulations of the droplets pinned on an oscillating substrate and the liquid bridge being stretched between the substrates. The latter simulations indicate the robustness of the pinning scheme in the presence of dynamic forcing of the substrate as well as multiple substrates. Reasonable agreement was observed between the results from the ISPH simulations and the experimental measurements for several of the cases. In our future work, the pinning model will be used to simulate the wetting of liquid droplets on topographically complex substrates. Additionally, simulation of thermocapillary effects due to surface tension gradients and resulting convection near the pinning region is also an interesting direction for future research. 

\appendix
\section{Derivation of the Young-Laplace equation for a pinned droplet under the influence of gravity. \label{appendix:YL_equation}}
\begin{figure}[h!]
    \centering
    \begin{subfigure}[b]{0.49\textwidth}
        \centering
        \includegraphics[width=2in]{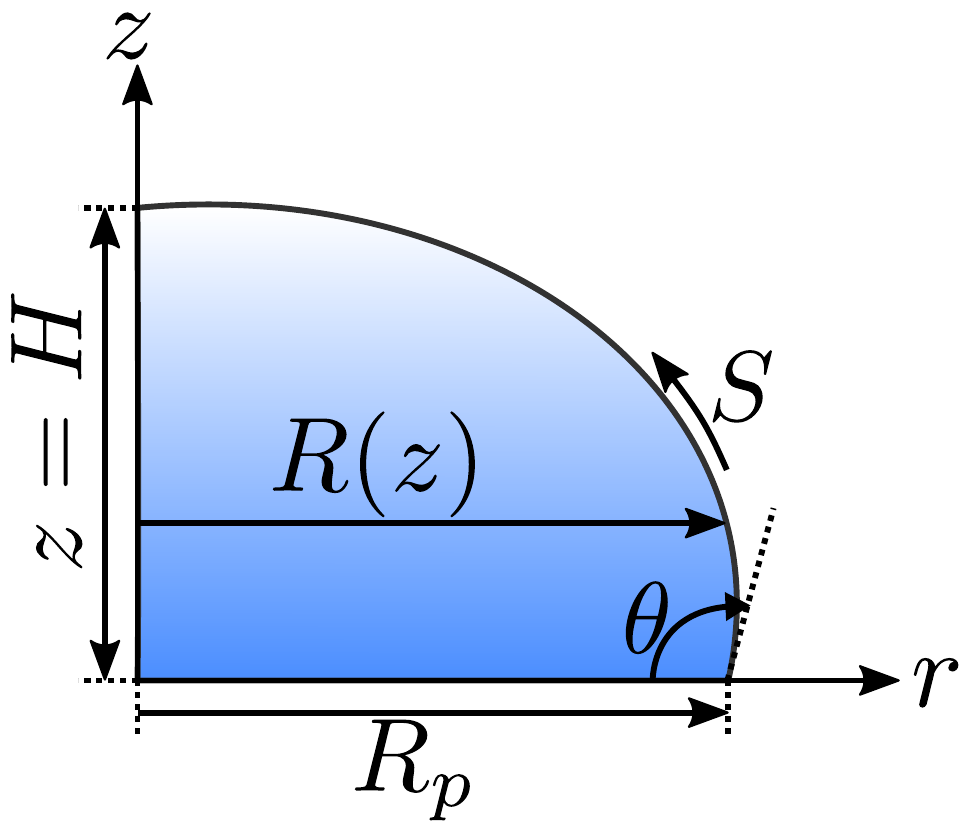}
        \caption{}
        \label{fig:YL_droplet_schematic}
    \end{subfigure}
    \begin{subfigure}[b]{0.49\textwidth}
        \centering
        \includegraphics[width=2.5in]{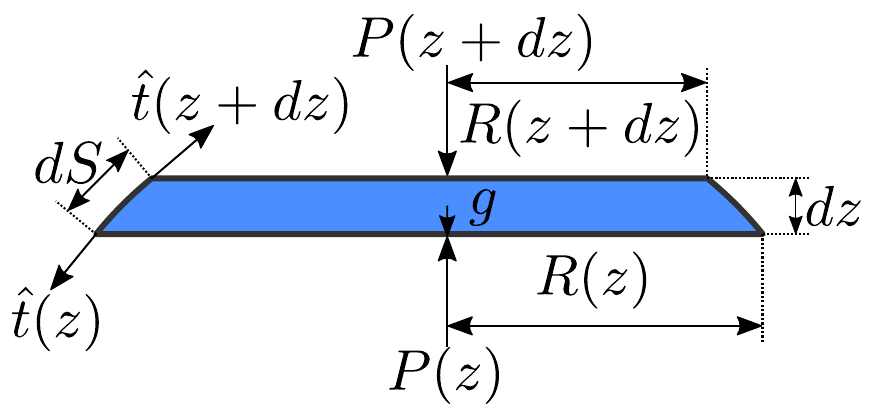}
        \caption{}
        \label{fig:YL_droplet_element}
    \end{subfigure}
    \caption{Schematic used for derivation of Young-Laplace equation.}
    \label{fig:droplet_element}
\end{figure}

Let the profile of the droplet be defined by $\mathcal{\tilde{X}}=(R(z),z)$ (Fig. \ref{fig:droplet_element}).
The differential arc length ($dS$) can be written as:
\begin{eqnarray}
    S' &= \sqrt{R'^2+1}.
\end{eqnarray}
The tangent($\hat{t}(z)$) is given as:
\begin{eqnarray}
    \hat{t}(z) &=& \frac{d\mathcal{\tilde{X}}/dz}{dS/dz}
    = \frac{R'(z)\hat{r}+\hat{k}}{\sqrt{R'+1}}.
    \label{eq:tangent}
\end{eqnarray}
Considering a differential fluid element of thickness $dz$ and balancing the forces acting on it:
The hydrostatic pressure force acting on the differential fluid element is given as $dP = \rho g dz$.\\
\begin{multline}
    \left[ \sigma \hat{t}(z+dz)2\pi R(z+dz) - \sigma\hat{t}(z)2\pi R(z) \right]\cdot\hat{k} + P(z)\pi R^2(z) \\- P(z+dz)\pi R^2 (z+dz) - \rho g \pi R^2(z)dz = 0.
    \label{eq:force_balance}
\end{multline}
Expanding, simplifying and neglecting the $dz^2$ term in eq.\ref{eq:force_balance}, we get:
\begin{equation}
    2\sigma\left[\hat{t}R' dz + \hat{t}' Rdz\right]\cdot \hat{k} - 2PRR'dz = 0.
    \label{eq:force_bal_simplified}
\end{equation}
Eq. \ref{eq:force_bal_simplified} can be simplified as:
\begin{equation}
    \frac{1}{\sqrt{R'^2 + 1}}-\frac{RR''}{\left(R'^2 +1\right)^{3/2}} = \frac{PR}{\sigma}
    \label{eq:static_gov_eq}
\end{equation}

 
 We know that pressure at any position($z$) can be decomposed into the hydrostatic pressure and the curvature \cite{vafaei2005theoretical} as:
 \begin{equation}
     P = P{(z)} + \rho g (H-z).
     \label{eq:Appendix_pressure}
 \end{equation}
 Eq. \ref{eq:static_gov_eq} can be generalized for the droplet as:

 \begin{multline}
\frac{1}{\sqrt{R'(z)^2 + 1}}
- \frac{R(z)R''(z)}{\left(R'(z)^2 + 1\right)^{3/2}} \\
= \frac{R(z)}{\sigma} \left(
   \Bigg[
      \frac{\sigma}{R(z)}
      \left(
        \frac{1}{\sqrt{(R'(z))^2+1}}
        + \frac{R(z)R''(z)}{\big((R'(z))^2+1\big)^{3/2}}
      \right)
   \Bigg]_{z=H}
   + \rho g (H-z)
\right)
\label{eq:static_gov_eq_modified}
\end{multline}

Using the set of dimensionless variables
\begin{equation*}
R^*=\frac{R}{R_p} \text{, } z^*=\frac{z}{R_p} \text{, } \frac{dR^*}{dz*}=\frac{dR}{dz}\text{, }\frac{d^2R^*}{{dz^*}^2} = R_p\frac{d^2R}{dz^2} \text{, and } Bo = \frac{\rho g R_p^2}{\sigma},
\end{equation*}
Eq. \ref{eq:static_gov_eq_modified} is non-dimensionalized as:
 \begin{equation}
     {R^*}'' = \frac{{R^*}'^2 +1}{{R^*}}-\left({P^*}(H^*) + Bo(H^* - z^*)\right)({R^*}'^2 + 1)^{3/2}
     \label{eq:static_gov_eq_ND_final}
 \end{equation}
 Eq. \ref{eq:static_gov_eq_ND_final} is a second-order ODE that shall be solved with the following initial and boundary conditions:
 \begin{eqnarray*}
     \text{At }z^*=0\text{, } {R^*}(0) = 1 \text{, and }{R^*}'(0) = cot(\theta)\\
     \text{At }z^*=H^*\text{, } {R^*}(H^*) \to 0.
 \end{eqnarray*}
 Using the above conditions for a given $Bo$, $R^* \to 0$ is targeted for an initial value of $\theta$ using shooting method.

\section{Derivation of the Young-Laplace equation for a pinned droplet under the influence of gravity held on a vertical cylinder. \label{appendix:YL_equation_cylinder}}
\begin{figure}[h!]
    \centering
        \centering
        \includegraphics[width=2in]{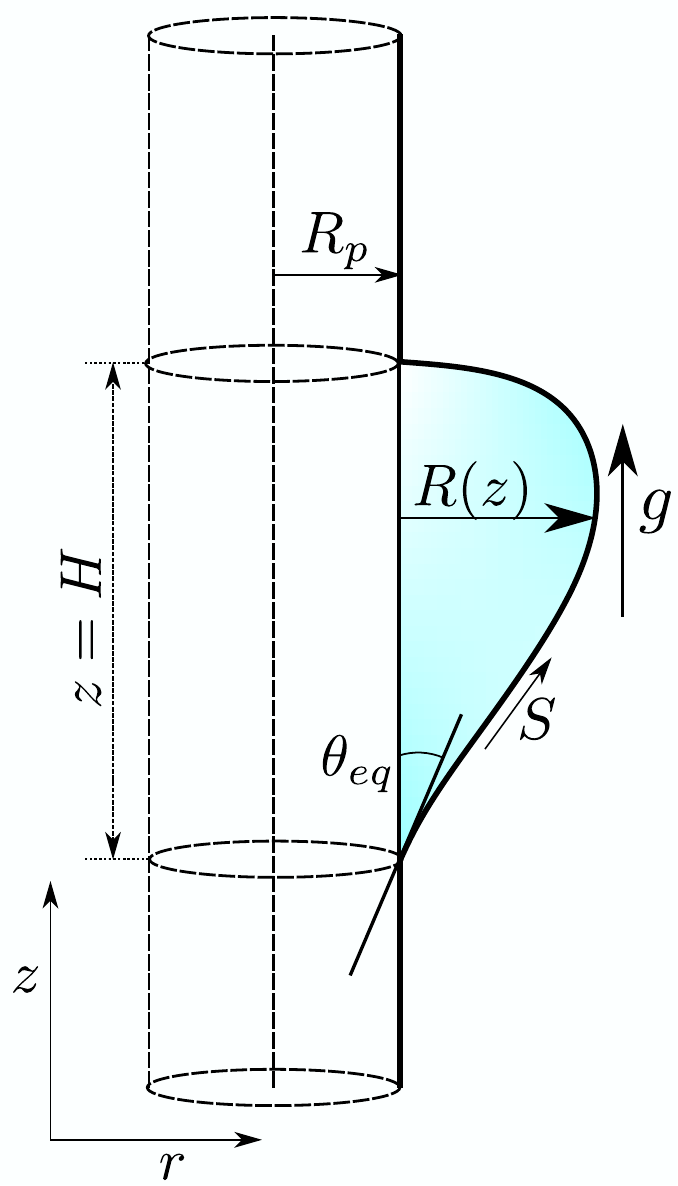}
    \caption{Schematic used for derivation of Young-Laplace equation.}
    \label{fig:sch_cyl_der}
\end{figure}
Consider a droplet is held on a vertical with gravity acting in the z-direction as shown in Fig \ref{fig:sch_cyl_der}.
Making the sign of gravity term $\rho g \pi R(z)^2dz$ in Eq. \ref{eq:force_balance} positive, rest ramains same till Eq. \ref{eq:force_bal_simplified}.
 
 As discussed pressure at any position($z$) can be decomposed into the hydrostatic pressure and the curvature \cite{vafaei2005theoretical} as:
 \begin{equation}
     P = P{(z)} + \rho gz.
     \label{eq:Appendix_pressure_v}
 \end{equation}
 Eq. \ref{eq:static_gov_eq} can be generalized for the droplet as:
 \begin{multline}
\frac{1}{\sqrt{R'(z)^2 + 1}}
- \frac{R(z)R''(z)}{\left(R'(z)^2 + 1\right)^{3/2}} \\
= \frac{R(z)}{\sigma} \left(
   \Bigg[
      \frac{\sigma}{R(z)}
      \left(
        \frac{1}{\sqrt{R'(z)^2+1}}
        + \frac{R(z)R''(z)}{(R'(z)^2+1)^{3/2}}
      \right)
   \Bigg]_{z=0}
   + \rho g z
\right)
\label{eq:static_gov_eq_modified_v}
\end{multline}

Using the set of dimensionless variables
\begin{equation*}
R^*=\frac{R}{R_p} \text{, } z^*=\frac{z}{R_p} \text{, } \frac{dR^*}{dz*}=\frac{dR}{dz}\text{, }\frac{d^2R^*}{{dz^*}^2} = R_p\frac{d^2R}{dz^2} \text{, and } Bo = \frac{\rho g R_p^2}{\sigma},
\end{equation*}
Eq. \ref{eq:static_gov_eq_modified_v} is non-dimensionalized as:
 \begin{equation}
     {R^*}'' = \frac{{R^*}'^2 +1}{{R^*}}-\left({P^*}(z^*=0) + z^*Bo \right)({R^*}'^2 + 1)^{3/2}
     \label{eq:static_gov_eq_ND_final_v}
 \end{equation}
 Eq. \ref{eq:static_gov_eq_ND_final_v} is a second-order ODE that shall be solved with the following initial and boundary conditions:
 \begin{eqnarray*}
     \text{At }z^*=0\text{, } {R^*}(0) = 1 \text{, and }{R^*}'(0) = tan(\theta)\\
     \text{At }z^*=H^*\text{, } {R^*}(H^*) \to 1.
 \end{eqnarray*}
 Using the above conditions for a given $Bo$, $R^* \to 0$ is targeted for an initial value of $\theta$ using shooting method.

\section{\label{app:pseudocode} Pseudo code for pinning scheme implementation }
\begin{algorithm}[H]
\caption{Pinning scheme implementation}
\begin{algorithmic}[1]
\REQUIRE Input parameters: Pinning curve $\mathcal{C}(X(s),Y(s),Z(s))$
    \STATE Identify particles of near substrate($\Omega_{ns}$): $a\in \Omega_{ns}$ if $a \in \Omega_l$ and $\mathcal{N}_a \in \Omega_s$
\STATE Initialize $\vec{\mathbf{n}}_a^{(p)} \leftarrow \mathbf{0}$
    \IF{$a\in \Omega_{ns} \text{ and } S_a\leq 0.95$}
        \STATE Compute $\mathbf{r}_{pa}$ such that $|\mathbf{r}_{pa}|$ is the shortest distance between $a$ and $\mathcal{C}(X(s),Y(s),Z(s))$.
        \STATE Compute $\hat{\mathbf{n}}_a^{(p)}$ using eq. \ref{eq:pin_treatment}
        \STATE Compute $\theta_i$ using eq. \ref{eq:instantaneous_CA}
        \IF{$\theta_i \leq \theta_{min}$}
            \STATE Compute $\hat{\mathbf{n}}_a^{\infty}$ using eq. \ref{eq:CA_normal}.
            \STATE Compute $\hat{\mathbf{n}}_{\theta_{min}}$ using eq. \ref{eq:de/wetting_normal} 
            \STATE Update $\hat{\mathbf{n}}_a^{(p)} = \hat{\mathbf{n}}_{\theta_{min}}$
        \ELSIF{$\theta_i \geq \theta_{max}$}
            \STATE Compute $\hat{\mathbf{n}}_a^{\infty}$ using eq. \ref{eq:CA_normal}.
            \STATE Compute $\hat{\mathbf{n}}_{\theta_{max}}$ using eq. \ref{eq:de/wetting_normal}
        \STATE Update $\hat{\mathbf{n}}_a^{(p)} = \hat{\mathbf{n}}_{\theta_{max}}$
        \ENDIF
    \ENDIF
    
    \IF{$a\in \Omega_{ns} \text{ and } S_a > 0.95$}
        \STATE Compute $\hat{\mathbf{n}}_a^{(p)}$ as eq. \ref{eqn:inside_sph_sum}
    \ELSE
        \STATE $\textbf{return }$ $\hat{\mathbf{n}}_a$
    \ENDIF
    \STATE Update $\hat{\mathbf{n}}_a^{(p)}$ as eq. \ref{eq:blending}
    \STATE Update $\hat{\mathbf{n}}_a=\hat{\mathbf{n}}_a^{(p)}$
    \IF{$a\in \Omega_s\text{ and } b \in \Omega_{ns}$}
        \STATE Assign $\hat{\mathbf{n}}_a$ as eq. \ref{eq:solid_assign_normal}
    \ENDIF

    \STATE Compute $\tilde{\kappa}_a = \frac{1}{2} \sum_b \frac{m_b}{\rho_b} \left(\hat{\mathbf{n}}_b - \hat{\mathbf{n}}_a\right) \cdot \tilde{\nabla}W_{ab}$
    \IF{$a\in\Omega_{ns}$}
    \STATE Compute $ \tilde{\kappa}_a = \frac{\sum_{b \in \Omega_{csf}} \frac{m_b}{\rho_b} \kappa_{b}W_{ab}}{\sum_{b \in \Omega_{csf}} \frac{m_b}{\rho_b}W_{ab}}$
    \ENDIF
    
\STATE Compute acceleration due to surface tension $(\mathbf{f}_a^s)$ as eq. \ref{eq:surface_tension_Blank}

\STATE $\textbf{return }$ $\hat{\mathbf{n}}_a$
\end{algorithmic}
\end{algorithm}

%
\section*{Acknowledgements}
The work is supported by the Science Education and Research Boar (SERB)
through the Core Research Grant (CRG) number CRG/2022/008634, the
Startup Research Grant (SRG) number SRG/2022/000436 and the Ministry of Shipping and Inland Waterways, India.

 \bibliographystyle{elsarticle-num-names} 
 \bibliography{bibliography}





\end{document}